\documentclass[mnsc,nonblindrev]{informs3rad}
\newif\ifMS
\MStrue

\OneAndAHalfSpacedXI
\TheoremsNumberedThrough 
\EquationsNumberedThrough
\usepackage{xfrac}
\usepackage{xcolor}
\usepackage{bbm}
\usepackage{booktabs}
\usepackage{subfig}
\usepackage[ruled,vlined,linesnumbered]{algorithm2e}
\SetNoFillComment
\usepackage[breakable, skins]{tcolorbox}
\DeclareRobustCommand{\mybox}[2][gray!10]{%
\begin{tcolorbox}[
        left=0.5pt,
        right=0.5pt,
        top=0.5pt,
        bottom=0.5pt,
        colback=#1,
        colframe=#1,
        width=\dimexpr\textwidth\relax, 
        enlarge left by=1mm,
        boxsep=3pt,
        arc=2pt,outer arc=2pt,
        code={\OneAndAHalfSpacedXII}
        ]
        #2
        
\end{tcolorbox}
}
\usepackage{csquotes}
\makeatletter
\renewenvironment*{displayquote}
  {\begingroup\setlength{\leftmargini}{0.2cm}\csq@getcargs{\csq@bdquote{}{}}}
  {\csq@edquote\endgroup}
\makeatother

\usepackage{mathtools}
\definecolor{cornellred}{rgb}{0.7, 0.11, 0.11}
\definecolor{maroon}{rgb}{0.52, 0, 0}
\definecolor{dgreen}{rgb}{0.0, 0.5, 0.0}
\definecolor{ballblue}{rgb}{0.13, 0.67, 0.8}
\definecolor{royalblue(web)}{rgb}{0.25, 0.41, 0.88}
\definecolor{bleudefrance}{rgb}{0.19, 0.55, 0.91}
\definecolor{royalazure}{rgb}{0.0, 0.22, 0.66}
\usepackage{hyperref}
\hypersetup{
	colorlinks = true,
	linkcolor=royalazure,
	urlcolor=royalazure,
	citecolor=royalazure,
	linkbordercolor = {white}
}
\usepackage{cleveref}
\usepackage{enumitem}
\usepackage{multirow}
\usepackage{pgf,tikz,pgfplots}
\pgfplotsset{compat=1.15}
\usetikzlibrary{arrows}
\usepackage{pgfplots}
\usetikzlibrary{patterns}
\usepgfplotslibrary{fillbetween}
\usetikzlibrary{intersections}

\usetikzlibrary{arrows, decorations.markings}

\tikzstyle{vecArrow} = [thick, decoration={markings,mark=at position
	1 with {\arrow[semithick]{open triangle 60}}},
	double distance=1.4pt, shorten >= 5.5pt,
	preaction = {decorate},
postaction = {draw,line width=1.4pt, white,shorten >= 4.5pt}]
\tikzstyle{innerWhite} = [semithick, white,line width=1.4pt, shorten >= 4.5pt]

	\usepackage[authoryear,round]{natbib}

\ifMS
\else
	\DeclareMathOperator{\argmax}{argmax}
	
\fi

\newcommand{\prob}[2][]{\text{\bf Pr}\ifthenelse{\not\equal{}{#1}}{_{#1}}{}\!\left[{\def\givenn{\middle|}#2}\right]}
\newcommand{\expect}[2][]{\text{\bf E}\ifthenelse{\not\equal{}{#1}}{_{#1}}{}\!\left[{\def\givenn{\middle|}#2}\right]}

\newcommand{\indicator}[1]{{\mathbbm{1}\left\{ #1 \right\}}}

\newcommand{\condition}{\,\mid\,}

	\DeclarePairedDelimiterX{\set}[1]\{\}{#1}
	\let\Pr\relax
	\DeclarePairedDelimiterXPP{\Pr}[1]{\mathbb{P}}[]{}{#1}
	\DeclarePairedDelimiterXPP{\Ex}[1]{\mathbb{E}}[]{}{#1}

\newcolumntype{P}[1]{>{\centering\arraybackslash}c{#1}}

\newcommand*{\rom}[1]{\expandafter\romannumeral #1}
\newcommand{\Rom}[1]{\uppercase\expandafter{\romannumeral #1\relax}}

\newcommand{\primed}{^\dagger}
\newcommand{\doubleprimed}{^\ddagger}

\newcommand{\weight}{w}
\newcommand{\weightHat}{\hat{\weight}}

\newcommand{\rider}{i}
\newcommand{\Rider}{U}

\newcommand{\driver}{j}
\newcommand{\Driver}{V}

\newcommand{\Driveri}{\Driver^{(\mspair)}}

\newcommand{\xbf}{\mathbf{x}}
\newcommand{\zbf}{\mathbf{z}}
\newcommand{\ybf}{\mathbf{y}}
\newcommand{\lambdabf}{\boldsymbol{\lambda}}
\newcommand{\thetabf}{\boldsymbol{\theta}}
\newcommand{\gammabf}{\boldsymbol{\gamma}}

\newcommand{\Lag}{\mathcal{L}}

\newcommand{\N}{\mathbb{N}}

\newcommand{\mspair}{l}

\newcommand{\M}{N}
\newcommand{\permu}{\pi}

\newcommand{\lusage}{\xi}

\newcommand{\PRMCA}{\texttt{PR-MCA}}

\newcommand{\config}{c}
\newcommand{\Config}{C}
\newcommand{\threeindex}{\rider,\config,\driver}
\newcommand{\wi}{\tau}
\newcommand{\totalweight}{T}
\newcommand{\twoindex}{\driver,\wi}
\newcommand{\comsumption}{\eta}
\newcommand{\alloc}{\xi}
\newcommand{\alloci}{\alloc_{\threeindex}}
\newcommand{\allocs}{\boldsymbol{\alloc}}
\newcommand{\pibf}{\boldsymbol{\pi}}
\newcommand{\mubf}{\boldsymbol{\mu}}
\newcommand{\phibf}{\boldsymbol{\phi}}
\newcommand{\psibf}{\boldsymbol{\psi}}
\newcommand{\iotabf}{\boldsymbol{\iota}}
\newcommand{\lambdai}{\lambda_{\rider}}
\newcommand{\thetai}{\theta_{\driver}}
\newcommand{\mui}{\mu_{\threeindex}}
\newcommand{\phii}{\phi_{\rider,\config}}
\newcommand{\psii}{\psi_{\threeindex}}
\newcommand{\starred}{^*}
\newcommand{\ked}{^{(k)}}
\newcommand{\KED}{^{(K)}}
\newcommand{\sed}{^{(s)}}

\newcommand{\kminused}{^{(k-1)}}
\newcommand{\KMINUSED}{^{(K-1)}}
\newcommand{\sminused}{^{(s-1)}}
\newcommand{\alphabf}{\boldsymbol{\alpha}}
\newcommand{\betabf}{\boldsymbol{\beta}}
\newcommand{\comsumptioni}{\comsumption_{\twoindex}\kminused}
\newcommand{\taup}{{\tau\primed}}
\newcommand{\taupp}{{\tau\doubleprimed}}
\newcommand{\Eta}{H}
\newcommand{\cumcomsumption}{\Eta}
\newcommand{\cumcomsumptioni}{\cumcomsumption_{\twoindex}}

\newcommand{\ALG}{\texttt{ALG}}

\newcommand{\weightbf}{\mathbf{\weight}}

\newcommand{\revcolor}[1]{{\color{black}#1}}
\newcommand{\ydrevcolor}[1]{{\color{black}#1}}

\MANUSCRIPTNO{MS-RMA-21-02783}

\begin{document}
%%%%%%%%%%%%%%%%

%\setcounter{page}{1}
%\DOI{10.1287/xxxx.0000.0000}%

\RUNAUTHOR{Feng and Niazadeh}

\RUNTITLE{Batching and Optimal Multi-stage Bipartite Allocations}

% \TITLE{Bundling Information Goods of Decreasing Value}
\TITLE{Batching and Optimal Multi-stage Bipartite Allocations}

\ARTICLEAUTHORS{%
\AUTHOR{Yiding Feng}
\AFF{Microsoft Research New England, Cambridge, MA, \EMAIL{yidingfeng@microsoft.com}}
\AUTHOR{Rad Niazadeh}
\AFF{University of Chicago Booth School of Business, Chicago, IL, \EMAIL{rad.niazadeh@chicagobooth.edu}}
} 

\ABSTRACT{%
    In several applications of real-time matching of demand to supply in online marketplaces, the platform allows for some latency to batch the demand and improve the efficiency of the resulting matching. Motivated by these applications, we study the optimal trade-off between batching and inefficiency in the context of designing robust online allocations. As our base model, we consider $K$-stage variants of the classic vertex weighted bipartite b-matching in the adversarial setting, 
    where online vertices arrive stage-wise and in $K$ batches --- in contrast to online arrival. Our main result for this problem is an optimal $\left(1-(1-1/K)^K\right)$-competitive (fractional) matching algorithm, improving the classic $(1-1/e)$ competitive ratio bound known for its online variant~\citep{MSVV-07,AGKM-11}. \revcolor{We also extend this result to the rich model of multi-stage configuration allocation with free-disposals~\citep{DHKMQ-16}, which is motivated by the display advertising application in the context of video streaming platforms.}

    Our main technique at high-level is developing algorithmic tools to vary the trade-off between ``greedy-ness'' and ``hedging" of the matching algorithm across stages. We rely on a particular family of convex-programming based matchings that distribute the demand in a specifically balanced way among supply in different stages, while carefully modifying the balancedness of the resulting matching across stages. More precisely, we identify a sequence of \emph{polynomials with decreasing degrees} to be used as strictly concave regularizers of the maximum weight matching linear program to form these convex programs. At each stage, our fractional multi-stage algorithm returns the corresponding regularized optimal solution as the matching of this stage (by solving the convex program). By providing structural decomposition of the underlying graph using the optimal solutions of these convex programs and recursively connecting the regularizers together, we develop a new multi-stage primal-dual framework to analyze the competitive ratio of this algorithm. We further show this algorithm is optimal competitive, even in the unweighted case, by providing an upper-bound instance in which no online algorithm obtains a competitive ratio better than $\left(1-(1-1/K)^K\right)$. \revcolor{For the extension to multi-stage configuration allocation, we introduce a novel extension of our regularized convex program that provides separate regularization at different "price levels". Despite the lack of a relevant graph decomposition in this extension, in contrast to our base model, we show how we can directly use convex duality to set up a primal-dual analysis framework for our new algorithm.}
}%

%\KEYWORDS{deterministic inventory theory; infinite linear programming duality;
%  existence of optimal policies; semi-Markov decision process; cyclic schedule}

%\KEYWORDS{butter, margarine, silliness} 

\maketitle
%%%%%%%%%%%%%%%%%%%%%%%%%%%%%%%%%%%%%%%%%%%%%%%%%%%%%%%%%%%%%%%%%%%%%%

% Samples of sectioning (and labeling) in MNSC
% NOTE: (1) \section and \subsection do NOT end with a period
%       (2) \subsubsection and lower need end punctuation
%       (3) capitalization is as shown (title style).
%
%\section{Introduction.}\label{intro} %%1.
%\subsection{Duality and the Classical EOQ Problem.}\label{class-EOQ} %% 1.1.
%\subsection{Outline.}\label{outline1} %% 1.2.
%\subsubsection{Cyclic Schedules for the General Deterministic SMDP.}
%  \label{cyclic-schedules} %% 1.2.1
%\section{Problem Description.}\label{problemdescription} %% 2.

% Text of your paper here

\newpage

\newcommand{\cratio}[1]{\Gamma(#1)}
\newcommand{\cratioexp}{1-\left(1-\frac{1}{K}\right)^K}
\section{Introduction}
\label{sec:intro}
\revcolor{
With the pervasiveness of online platforms and web-based services, internet advertising has grown to become a major component of today's economy. Worldwide spending of internet advertising stood at an estimated 521 billion U.S. dollars in 2021 --- a figure that is forecast to constantly increase in the coming years, reaching a total of 876.1 billion U.S. dollars by 2026.\footnote{See \url{https://www.statista.com/statistics/237974/online-advertising-spending-worldwide/}.} The combination of the economic impact and the inherent flexibility of internet advertising has helped with its prevalence as a modern market. At the same time, it has also given rise to numerous revenue management challenges and paradigms, some of which pertain to real-time algorithmic allocation of inventories of ad opportunities to advertisers in various search and display internet advertising scenarios, e.g., see \citet{MSVV-07,FKMMP-09,BFMM-14,DHKMQ-16}. 

As digital consumption behaviors change, and new technologies provide new ways for users to interact with social media platforms and publishers, traditional forms of real-time internet ad allocations  also evolve. A prominent example of this sort, which is one of the main motivating application behind our paper, is \emph{in-stream video advertising}. Nowadays, in-stream video advertising is rapidly growing~\footnote{According to \citet{Deloitte}, 80\% of U.S. consumers are now paying for streaming video services, up from 73\% before the COVID-19 pandemic began. Using ad spending as an indicator, video advertising will likely continue to grow for the foreseeable future. Video ad spending in the U.S. is increased from 8.92 billion U.S. dollars in 2017 to 55.34 billion U.S. dollars in 2021; it is also projected to reach as high as 78.45 billion U.S. dollars by 2023. See \url{https://www.statista.com/statistics/256272/digital-video-advertising-spending-in-the-us/}.} and is playing an essential role in the business model of online video sharing platforms such as YouTube and Vimeo, on-demand/live streaming platforms such as Netflix, Roku, and Twitch, and even social video streaming platforms such as TikTok. In this context, when a user starts watching a new video session (e.g., on YouTube), an advertising opportunity to display video-ads, and hence allocating user impressions to advertisers and charging them based on per-impression rates, is created. Video-ads refer to all possible ad formats in this context such as in-stream advertising videos of different lengths, as well as the banner-like overlay and text- or image-based ads that appear over a video. See \Cref{fig:ad-formats} for specific video-ad formats that are currently used in YouTube.\footnote{See \url{https://support.google.com/youtube/answer/2375464?hl=en}.}
Given these ad format varieties, the real-time ad allocation for a sequence of arriving users is done by displaying ads of one or multiple advertisers to each arriving user (or equivalently, each video session). It is important to add that such a dynamic environment is highly non-stationary (in terms of users that start watching videos) and hence enabling the platform to run a robust and prior free ad allocation policy is a common practice.

\captionsetup[subfloat]{captionskip=0pt}
\begin{figure}[hbt!]
	\centering
	\subfloat[
   Skippable in-stream ads
	]{
\includegraphics[scale=0.32]{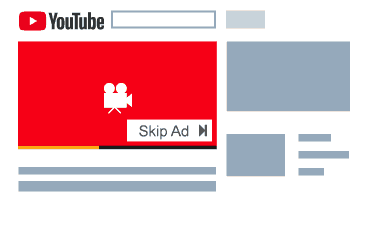}
	}
	\subfloat[Non-skippable in-stream ads]{
\includegraphics[scale=0.32]{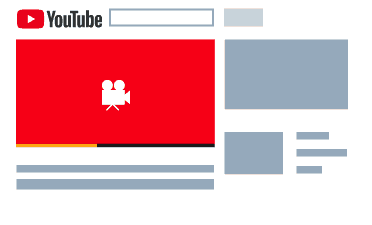}
	}
	\subfloat[
   Bumper short ads
	]
	{
\includegraphics[scale=0.32]{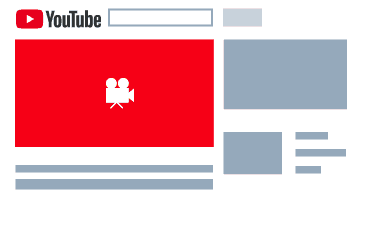}
	}
	\\
	\centering
	\subfloat[
   Overlay ads
	]{
\includegraphics[scale=0.34]{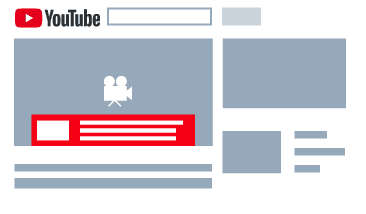}
	}
	\subfloat[
   Clickable sponsored cards
	]{
\includegraphics[scale=0.34]{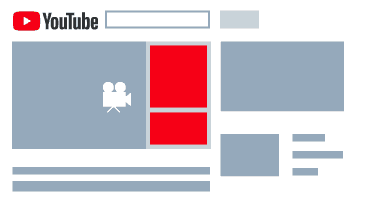}
	}
	\vspace{1mm}
	\caption{
Different formats of video advertising in YouTube streaming; These formats can be used at different time-stamps of a YouTube video, and differ in their lengths, designs, and how users interact with them.}
	\label{fig:ad-formats}
\end{figure}

While in-stream video advertising is a real-time application, it still admits a natural form of allowance for latency in displaying ads to the users. In particular, as several of the video-ads show up in the middle (or at the end) of users' video sessions, an ad allocator can delay the time of its decision for a session (or the current chunk of a long video session) from the beginning of the video to the first time-stamp that a video-ad is required to be displayed. This flexibility provides an immense opportunity to \emph{batch} the arriving video sessions, where a batch consists of the most recent group of users that their ad allocation decisions can be made in the same time frame. Given these batches, the goal of the ad allocator is to make allocation decisions as each batch of users arrives --- in contrast to fully online decisions as each single user arrives --- in order to maximize the revenue from the advertisers. This batching technique is indeed practiced in the web-based video advertising industry~\citep{personal}.

Besides video advertising, there are also other resource allocation applications with allowance for latency, e.g., matching delivery requests to dispatching centers in Amazon or matching user jobs to computing servers in Azure cloud computing platform. Even in fast-paced applications such as sponsored search, the ad decisions for the arriving search queries can still be delayed in the order of milliseconds, and hence batching is possible. In all of these applications, while dividing the entire real-time sequence of demand requests into less number of batches (of larger sizes) requires more latency, it clearly helps the platform to make improved allocation decisions in terms of its objective as it has more information about the sequence of arriving demand requests.\footnote{To see numbers in practice, for example, ridesharing platforms such as Uber~\citep{Uber}, Lyft~\citep{Lyft} and DiDi~\citep{ZHMWZFGY-17} have reported substantial efficiency improvements due to shifting from match-as-you-go to batching-and-matching. Same results are reported in other corners of gig-economy such as cloud computing markets~\citep{AWS} and online food delivery~\citep{joshi2021batching}. Notably, batching not only helps with improving the efficiency, it can also help with reducing non-operational forms of cost such as waste and environmental cost~\citep{AWSwaste}.}}

\vspace{2mm}
\revcolor{\noindent{\textbf{Our Theoretical Lens~}} Motivated by the batching-and-allocation paradigm in practical applications such as video advertising, and also to theoretically characterize the the inherent trade-off between batching and inefficiency in robust and prior-free dynamic allocations, we revisit the literature on (adversarial) online bipartite allocations, originated from the seminal work of \cite{KVV-90}, where the goal is to allocate the arriving demand vertices (online side) to the available supply vertices (offline side). We study several canonical settings in this literature under the \emph{multi-stage} arrival model, where the online vertices are partitioned into $K>0$ fixed number of batches and arrive over $K$ stages (one batch per each stage). In this model, once a batch arrives, every information in the problem instance related to this batch (which depends on the problem's specifications) is revealed to the multi-stage allocation algorithm. The algorithm then makes irrevocable stage-wise decisions by allocating the arriving batch of vertices to the offline side at each stage.

As our base model, we consider the \emph{multi-stage vertex weighted b-matching}~\citep{AGKM-11,DJK-13}. In this simple problem, online vertices arrive in $K>0$ batches. Then the multi-stage algorithm matches the arriving batches to the offline side while respecting the capacity constraints on the offline side. The goal is to maximize the total weight of the matching. 

Motivated by the video advertising application, we also consider an extension model called \emph{multi-stage configuration allocation}~\citep{DHKMQ-16}, which generalizes a rich set of multi-stage resource allocation problems including vertex-weighted b-matching, budgeted allocation (also known as ``AdWords''~\citep{MSVV-07}), and edge-weighted online matching with free-disposal~\citep{feldman2009online}. In this problem, we have online users (each corresponding to a new video session) that arrive in $K>0$ batches and offline advertisers that have ads to be displayed to these users so that they receive impressions from the users. Once a batch arrives, the multi-stage allocation algorithm assigns a feasible ``configuration'' to each user in that batch. Here, a configuration can refer to an arrangement of video-ads of various lengths and formats for a subset of advertisers, showing up at different time-stamps of the user's video session.\footnote{It can also be the case that a long video session is divided into chunks, and then each chunk is assigned a separate configuration, i.e., is treated as a new video session.}  Each assigned configuration yields a certain number of impressions to each advertiser and charges the advertiser at a certain price per impression. Depending on advertisers' long-term contracts, each advertiser is willing to pay for a certain number of impressions during the decision making horizon. Given this model, the goal is to maximize the revenue (i.e., total collected prices) in the finite decision making horizon. Particularly important in this model, the set of feasible configurations can model various complex user-level constraints.\footnote{\revcolor{To name a few, our model can capture exclusion (e.g., competing advertisers can not be shown in the same video), frequency capping (e.g., similar ads should not be shown back to back in the same video), and bundling (e.g., when advertisers require that all or none of a set of related ads be shown in the same video).}} }

\revcolor{We measure the performance of our algorithms by their \emph{competitive ratio}, which is the worst-case ratio between the expected objective value of the algorithm and the in-hindsight optimal solution --- also known as the optimal offline benchmark. Considering different degrees of batching by changing the number of batches $K$ and their sizes, one extreme regime of our problem is the fully offline setting when there is only a single large batch containing all the demand vertices. In this extreme regime one can achieve the performance of the optimum offline benchmark. The other extreme regime is the fully online setting when each online demand vertex arrives essentially as a batch of size one (no batching). Note that any competitive ratio lower-bound for the fully online setting establishes a lower-bound for the multi-stage setting, as any algorithm in the fully online setting can be used in the multi-stage setting by simply considering any arbitrary (synthetic) arrival order over vertices in each batch. We now ask the following research question in this paper:

\vspace{2mm}
\begin{displayquote}
\emph{In the middle-ground multi-stage regime of bipartite allocation problems with batch arrivals (in contrast to fully online and offline regimes), when we have $K>0$ number of batches, what are the optimal competitive ratios achievable by multi-stage algorithms in these problems?}
\end{displayquote}
\vspace{2mm}}

\revcolor{\smallskip
\noindent{\textbf{Our Contributions~}}
 The main contribution of this paper is the following result.
 
\vspace{2mm}
\mybox{
\begin{displayquote}
{\textbf{[Main Result]}~\emph{For the multi-stage vertex weighted matching (resp. multi-stage configuration allocation) problem with $K$ batches, we propose \Cref{alg:b-matching} (resp. \Cref{alg:opt}) that achieves a competitive ratio of $\cratio{K}\triangleq\cratioexp$. Moreover, by proposing an upper-bound instance, we show that no $K$-stage algorithm, fractional or integral, can have a competitive ratio better than $\cratio{K}$ in these problems.}}
\end{displayquote}}
\vspace{1mm}
\noindent The competitive ratio function $\cratio{K}\in(1-1/e,1]$ is monotone decreasing in $K$. It is equal to $1$ at $K=1$
and converges to $1-1/e$ as $K$ goes to infinity.\footnote{\revcolor{In fact, a simple asymptotic expansion shows that $\cratio{K}=\left(1-\frac{1}{e}\right)+\frac{1}{2eK}+O\left(\frac{1}{K^2}\right)$.}}. For the special case of two-stage vertex weighted matching ($K=2$), our algorithm is $3/4$-competitive, which matches the result of \cite{feng2020two}. Also, for three stages ($K=3$), it has a competitive ratio of $19/27\approx 0.704$. To the best of our knowledge, this is the first bound beating $1-1/e\approx 0.63$ in the $K$-stage bipartite matching problems (weighted or unweighted) for any $K>2$. While our results pertains to fractional allocations, they can easily be extended to the integral version under the large budget assumption. In fact, in the multi-stage configuration allocation problem (and all of its special cases) one can use a simple independent rounding schemes after slightly decreasing the capacities, and show the resulting multi-stage integral algorithm has a competitive ratio no smaller than $\cratio{K}-O\left(\sqrt{\sfrac{\log(B_{\text{min}})}{B_{\text{min}}}}\right)$ using standard concentration bounds (e.g., see \cite{DJSW-11} for technical details). Here, $B_{\text{min}}$ should be interpreted as the minimum capacity/required number of impressions of each offline node/advertiser.}

% We further extend our result to the fractional multi-stage vertex weighted b-matching and multi-stage budgeted allocation problems. We obtain these results by proposing natural adaptations of the previous algorithm for vertex weighted matching (\Cref{alg:b-matching integral} and \Cref{alg:adwords fractional}).  Following the same approach as in the vertex weighted matching problem, we show these adaptations are also $\cratio{K}$-competitive for these generalizations/variations of our base problem. Switching to integral allocations, we also design a simple rounding scheme for multi-stage vertex weighted b-matching, and show the resulting multi-stage integral algorithm obtains a competitive ratio no smaller than $\cratio{K}-O\left(\sfrac{1}{B_{\textrm{min}}}\right)$, where $B_{\text{min}}$ is the minimum capacity of an offline node. Finally, we design a different yet another simple rounding scheme for multi-stage budgeted allocation, and show the resulting multi-stage integral algorithm has a competitive ratio no smaller than $\cratio{K}-O\left(\sqrt{\sfrac{\log(B_{\text{min}})}{B_{\text{min}}}}\right)$. Here, $B_{\text{min}}$ should be interpreted as the maximum budget over bid ratio. See \Cref{tab:summary} for a summary of our results and how they are compared with the previous work.
% \input{tex/result-table}

\smallskip
\noindent{\textbf{Overview of the Main Technique~}} 
\revcolor{Considering the multi-stage vertex weighted matching problem first,} any greedy algorithm allocates the arriving batch of demand requests to the available pool of offline supply at each stage $k\in[K]$, so as to maximize the matching weight of the current stage. Such an allocation possibly suffers from the lack of enough \emph{balancedness} on the supply side --- a property that can help the algorithm to remain competitive against the adversary in future stages. To fix this issue, we replace greedy with a more balanced allocation that (i) takes into account the matching weight of each stage, and (ii) simultaneously provides \emph{hedging} against future stages in a controlled fashion.
For the special case of online arrivals, this simple idea is the essence of the celebrated BALANCE algorithm~\citep{AGKM-11}, which is optimal competitive under online arrivals. 

To systematically extend the above balanced matching idea under the online setting to the multi-stage setting --- where the algorithm can basically match multiple vertices of the arriving batch to the offline side --- we propose using specific \emph{stage-dependent convex programs} and output their optimal solution in each stage as the fractional matching of that stage. Given the vector of remaining capacities at the beginning of each stage, the convex program for this stage essentially finds a feasible fractional allocation that maximizes a regularized version of the total weight objective function. In particular, in order to favor more balanced fractional allocations, we choose our regularizers to be \emph{specific} (strictly) concave functions of the final matching degrees (or allocation levels) of the offline vertices. More precisely, we use the primitive function $F_k(x)\triangleq\int_{0}^{x}f_k(y)dy$ to map the normalized fractional degree (or normalized allocation level) of each offline vertex to a penalty term in $[0,1]$ at stage $k$. We then subtract the weighted sum of these penalties from the total weight objective function to construct our regularized objective function in that stage; see the convex program \ref{eq:concave-k-bmatching} used in \Cref{alg:b-matching}.

Considering the above algorithmic construct, one major question is how to pick these strictly concave regularizers to achieve our desired $\cratio{K}$ competitive ratio in $K$ stages. Our intuition suggests that an appropriate multi-stage algorithm requires hedging more in earlier stages and less in later stages, e.g., in the extreme case of the last stage, picking a maximum weight matching is optimal and no balancedness is required. To obtain the desired variation in the balancedness of our allocation, i.e., having a more balanced allocation early on and gradually pushing it towards a greedy allocation in later stages, we aim to incorporate ``more intense'' regularization at first and then gradually shift towards no regularization in the course of $K$ stages. 

\begin{figure}[hbt!]
	\centering
	\subfloat[
    $k = K$
	]{
	\begin{tikzpicture}\begin{axis}[height=4.5cm,width=4.5cm,xmin=0,xmax=1,ymin=0,ymax=1,xtick={0,1},xticklabels={0,1},ytick={0,1},yticklabels={0,1},]
\addplot [draw=black, line width=0.4mm, densely dashed] coordinates {
(0.0050, 0.3697)
(0.0100, 0.3716)
(0.0150, 0.3734)
(0.0200, 0.3753)
(0.0250, 0.3772)
(0.0300, 0.3791)
(0.0350, 0.3810)
(0.0400, 0.3829)
(0.0450, 0.3848)
(0.0500, 0.3867)
(0.0550, 0.3887)
(0.0600, 0.3906)
(0.0650, 0.3926)
(0.0700, 0.3946)
(0.0750, 0.3965)
(0.0800, 0.3985)
(0.0850, 0.4005)
(0.0900, 0.4025)
(0.0950, 0.4045)
(0.1000, 0.4066)
(0.1050, 0.4086)
(0.1100, 0.4107)
(0.1150, 0.4127)
(0.1200, 0.4148)
(0.1250, 0.4169)
(0.1300, 0.4190)
(0.1350, 0.4211)
(0.1400, 0.4232)
(0.1450, 0.4253)
(0.1500, 0.4274)
(0.1550, 0.4296)
(0.1600, 0.4317)
(0.1650, 0.4339)
(0.1700, 0.4360)
(0.1750, 0.4382)
(0.1800, 0.4404)
(0.1850, 0.4426)
(0.1900, 0.4449)
(0.1950, 0.4471)
(0.2000, 0.4493)
(0.2050, 0.4516)
(0.2100, 0.4538)
(0.2150, 0.4561)
(0.2200, 0.4584)
(0.2250, 0.4607)
(0.2300, 0.4630)
(0.2350, 0.4653)
(0.2400, 0.4677)
(0.2450, 0.4700)
(0.2500, 0.4724)
(0.2550, 0.4747)
(0.2600, 0.4771)
(0.2650, 0.4795)
(0.2700, 0.4819)
(0.2750, 0.4843)
(0.2800, 0.4868)
(0.2850, 0.4892)
(0.2900, 0.4916)
(0.2950, 0.4941)
(0.3000, 0.4966)
(0.3050, 0.4991)
(0.3100, 0.5016)
(0.3150, 0.5041)
(0.3200, 0.5066)
(0.3250, 0.5092)
(0.3300, 0.5117)
(0.3350, 0.5143)
(0.3400, 0.5169)
(0.3450, 0.5194)
(0.3500, 0.5220)
(0.3550, 0.5247)
(0.3600, 0.5273)
(0.3650, 0.5299)
(0.3700, 0.5326)
(0.3750, 0.5353)
(0.3800, 0.5379)
(0.3850, 0.5406)
(0.3900, 0.5434)
(0.3950, 0.5461)
(0.4000, 0.5488)
(0.4050, 0.5516)
(0.4100, 0.5543)
(0.4150, 0.5571)
(0.4200, 0.5599)
(0.4250, 0.5627)
(0.4300, 0.5655)
(0.4350, 0.5684)
(0.4400, 0.5712)
(0.4450, 0.5741)
(0.4500, 0.5769)
(0.4550, 0.5798)
(0.4600, 0.5827)
(0.4650, 0.5857)
(0.4700, 0.5886)
(0.4750, 0.5916)
(0.4800, 0.5945)
(0.4850, 0.5975)
(0.4900, 0.6005)
(0.4950, 0.6035)
(0.5000, 0.6065)
(0.5050, 0.6096)
(0.5100, 0.6126)
(0.5150, 0.6157)
(0.5200, 0.6188)
(0.5250, 0.6219)
(0.5300, 0.6250)
(0.5350, 0.6281)
(0.5400, 0.6313)
(0.5450, 0.6344)
(0.5500, 0.6376)
(0.5550, 0.6408)
(0.5600, 0.6440)
(0.5650, 0.6473)
(0.5700, 0.6505)
(0.5750, 0.6538)
(0.5800, 0.6570)
(0.5850, 0.6603)
(0.5900, 0.6637)
(0.5950, 0.6670)
(0.6000, 0.6703)
(0.6050, 0.6737)
(0.6100, 0.6771)
(0.6150, 0.6805)
(0.6200, 0.6839)
(0.6250, 0.6873)
(0.6300, 0.6907)
(0.6350, 0.6942)
(0.6400, 0.6977)
(0.6450, 0.7012)
(0.6500, 0.7047)
(0.6550, 0.7082)
(0.6600, 0.7118)
(0.6650, 0.7153)
(0.6700, 0.7189)
(0.6750, 0.7225)
(0.6800, 0.7261)
(0.6850, 0.7298)
(0.6900, 0.7334)
(0.6950, 0.7371)
(0.7000, 0.7408)
(0.7050, 0.7445)
(0.7100, 0.7483)
(0.7150, 0.7520)
(0.7200, 0.7558)
(0.7250, 0.7596)
(0.7300, 0.7634)
(0.7350, 0.7672)
(0.7400, 0.7711)
(0.7450, 0.7749)
(0.7500, 0.7788)
(0.7550, 0.7827)
(0.7600, 0.7866)
(0.7650, 0.7906)
(0.7700, 0.7945)
(0.7750, 0.7985)
(0.7800, 0.8025)
(0.7850, 0.8065)
(0.7900, 0.8106)
(0.7950, 0.8146)
(0.8000, 0.8187)
(0.8050, 0.8228)
(0.8100, 0.8270)
(0.8150, 0.8311)
(0.8200, 0.8353)
(0.8250, 0.8395)
(0.8300, 0.8437)
(0.8350, 0.8479)
(0.8400, 0.8521)
(0.8450, 0.8564)
(0.8500, 0.8607)
(0.8550, 0.8650)
(0.8600, 0.8694)
(0.8650, 0.8737)
(0.8700, 0.8781)
(0.8750, 0.8825)
(0.8800, 0.8869)
(0.8850, 0.8914)
(0.8900, 0.8958)
(0.8950, 0.9003)
(0.9000, 0.9048)
(0.9050, 0.9094)
(0.9100, 0.9139)
(0.9150, 0.9185)
(0.9200, 0.9231)
(0.9250, 0.9277)
(0.9300, 0.9324)
(0.9350, 0.9371)
(0.9400, 0.9418)
(0.9450, 0.9465)
(0.9500, 0.9512)
(0.9550, 0.9560)
(0.9600, 0.9608)
(0.9650, 0.9656)
(0.9700, 0.9704)
(0.9750, 0.9753)
(0.9800, 0.9802)
(0.9850, 0.9851)
(0.9900, 0.9900)
(0.9950, 0.9950)
(1.0000, 1.0000)
};
\addplot [draw=black, line width=0.8mm] coordinates {
(0.0050, 0.0000)
(0.0100, 0.0000)
(0.0150, 0.0000)
(0.0200, 0.0000)
(0.0250, 0.0000)
(0.0300, 0.0000)
(0.0350, 0.0000)
(0.0400, 0.0000)
(0.0450, 0.0000)
(0.0500, 0.0000)
(0.0550, 0.0000)
(0.0600, 0.0000)
(0.0650, 0.0000)
(0.0700, 0.0000)
(0.0750, 0.0000)
(0.0800, 0.0000)
(0.0850, 0.0000)
(0.0900, 0.0000)
(0.0950, 0.0000)
(0.1000, 0.0000)
(0.1050, 0.0000)
(0.1100, 0.0000)
(0.1150, 0.0000)
(0.1200, 0.0000)
(0.1250, 0.0000)
(0.1300, 0.0000)
(0.1350, 0.0000)
(0.1400, 0.0000)
(0.1450, 0.0000)
(0.1500, 0.0000)
(0.1550, 0.0000)
(0.1600, 0.0000)
(0.1650, 0.0000)
(0.1700, 0.0000)
(0.1750, 0.0000)
(0.1800, 0.0000)
(0.1850, 0.0000)
(0.1900, 0.0000)
(0.1950, 0.0000)
(0.2000, 0.0000)
(0.2050, 0.0000)
(0.2100, 0.0000)
(0.2150, 0.0000)
(0.2200, 0.0000)
(0.2250, 0.0000)
(0.2300, 0.0000)
(0.2350, 0.0000)
(0.2400, 0.0000)
(0.2450, 0.0000)
(0.2500, 0.0000)
(0.2550, 0.0000)
(0.2600, 0.0000)
(0.2650, 0.0000)
(0.2700, 0.0000)
(0.2750, 0.0000)
(0.2800, 0.0000)
(0.2850, 0.0000)
(0.2900, 0.0000)
(0.2950, 0.0000)
(0.3000, 0.0000)
(0.3050, 0.0000)
(0.3100, 0.0000)
(0.3150, 0.0000)
(0.3200, 0.0000)
(0.3250, 0.0000)
(0.3300, 0.0000)
(0.3350, 0.0000)
(0.3400, 0.0000)
(0.3450, 0.0000)
(0.3500, 0.0000)
(0.3550, 0.0000)
(0.3600, 0.0000)
(0.3650, 0.0000)
(0.3700, 0.0000)
(0.3750, 0.0000)
(0.3800, 0.0000)
(0.3850, 0.0000)
(0.3900, 0.0000)
(0.3950, 0.0000)
(0.4000, 0.0000)
(0.4050, 0.0000)
(0.4100, 0.0000)
(0.4150, 0.0000)
(0.4200, 0.0000)
(0.4250, 0.0000)
(0.4300, 0.0000)
(0.4350, 0.0000)
(0.4400, 0.0000)
(0.4450, 0.0000)
(0.4500, 0.0000)
(0.4550, 0.0000)
(0.4600, 0.0000)
(0.4650, 0.0000)
(0.4700, 0.0000)
(0.4750, 0.0000)
(0.4800, 0.0000)
(0.4850, 0.0000)
(0.4900, 0.0000)
(0.4950, 0.0000)
(0.5000, 0.0000)
(0.5050, 0.0000)
(0.5100, 0.0000)
(0.5150, 0.0000)
(0.5200, 0.0000)
(0.5250, 0.0000)
(0.5300, 0.0000)
(0.5350, 0.0000)
(0.5400, 0.0000)
(0.5450, 0.0000)
(0.5500, 0.0000)
(0.5550, 0.0000)
(0.5600, 0.0000)
(0.5650, 0.0000)
(0.5700, 0.0000)
(0.5750, 0.0000)
(0.5800, 0.0000)
(0.5850, 0.0000)
(0.5900, 0.0000)
(0.5950, 0.0000)
(0.6000, 0.0000)
(0.6050, 0.0000)
(0.6100, 0.0000)
(0.6150, 0.0000)
(0.6200, 0.0000)
(0.6250, 0.0000)
(0.6300, 0.0000)
(0.6350, 0.0000)
(0.6400, 0.0000)
(0.6450, 0.0000)
(0.6500, 0.0000)
(0.6550, 0.0000)
(0.6600, 0.0000)
(0.6650, 0.0000)
(0.6700, 0.0000)
(0.6750, 0.0000)
(0.6800, 0.0000)
(0.6850, 0.0000)
(0.6900, 0.0000)
(0.6950, 0.0000)
(0.7000, 0.0000)
(0.7050, 0.0000)
(0.7100, 0.0000)
(0.7150, 0.0000)
(0.7200, 0.0000)
(0.7250, 0.0000)
(0.7300, 0.0000)
(0.7350, 0.0000)
(0.7400, 0.0000)
(0.7450, 0.0000)
(0.7500, 0.0000)
(0.7550, 0.0000)
(0.7600, 0.0000)
(0.7650, 0.0000)
(0.7700, 0.0000)
(0.7750, 0.0000)
(0.7800, 0.0000)
(0.7850, 0.0000)
(0.7900, 0.0000)
(0.7950, 0.0000)
(0.8000, 0.0000)
(0.8050, 0.0000)
(0.8100, 0.0000)
(0.8150, 0.0000)
(0.8200, 0.0000)
(0.8250, 0.0000)
(0.8300, 0.0000)
(0.8350, 0.0000)
(0.8400, 0.0000)
(0.8450, 0.0000)
(0.8500, 0.0000)
(0.8550, 0.0000)
(0.8600, 0.0000)
(0.8650, 0.0000)
(0.8700, 0.0000)
(0.8750, 0.0000)
(0.8800, 0.0000)
(0.8850, 0.0000)
(0.8900, 0.0000)
(0.8950, 0.0000)
(0.9000, 0.0000)
(0.9050, 0.0000)
(0.9100, 0.0000)
(0.9150, 0.0000)
(0.9200, 0.0000)
(0.9250, 0.0000)
(0.9300, 0.0000)
(0.9350, 0.0000)
(0.9400, 0.0000)
(0.9450, 0.0000)
(0.9500, 0.0000)
(0.9550, 0.0000)
(0.9600, 0.0000)
(0.9650, 0.0000)
(0.9700, 0.0000)
(0.9750, 0.0000)
(0.9800, 0.0000)
(0.9850, 0.0000)
(0.9900, 0.0000)
(0.9950, 0.0000)
(1.0000, 0.0000)
};
\addplot [only marks,mark=*] coordinates { (1,1) };
\addplot [only marks,mark=*,mark options={ fill=white}]coordinates { (1,0) };
\end{axis}\end{tikzpicture}
	}
	\subfloat[
    $k = K - 1$
	]{
	\begin{tikzpicture}\begin{axis}[height=4.5cm,width=4.5cm,xmin=0,xmax=1,ymin=0,ymax=1,xtick={0,1},xticklabels={0,1},ytick={0,1},yticklabels={0,1},]
\addplot [draw=black, line width=0.5mm] coordinates {
(0.0050, 0.0050)
(0.0100, 0.0100)
(0.0150, 0.0150)
(0.0200, 0.0200)
(0.0250, 0.0250)
(0.0300, 0.0300)
(0.0350, 0.0350)
(0.0400, 0.0400)
(0.0450, 0.0450)
(0.0500, 0.0500)
(0.0550, 0.0550)
(0.0600, 0.0600)
(0.0650, 0.0650)
(0.0700, 0.0700)
(0.0750, 0.0750)
(0.0800, 0.0800)
(0.0850, 0.0850)
(0.0900, 0.0900)
(0.0950, 0.0950)
(0.1000, 0.1000)
(0.1050, 0.1050)
(0.1100, 0.1100)
(0.1150, 0.1150)
(0.1200, 0.1200)
(0.1250, 0.1250)
(0.1300, 0.1300)
(0.1350, 0.1350)
(0.1400, 0.1400)
(0.1450, 0.1450)
(0.1500, 0.1500)
(0.1550, 0.1550)
(0.1600, 0.1600)
(0.1650, 0.1650)
(0.1700, 0.1700)
(0.1750, 0.1750)
(0.1800, 0.1800)
(0.1850, 0.1850)
(0.1900, 0.1900)
(0.1950, 0.1950)
(0.2000, 0.2000)
(0.2050, 0.2050)
(0.2100, 0.2100)
(0.2150, 0.2150)
(0.2200, 0.2200)
(0.2250, 0.2250)
(0.2300, 0.2300)
(0.2350, 0.2350)
(0.2400, 0.2400)
(0.2450, 0.2450)
(0.2500, 0.2500)
(0.2550, 0.2550)
(0.2600, 0.2600)
(0.2650, 0.2650)
(0.2700, 0.2700)
(0.2750, 0.2750)
(0.2800, 0.2800)
(0.2850, 0.2850)
(0.2900, 0.2900)
(0.2950, 0.2950)
(0.3000, 0.3000)
(0.3050, 0.3050)
(0.3100, 0.3100)
(0.3150, 0.3150)
(0.3200, 0.3200)
(0.3250, 0.3250)
(0.3300, 0.3300)
(0.3350, 0.3350)
(0.3400, 0.3400)
(0.3450, 0.3450)
(0.3500, 0.3500)
(0.3550, 0.3550)
(0.3600, 0.3600)
(0.3650, 0.3650)
(0.3700, 0.3700)
(0.3750, 0.3750)
(0.3800, 0.3800)
(0.3850, 0.3850)
(0.3900, 0.3900)
(0.3950, 0.3950)
(0.4000, 0.4000)
(0.4050, 0.4050)
(0.4100, 0.4100)
(0.4150, 0.4150)
(0.4200, 0.4200)
(0.4250, 0.4250)
(0.4300, 0.4300)
(0.4350, 0.4350)
(0.4400, 0.4400)
(0.4450, 0.4450)
(0.4500, 0.4500)
(0.4550, 0.4550)
(0.4600, 0.4600)
(0.4650, 0.4650)
(0.4700, 0.4700)
(0.4750, 0.4750)
(0.4800, 0.4800)
(0.4850, 0.4850)
(0.4900, 0.4900)
(0.4950, 0.4950)
(0.5000, 0.5000)
(0.5050, 0.5050)
(0.5100, 0.5100)
(0.5150, 0.5150)
(0.5200, 0.5200)
(0.5250, 0.5250)
(0.5300, 0.5300)
(0.5350, 0.5350)
(0.5400, 0.5400)
(0.5450, 0.5450)
(0.5500, 0.5500)
(0.5550, 0.5550)
(0.5600, 0.5600)
(0.5650, 0.5650)
(0.5700, 0.5700)
(0.5750, 0.5750)
(0.5800, 0.5800)
(0.5850, 0.5850)
(0.5900, 0.5900)
(0.5950, 0.5950)
(0.6000, 0.6000)
(0.6050, 0.6050)
(0.6100, 0.6100)
(0.6150, 0.6150)
(0.6200, 0.6200)
(0.6250, 0.6250)
(0.6300, 0.6300)
(0.6350, 0.6350)
(0.6400, 0.6400)
(0.6450, 0.6450)
(0.6500, 0.6500)
(0.6550, 0.6550)
(0.6600, 0.6600)
(0.6650, 0.6650)
(0.6700, 0.6700)
(0.6750, 0.6750)
(0.6800, 0.6800)
(0.6850, 0.6850)
(0.6900, 0.6900)
(0.6950, 0.6950)
(0.7000, 0.7000)
(0.7050, 0.7050)
(0.7100, 0.7100)
(0.7150, 0.7150)
(0.7200, 0.7200)
(0.7250, 0.7250)
(0.7300, 0.7300)
(0.7350, 0.7350)
(0.7400, 0.7400)
(0.7450, 0.7450)
(0.7500, 0.7500)
(0.7550, 0.7550)
(0.7600, 0.7600)
(0.7650, 0.7650)
(0.7700, 0.7700)
(0.7750, 0.7750)
(0.7800, 0.7800)
(0.7850, 0.7850)
(0.7900, 0.7900)
(0.7950, 0.7950)
(0.8000, 0.8000)
(0.8050, 0.8050)
(0.8100, 0.8100)
(0.8150, 0.8150)
(0.8200, 0.8200)
(0.8250, 0.8250)
(0.8300, 0.8300)
(0.8350, 0.8350)
(0.8400, 0.8400)
(0.8450, 0.8450)
(0.8500, 0.8500)
(0.8550, 0.8550)
(0.8600, 0.8600)
(0.8650, 0.8650)
(0.8700, 0.8700)
(0.8750, 0.8750)
(0.8800, 0.8800)
(0.8850, 0.8850)
(0.8900, 0.8900)
(0.8950, 0.8950)
(0.9000, 0.9000)
(0.9050, 0.9050)
(0.9100, 0.9100)
(0.9150, 0.9150)
(0.9200, 0.9200)
(0.9250, 0.9250)
(0.9300, 0.9300)
(0.9350, 0.9350)
(0.9400, 0.9400)
(0.9450, 0.9450)
(0.9500, 0.9500)
(0.9550, 0.9550)
(0.9600, 0.9600)
(0.9650, 0.9650)
(0.9700, 0.9700)
(0.9750, 0.9750)
(0.9800, 0.9800)
(0.9850, 0.9850)
(0.9900, 0.9900)
(0.9950, 0.9950)
(1.0000, 1.0000)
};
\addplot [draw=black, line width=0.4mm, densely dashed] coordinates {
(0.0050, 0.3697)
(0.0100, 0.3716)
(0.0150, 0.3734)
(0.0200, 0.3753)
(0.0250, 0.3772)
(0.0300, 0.3791)
(0.0350, 0.3810)
(0.0400, 0.3829)
(0.0450, 0.3848)
(0.0500, 0.3867)
(0.0550, 0.3887)
(0.0600, 0.3906)
(0.0650, 0.3926)
(0.0700, 0.3946)
(0.0750, 0.3965)
(0.0800, 0.3985)
(0.0850, 0.4005)
(0.0900, 0.4025)
(0.0950, 0.4045)
(0.1000, 0.4066)
(0.1050, 0.4086)
(0.1100, 0.4107)
(0.1150, 0.4127)
(0.1200, 0.4148)
(0.1250, 0.4169)
(0.1300, 0.4190)
(0.1350, 0.4211)
(0.1400, 0.4232)
(0.1450, 0.4253)
(0.1500, 0.4274)
(0.1550, 0.4296)
(0.1600, 0.4317)
(0.1650, 0.4339)
(0.1700, 0.4360)
(0.1750, 0.4382)
(0.1800, 0.4404)
(0.1850, 0.4426)
(0.1900, 0.4449)
(0.1950, 0.4471)
(0.2000, 0.4493)
(0.2050, 0.4516)
(0.2100, 0.4538)
(0.2150, 0.4561)
(0.2200, 0.4584)
(0.2250, 0.4607)
(0.2300, 0.4630)
(0.2350, 0.4653)
(0.2400, 0.4677)
(0.2450, 0.4700)
(0.2500, 0.4724)
(0.2550, 0.4747)
(0.2600, 0.4771)
(0.2650, 0.4795)
(0.2700, 0.4819)
(0.2750, 0.4843)
(0.2800, 0.4868)
(0.2850, 0.4892)
(0.2900, 0.4916)
(0.2950, 0.4941)
(0.3000, 0.4966)
(0.3050, 0.4991)
(0.3100, 0.5016)
(0.3150, 0.5041)
(0.3200, 0.5066)
(0.3250, 0.5092)
(0.3300, 0.5117)
(0.3350, 0.5143)
(0.3400, 0.5169)
(0.3450, 0.5194)
(0.3500, 0.5220)
(0.3550, 0.5247)
(0.3600, 0.5273)
(0.3650, 0.5299)
(0.3700, 0.5326)
(0.3750, 0.5353)
(0.3800, 0.5379)
(0.3850, 0.5406)
(0.3900, 0.5434)
(0.3950, 0.5461)
(0.4000, 0.5488)
(0.4050, 0.5516)
(0.4100, 0.5543)
(0.4150, 0.5571)
(0.4200, 0.5599)
(0.4250, 0.5627)
(0.4300, 0.5655)
(0.4350, 0.5684)
(0.4400, 0.5712)
(0.4450, 0.5741)
(0.4500, 0.5769)
(0.4550, 0.5798)
(0.4600, 0.5827)
(0.4650, 0.5857)
(0.4700, 0.5886)
(0.4750, 0.5916)
(0.4800, 0.5945)
(0.4850, 0.5975)
(0.4900, 0.6005)
(0.4950, 0.6035)
(0.5000, 0.6065)
(0.5050, 0.6096)
(0.5100, 0.6126)
(0.5150, 0.6157)
(0.5200, 0.6188)
(0.5250, 0.6219)
(0.5300, 0.6250)
(0.5350, 0.6281)
(0.5400, 0.6313)
(0.5450, 0.6344)
(0.5500, 0.6376)
(0.5550, 0.6408)
(0.5600, 0.6440)
(0.5650, 0.6473)
(0.5700, 0.6505)
(0.5750, 0.6538)
(0.5800, 0.6570)
(0.5850, 0.6603)
(0.5900, 0.6637)
(0.5950, 0.6670)
(0.6000, 0.6703)
(0.6050, 0.6737)
(0.6100, 0.6771)
(0.6150, 0.6805)
(0.6200, 0.6839)
(0.6250, 0.6873)
(0.6300, 0.6907)
(0.6350, 0.6942)
(0.6400, 0.6977)
(0.6450, 0.7012)
(0.6500, 0.7047)
(0.6550, 0.7082)
(0.6600, 0.7118)
(0.6650, 0.7153)
(0.6700, 0.7189)
(0.6750, 0.7225)
(0.6800, 0.7261)
(0.6850, 0.7298)
(0.6900, 0.7334)
(0.6950, 0.7371)
(0.7000, 0.7408)
(0.7050, 0.7445)
(0.7100, 0.7483)
(0.7150, 0.7520)
(0.7200, 0.7558)
(0.7250, 0.7596)
(0.7300, 0.7634)
(0.7350, 0.7672)
(0.7400, 0.7711)
(0.7450, 0.7749)
(0.7500, 0.7788)
(0.7550, 0.7827)
(0.7600, 0.7866)
(0.7650, 0.7906)
(0.7700, 0.7945)
(0.7750, 0.7985)
(0.7800, 0.8025)
(0.7850, 0.8065)
(0.7900, 0.8106)
(0.7950, 0.8146)
(0.8000, 0.8187)
(0.8050, 0.8228)
(0.8100, 0.8270)
(0.8150, 0.8311)
(0.8200, 0.8353)
(0.8250, 0.8395)
(0.8300, 0.8437)
(0.8350, 0.8479)
(0.8400, 0.8521)
(0.8450, 0.8564)
(0.8500, 0.8607)
(0.8550, 0.8650)
(0.8600, 0.8694)
(0.8650, 0.8737)
(0.8700, 0.8781)
(0.8750, 0.8825)
(0.8800, 0.8869)
(0.8850, 0.8914)
(0.8900, 0.8958)
(0.8950, 0.9003)
(0.9000, 0.9048)
(0.9050, 0.9094)
(0.9100, 0.9139)
(0.9150, 0.9185)
(0.9200, 0.9231)
(0.9250, 0.9277)
(0.9300, 0.9324)
(0.9350, 0.9371)
(0.9400, 0.9418)
(0.9450, 0.9465)
(0.9500, 0.9512)
(0.9550, 0.9560)
(0.9600, 0.9608)
(0.9650, 0.9656)
(0.9700, 0.9704)
(0.9750, 0.9753)
(0.9800, 0.9802)
(0.9850, 0.9851)
(0.9900, 0.9900)
(0.9950, 0.9950)
(1.0000, 1.0000)
};
\end{axis}\end{tikzpicture}
	}
	\subfloat[
    $k = K - 2$
	]
	{
	\begin{tikzpicture}\begin{axis}[height=4.5cm,width=4.5cm,xmin=0,xmax=1,ymin=0,ymax=1,xtick={0,1},xticklabels={0,1},ytick={0,1},yticklabels={0,1},]
\addplot [draw=black, line width=0.5mm] coordinates {
(0.0050, 0.2525)
(0.0100, 0.2550)
(0.0150, 0.2576)
(0.0200, 0.2601)
(0.0250, 0.2627)
(0.0300, 0.2652)
(0.0350, 0.2678)
(0.0400, 0.2704)
(0.0450, 0.2730)
(0.0500, 0.2756)
(0.0550, 0.2783)
(0.0600, 0.2809)
(0.0650, 0.2836)
(0.0700, 0.2862)
(0.0750, 0.2889)
(0.0800, 0.2916)
(0.0850, 0.2943)
(0.0900, 0.2970)
(0.0950, 0.2998)
(0.1000, 0.3025)
(0.1050, 0.3053)
(0.1100, 0.3080)
(0.1150, 0.3108)
(0.1200, 0.3136)
(0.1250, 0.3164)
(0.1300, 0.3192)
(0.1350, 0.3221)
(0.1400, 0.3249)
(0.1450, 0.3278)
(0.1500, 0.3306)
(0.1550, 0.3335)
(0.1600, 0.3364)
(0.1650, 0.3393)
(0.1700, 0.3422)
(0.1750, 0.3452)
(0.1800, 0.3481)
(0.1850, 0.3511)
(0.1900, 0.3540)
(0.1950, 0.3570)
(0.2000, 0.3600)
(0.2050, 0.3630)
(0.2100, 0.3660)
(0.2150, 0.3691)
(0.2200, 0.3721)
(0.2250, 0.3752)
(0.2300, 0.3782)
(0.2350, 0.3813)
(0.2400, 0.3844)
(0.2450, 0.3875)
(0.2500, 0.3906)
(0.2550, 0.3938)
(0.2600, 0.3969)
(0.2650, 0.4001)
(0.2700, 0.4032)
(0.2750, 0.4064)
(0.2800, 0.4096)
(0.2850, 0.4128)
(0.2900, 0.4160)
(0.2950, 0.4193)
(0.3000, 0.4225)
(0.3050, 0.4258)
(0.3100, 0.4290)
(0.3150, 0.4323)
(0.3200, 0.4356)
(0.3250, 0.4389)
(0.3300, 0.4422)
(0.3350, 0.4456)
(0.3400, 0.4489)
(0.3450, 0.4523)
(0.3500, 0.4556)
(0.3550, 0.4590)
(0.3600, 0.4624)
(0.3650, 0.4658)
(0.3700, 0.4692)
(0.3750, 0.4727)
(0.3800, 0.4761)
(0.3850, 0.4796)
(0.3900, 0.4830)
(0.3950, 0.4865)
(0.4000, 0.4900)
(0.4050, 0.4935)
(0.4100, 0.4970)
(0.4150, 0.5006)
(0.4200, 0.5041)
(0.4250, 0.5077)
(0.4300, 0.5112)
(0.4350, 0.5148)
(0.4400, 0.5184)
(0.4450, 0.5220)
(0.4500, 0.5256)
(0.4550, 0.5293)
(0.4600, 0.5329)
(0.4650, 0.5366)
(0.4700, 0.5402)
(0.4750, 0.5439)
(0.4800, 0.5476)
(0.4850, 0.5513)
(0.4900, 0.5550)
(0.4950, 0.5588)
(0.5000, 0.5625)
(0.5050, 0.5663)
(0.5100, 0.5700)
(0.5150, 0.5738)
(0.5200, 0.5776)
(0.5250, 0.5814)
(0.5300, 0.5852)
(0.5350, 0.5891)
(0.5400, 0.5929)
(0.5450, 0.5968)
(0.5500, 0.6006)
(0.5550, 0.6045)
(0.5600, 0.6084)
(0.5650, 0.6123)
(0.5700, 0.6162)
(0.5750, 0.6202)
(0.5800, 0.6241)
(0.5850, 0.6281)
(0.5900, 0.6320)
(0.5950, 0.6360)
(0.6000, 0.6400)
(0.6050, 0.6440)
(0.6100, 0.6480)
(0.6150, 0.6521)
(0.6200, 0.6561)
(0.6250, 0.6602)
(0.6300, 0.6642)
(0.6350, 0.6683)
(0.6400, 0.6724)
(0.6450, 0.6765)
(0.6500, 0.6806)
(0.6550, 0.6848)
(0.6600, 0.6889)
(0.6650, 0.6931)
(0.6700, 0.6972)
(0.6750, 0.7014)
(0.6800, 0.7056)
(0.6850, 0.7098)
(0.6900, 0.7140)
(0.6950, 0.7183)
(0.7000, 0.7225)
(0.7050, 0.7268)
(0.7100, 0.7310)
(0.7150, 0.7353)
(0.7200, 0.7396)
(0.7250, 0.7439)
(0.7300, 0.7482)
(0.7350, 0.7526)
(0.7400, 0.7569)
(0.7450, 0.7613)
(0.7500, 0.7656)
(0.7550, 0.7700)
(0.7600, 0.7744)
(0.7650, 0.7788)
(0.7700, 0.7832)
(0.7750, 0.7877)
(0.7800, 0.7921)
(0.7850, 0.7966)
(0.7900, 0.8010)
(0.7950, 0.8055)
(0.8000, 0.8100)
(0.8050, 0.8145)
(0.8100, 0.8190)
(0.8150, 0.8236)
(0.8200, 0.8281)
(0.8250, 0.8327)
(0.8300, 0.8372)
(0.8350, 0.8418)
(0.8400, 0.8464)
(0.8450, 0.8510)
(0.8500, 0.8556)
(0.8550, 0.8603)
(0.8600, 0.8649)
(0.8650, 0.8696)
(0.8700, 0.8742)
(0.8750, 0.8789)
(0.8800, 0.8836)
(0.8850, 0.8883)
(0.8900, 0.8930)
(0.8950, 0.8978)
(0.9000, 0.9025)
(0.9050, 0.9073)
(0.9100, 0.9120)
(0.9150, 0.9168)
(0.9200, 0.9216)
(0.9250, 0.9264)
(0.9300, 0.9312)
(0.9350, 0.9361)
(0.9400, 0.9409)
(0.9450, 0.9458)
(0.9500, 0.9506)
(0.9550, 0.9555)
(0.9600, 0.9604)
(0.9650, 0.9653)
(0.9700, 0.9702)
(0.9750, 0.9752)
(0.9800, 0.9801)
(0.9850, 0.9851)
(0.9900, 0.9900)
(0.9950, 0.9950)
(1.0000, 1.0000)
};
\addplot [draw=black, line width=0.4mm, densely dashed] coordinates {
(0.0050, 0.3697)
(0.0100, 0.3716)
(0.0150, 0.3734)
(0.0200, 0.3753)
(0.0250, 0.3772)
(0.0300, 0.3791)
(0.0350, 0.3810)
(0.0400, 0.3829)
(0.0450, 0.3848)
(0.0500, 0.3867)
(0.0550, 0.3887)
(0.0600, 0.3906)
(0.0650, 0.3926)
(0.0700, 0.3946)
(0.0750, 0.3965)
(0.0800, 0.3985)
(0.0850, 0.4005)
(0.0900, 0.4025)
(0.0950, 0.4045)
(0.1000, 0.4066)
(0.1050, 0.4086)
(0.1100, 0.4107)
(0.1150, 0.4127)
(0.1200, 0.4148)
(0.1250, 0.4169)
(0.1300, 0.4190)
(0.1350, 0.4211)
(0.1400, 0.4232)
(0.1450, 0.4253)
(0.1500, 0.4274)
(0.1550, 0.4296)
(0.1600, 0.4317)
(0.1650, 0.4339)
(0.1700, 0.4360)
(0.1750, 0.4382)
(0.1800, 0.4404)
(0.1850, 0.4426)
(0.1900, 0.4449)
(0.1950, 0.4471)
(0.2000, 0.4493)
(0.2050, 0.4516)
(0.2100, 0.4538)
(0.2150, 0.4561)
(0.2200, 0.4584)
(0.2250, 0.4607)
(0.2300, 0.4630)
(0.2350, 0.4653)
(0.2400, 0.4677)
(0.2450, 0.4700)
(0.2500, 0.4724)
(0.2550, 0.4747)
(0.2600, 0.4771)
(0.2650, 0.4795)
(0.2700, 0.4819)
(0.2750, 0.4843)
(0.2800, 0.4868)
(0.2850, 0.4892)
(0.2900, 0.4916)
(0.2950, 0.4941)
(0.3000, 0.4966)
(0.3050, 0.4991)
(0.3100, 0.5016)
(0.3150, 0.5041)
(0.3200, 0.5066)
(0.3250, 0.5092)
(0.3300, 0.5117)
(0.3350, 0.5143)
(0.3400, 0.5169)
(0.3450, 0.5194)
(0.3500, 0.5220)
(0.3550, 0.5247)
(0.3600, 0.5273)
(0.3650, 0.5299)
(0.3700, 0.5326)
(0.3750, 0.5353)
(0.3800, 0.5379)
(0.3850, 0.5406)
(0.3900, 0.5434)
(0.3950, 0.5461)
(0.4000, 0.5488)
(0.4050, 0.5516)
(0.4100, 0.5543)
(0.4150, 0.5571)
(0.4200, 0.5599)
(0.4250, 0.5627)
(0.4300, 0.5655)
(0.4350, 0.5684)
(0.4400, 0.5712)
(0.4450, 0.5741)
(0.4500, 0.5769)
(0.4550, 0.5798)
(0.4600, 0.5827)
(0.4650, 0.5857)
(0.4700, 0.5886)
(0.4750, 0.5916)
(0.4800, 0.5945)
(0.4850, 0.5975)
(0.4900, 0.6005)
(0.4950, 0.6035)
(0.5000, 0.6065)
(0.5050, 0.6096)
(0.5100, 0.6126)
(0.5150, 0.6157)
(0.5200, 0.6188)
(0.5250, 0.6219)
(0.5300, 0.6250)
(0.5350, 0.6281)
(0.5400, 0.6313)
(0.5450, 0.6344)
(0.5500, 0.6376)
(0.5550, 0.6408)
(0.5600, 0.6440)
(0.5650, 0.6473)
(0.5700, 0.6505)
(0.5750, 0.6538)
(0.5800, 0.6570)
(0.5850, 0.6603)
(0.5900, 0.6637)
(0.5950, 0.6670)
(0.6000, 0.6703)
(0.6050, 0.6737)
(0.6100, 0.6771)
(0.6150, 0.6805)
(0.6200, 0.6839)
(0.6250, 0.6873)
(0.6300, 0.6907)
(0.6350, 0.6942)
(0.6400, 0.6977)
(0.6450, 0.7012)
(0.6500, 0.7047)
(0.6550, 0.7082)
(0.6600, 0.7118)
(0.6650, 0.7153)
(0.6700, 0.7189)
(0.6750, 0.7225)
(0.6800, 0.7261)
(0.6850, 0.7298)
(0.6900, 0.7334)
(0.6950, 0.7371)
(0.7000, 0.7408)
(0.7050, 0.7445)
(0.7100, 0.7483)
(0.7150, 0.7520)
(0.7200, 0.7558)
(0.7250, 0.7596)
(0.7300, 0.7634)
(0.7350, 0.7672)
(0.7400, 0.7711)
(0.7450, 0.7749)
(0.7500, 0.7788)
(0.7550, 0.7827)
(0.7600, 0.7866)
(0.7650, 0.7906)
(0.7700, 0.7945)
(0.7750, 0.7985)
(0.7800, 0.8025)
(0.7850, 0.8065)
(0.7900, 0.8106)
(0.7950, 0.8146)
(0.8000, 0.8187)
(0.8050, 0.8228)
(0.8100, 0.8270)
(0.8150, 0.8311)
(0.8200, 0.8353)
(0.8250, 0.8395)
(0.8300, 0.8437)
(0.8350, 0.8479)
(0.8400, 0.8521)
(0.8450, 0.8564)
(0.8500, 0.8607)
(0.8550, 0.8650)
(0.8600, 0.8694)
(0.8650, 0.8737)
(0.8700, 0.8781)
(0.8750, 0.8825)
(0.8800, 0.8869)
(0.8850, 0.8914)
(0.8900, 0.8958)
(0.8950, 0.9003)
(0.9000, 0.9048)
(0.9050, 0.9094)
(0.9100, 0.9139)
(0.9150, 0.9185)
(0.9200, 0.9231)
(0.9250, 0.9277)
(0.9300, 0.9324)
(0.9350, 0.9371)
(0.9400, 0.9418)
(0.9450, 0.9465)
(0.9500, 0.9512)
(0.9550, 0.9560)
(0.9600, 0.9608)
(0.9650, 0.9656)
(0.9700, 0.9704)
(0.9750, 0.9753)
(0.9800, 0.9802)
(0.9850, 0.9851)
(0.9900, 0.9900)
(0.9950, 0.9950)
(1.0000, 1.0000)
};
\end{axis}\end{tikzpicture}
	}
	\\
	\centering
	\subfloat[
    $k = K - 3$
	]{
	\begin{tikzpicture}\begin{axis}[height=4.5cm,width=4.5cm,xmin=0,xmax=1,ymin=0,ymax=1,xtick={0,1},xticklabels={0,1},ytick={0,1},yticklabels={0,1},]
\addplot [draw=black, line width=0.5mm] coordinates {
(0.0050, 0.2985)
(0.0100, 0.3008)
(0.0150, 0.3030)
(0.0200, 0.3053)
(0.0250, 0.3075)
(0.0300, 0.3098)
(0.0350, 0.3121)
(0.0400, 0.3144)
(0.0450, 0.3167)
(0.0500, 0.3191)
(0.0550, 0.3214)
(0.0600, 0.3238)
(0.0650, 0.3261)
(0.0700, 0.3285)
(0.0750, 0.3309)
(0.0800, 0.3333)
(0.0850, 0.3357)
(0.0900, 0.3381)
(0.0950, 0.3406)
(0.1000, 0.3430)
(0.1050, 0.3455)
(0.1100, 0.3479)
(0.1150, 0.3504)
(0.1200, 0.3529)
(0.1250, 0.3554)
(0.1300, 0.3579)
(0.1350, 0.3604)
(0.1400, 0.3630)
(0.1450, 0.3655)
(0.1500, 0.3681)
(0.1550, 0.3707)
(0.1600, 0.3732)
(0.1650, 0.3758)
(0.1700, 0.3785)
(0.1750, 0.3811)
(0.1800, 0.3837)
(0.1850, 0.3864)
(0.1900, 0.3890)
(0.1950, 0.3917)
(0.2000, 0.3944)
(0.2050, 0.3971)
(0.2100, 0.3998)
(0.2150, 0.4025)
(0.2200, 0.4052)
(0.2250, 0.4080)
(0.2300, 0.4107)
(0.2350, 0.4135)
(0.2400, 0.4163)
(0.2450, 0.4191)
(0.2500, 0.4219)
(0.2550, 0.4247)
(0.2600, 0.4275)
(0.2650, 0.4304)
(0.2700, 0.4332)
(0.2750, 0.4361)
(0.2800, 0.4390)
(0.2850, 0.4419)
(0.2900, 0.4448)
(0.2950, 0.4477)
(0.3000, 0.4506)
(0.3050, 0.4536)
(0.3100, 0.4565)
(0.3150, 0.4595)
(0.3200, 0.4625)
(0.3250, 0.4655)
(0.3300, 0.4685)
(0.3350, 0.4715)
(0.3400, 0.4746)
(0.3450, 0.4776)
(0.3500, 0.4807)
(0.3550, 0.4837)
(0.3600, 0.4868)
(0.3650, 0.4899)
(0.3700, 0.4930)
(0.3750, 0.4962)
(0.3800, 0.4993)
(0.3850, 0.5025)
(0.3900, 0.5056)
(0.3950, 0.5088)
(0.4000, 0.5120)
(0.4050, 0.5152)
(0.4100, 0.5184)
(0.4150, 0.5217)
(0.4200, 0.5249)
(0.4250, 0.5282)
(0.4300, 0.5314)
(0.4350, 0.5347)
(0.4400, 0.5380)
(0.4450, 0.5413)
(0.4500, 0.5447)
(0.4550, 0.5480)
(0.4600, 0.5514)
(0.4650, 0.5547)
(0.4700, 0.5581)
(0.4750, 0.5615)
(0.4800, 0.5649)
(0.4850, 0.5683)
(0.4900, 0.5718)
(0.4950, 0.5752)
(0.5000, 0.5787)
(0.5050, 0.5822)
(0.5100, 0.5857)
(0.5150, 0.5892)
(0.5200, 0.5927)
(0.5250, 0.5962)
(0.5300, 0.5998)
(0.5350, 0.6034)
(0.5400, 0.6069)
(0.5450, 0.6105)
(0.5500, 0.6141)
(0.5550, 0.6177)
(0.5600, 0.6214)
(0.5650, 0.6250)
(0.5700, 0.6287)
(0.5750, 0.6324)
(0.5800, 0.6361)
(0.5850, 0.6398)
(0.5900, 0.6435)
(0.5950, 0.6472)
(0.6000, 0.6510)
(0.6050, 0.6547)
(0.6100, 0.6585)
(0.6150, 0.6623)
(0.6200, 0.6661)
(0.6250, 0.6699)
(0.6300, 0.6738)
(0.6350, 0.6776)
(0.6400, 0.6815)
(0.6450, 0.6854)
(0.6500, 0.6892)
(0.6550, 0.6932)
(0.6600, 0.6971)
(0.6650, 0.7010)
(0.6700, 0.7050)
(0.6750, 0.7089)
(0.6800, 0.7129)
(0.6850, 0.7169)
(0.6900, 0.7209)
(0.6950, 0.7250)
(0.7000, 0.7290)
(0.7050, 0.7331)
(0.7100, 0.7371)
(0.7150, 0.7412)
(0.7200, 0.7453)
(0.7250, 0.7494)
(0.7300, 0.7536)
(0.7350, 0.7577)
(0.7400, 0.7619)
(0.7450, 0.7661)
(0.7500, 0.7703)
(0.7550, 0.7745)
(0.7600, 0.7787)
(0.7650, 0.7829)
(0.7700, 0.7872)
(0.7750, 0.7915)
(0.7800, 0.7957)
(0.7850, 0.8000)
(0.7900, 0.8044)
(0.7950, 0.8087)
(0.8000, 0.8130)
(0.8050, 0.8174)
(0.8100, 0.8218)
(0.8150, 0.8262)
(0.8200, 0.8306)
(0.8250, 0.8350)
(0.8300, 0.8395)
(0.8350, 0.8439)
(0.8400, 0.8484)
(0.8450, 0.8529)
(0.8500, 0.8574)
(0.8550, 0.8619)
(0.8600, 0.8664)
(0.8650, 0.8710)
(0.8700, 0.8756)
(0.8750, 0.8801)
(0.8800, 0.8847)
(0.8850, 0.8894)
(0.8900, 0.8940)
(0.8950, 0.8986)
(0.9000, 0.9033)
(0.9050, 0.9080)
(0.9100, 0.9127)
(0.9150, 0.9174)
(0.9200, 0.9221)
(0.9250, 0.9269)
(0.9300, 0.9316)
(0.9350, 0.9364)
(0.9400, 0.9412)
(0.9450, 0.9460)
(0.9500, 0.9508)
(0.9550, 0.9557)
(0.9600, 0.9605)
(0.9650, 0.9654)
(0.9700, 0.9703)
(0.9750, 0.9752)
(0.9800, 0.9801)
(0.9850, 0.9851)
(0.9900, 0.9900)
(0.9950, 0.9950)
(1.0000, 1.0000)
};
\addplot [draw=black, line width=0.4mm, densely dashed] coordinates {
(0.0050, 0.3697)
(0.0100, 0.3716)
(0.0150, 0.3734)
(0.0200, 0.3753)
(0.0250, 0.3772)
(0.0300, 0.3791)
(0.0350, 0.3810)
(0.0400, 0.3829)
(0.0450, 0.3848)
(0.0500, 0.3867)
(0.0550, 0.3887)
(0.0600, 0.3906)
(0.0650, 0.3926)
(0.0700, 0.3946)
(0.0750, 0.3965)
(0.0800, 0.3985)
(0.0850, 0.4005)
(0.0900, 0.4025)
(0.0950, 0.4045)
(0.1000, 0.4066)
(0.1050, 0.4086)
(0.1100, 0.4107)
(0.1150, 0.4127)
(0.1200, 0.4148)
(0.1250, 0.4169)
(0.1300, 0.4190)
(0.1350, 0.4211)
(0.1400, 0.4232)
(0.1450, 0.4253)
(0.1500, 0.4274)
(0.1550, 0.4296)
(0.1600, 0.4317)
(0.1650, 0.4339)
(0.1700, 0.4360)
(0.1750, 0.4382)
(0.1800, 0.4404)
(0.1850, 0.4426)
(0.1900, 0.4449)
(0.1950, 0.4471)
(0.2000, 0.4493)
(0.2050, 0.4516)
(0.2100, 0.4538)
(0.2150, 0.4561)
(0.2200, 0.4584)
(0.2250, 0.4607)
(0.2300, 0.4630)
(0.2350, 0.4653)
(0.2400, 0.4677)
(0.2450, 0.4700)
(0.2500, 0.4724)
(0.2550, 0.4747)
(0.2600, 0.4771)
(0.2650, 0.4795)
(0.2700, 0.4819)
(0.2750, 0.4843)
(0.2800, 0.4868)
(0.2850, 0.4892)
(0.2900, 0.4916)
(0.2950, 0.4941)
(0.3000, 0.4966)
(0.3050, 0.4991)
(0.3100, 0.5016)
(0.3150, 0.5041)
(0.3200, 0.5066)
(0.3250, 0.5092)
(0.3300, 0.5117)
(0.3350, 0.5143)
(0.3400, 0.5169)
(0.3450, 0.5194)
(0.3500, 0.5220)
(0.3550, 0.5247)
(0.3600, 0.5273)
(0.3650, 0.5299)
(0.3700, 0.5326)
(0.3750, 0.5353)
(0.3800, 0.5379)
(0.3850, 0.5406)
(0.3900, 0.5434)
(0.3950, 0.5461)
(0.4000, 0.5488)
(0.4050, 0.5516)
(0.4100, 0.5543)
(0.4150, 0.5571)
(0.4200, 0.5599)
(0.4250, 0.5627)
(0.4300, 0.5655)
(0.4350, 0.5684)
(0.4400, 0.5712)
(0.4450, 0.5741)
(0.4500, 0.5769)
(0.4550, 0.5798)
(0.4600, 0.5827)
(0.4650, 0.5857)
(0.4700, 0.5886)
(0.4750, 0.5916)
(0.4800, 0.5945)
(0.4850, 0.5975)
(0.4900, 0.6005)
(0.4950, 0.6035)
(0.5000, 0.6065)
(0.5050, 0.6096)
(0.5100, 0.6126)
(0.5150, 0.6157)
(0.5200, 0.6188)
(0.5250, 0.6219)
(0.5300, 0.6250)
(0.5350, 0.6281)
(0.5400, 0.6313)
(0.5450, 0.6344)
(0.5500, 0.6376)
(0.5550, 0.6408)
(0.5600, 0.6440)
(0.5650, 0.6473)
(0.5700, 0.6505)
(0.5750, 0.6538)
(0.5800, 0.6570)
(0.5850, 0.6603)
(0.5900, 0.6637)
(0.5950, 0.6670)
(0.6000, 0.6703)
(0.6050, 0.6737)
(0.6100, 0.6771)
(0.6150, 0.6805)
(0.6200, 0.6839)
(0.6250, 0.6873)
(0.6300, 0.6907)
(0.6350, 0.6942)
(0.6400, 0.6977)
(0.6450, 0.7012)
(0.6500, 0.7047)
(0.6550, 0.7082)
(0.6600, 0.7118)
(0.6650, 0.7153)
(0.6700, 0.7189)
(0.6750, 0.7225)
(0.6800, 0.7261)
(0.6850, 0.7298)
(0.6900, 0.7334)
(0.6950, 0.7371)
(0.7000, 0.7408)
(0.7050, 0.7445)
(0.7100, 0.7483)
(0.7150, 0.7520)
(0.7200, 0.7558)
(0.7250, 0.7596)
(0.7300, 0.7634)
(0.7350, 0.7672)
(0.7400, 0.7711)
(0.7450, 0.7749)
(0.7500, 0.7788)
(0.7550, 0.7827)
(0.7600, 0.7866)
(0.7650, 0.7906)
(0.7700, 0.7945)
(0.7750, 0.7985)
(0.7800, 0.8025)
(0.7850, 0.8065)
(0.7900, 0.8106)
(0.7950, 0.8146)
(0.8000, 0.8187)
(0.8050, 0.8228)
(0.8100, 0.8270)
(0.8150, 0.8311)
(0.8200, 0.8353)
(0.8250, 0.8395)
(0.8300, 0.8437)
(0.8350, 0.8479)
(0.8400, 0.8521)
(0.8450, 0.8564)
(0.8500, 0.8607)
(0.8550, 0.8650)
(0.8600, 0.8694)
(0.8650, 0.8737)
(0.8700, 0.8781)
(0.8750, 0.8825)
(0.8800, 0.8869)
(0.8850, 0.8914)
(0.8900, 0.8958)
(0.8950, 0.9003)
(0.9000, 0.9048)
(0.9050, 0.9094)
(0.9100, 0.9139)
(0.9150, 0.9185)
(0.9200, 0.9231)
(0.9250, 0.9277)
(0.9300, 0.9324)
(0.9350, 0.9371)
(0.9400, 0.9418)
(0.9450, 0.9465)
(0.9500, 0.9512)
(0.9550, 0.9560)
(0.9600, 0.9608)
(0.9650, 0.9656)
(0.9700, 0.9704)
(0.9750, 0.9753)
(0.9800, 0.9802)
(0.9850, 0.9851)
(0.9900, 0.9900)
(0.9950, 0.9950)
(1.0000, 1.0000)
};
\end{axis}\end{tikzpicture}
	}
	\subfloat[
    $k = K - 4$
	]{
	\begin{tikzpicture}\begin{axis}[height=4.5cm,width=4.5cm,xmin=0,xmax=1,ymin=0,ymax=1,xtick={0,1},xticklabels={0,1},ytick={0,1},yticklabels={0,1},]
\addplot [draw=black, line width=0.5mm] coordinates {
(0.0050, 0.3185)
(0.0100, 0.3206)
(0.0150, 0.3228)
(0.0200, 0.3249)
(0.0250, 0.3271)
(0.0300, 0.3293)
(0.0350, 0.3314)
(0.0400, 0.3336)
(0.0450, 0.3358)
(0.0500, 0.3380)
(0.0550, 0.3403)
(0.0600, 0.3425)
(0.0650, 0.3447)
(0.0700, 0.3470)
(0.0750, 0.3493)
(0.0800, 0.3515)
(0.0850, 0.3538)
(0.0900, 0.3561)
(0.0950, 0.3584)
(0.1000, 0.3608)
(0.1050, 0.3631)
(0.1100, 0.3654)
(0.1150, 0.3678)
(0.1200, 0.3702)
(0.1250, 0.3725)
(0.1300, 0.3749)
(0.1350, 0.3773)
(0.1400, 0.3797)
(0.1450, 0.3822)
(0.1500, 0.3846)
(0.1550, 0.3870)
(0.1600, 0.3895)
(0.1650, 0.3920)
(0.1700, 0.3945)
(0.1750, 0.3969)
(0.1800, 0.3995)
(0.1850, 0.4020)
(0.1900, 0.4045)
(0.1950, 0.4070)
(0.2000, 0.4096)
(0.2050, 0.4122)
(0.2100, 0.4147)
(0.2150, 0.4173)
(0.2200, 0.4199)
(0.2250, 0.4226)
(0.2300, 0.4252)
(0.2350, 0.4278)
(0.2400, 0.4305)
(0.2450, 0.4331)
(0.2500, 0.4358)
(0.2550, 0.4385)
(0.2600, 0.4412)
(0.2650, 0.4439)
(0.2700, 0.4466)
(0.2750, 0.4494)
(0.2800, 0.4521)
(0.2850, 0.4549)
(0.2900, 0.4577)
(0.2950, 0.4604)
(0.3000, 0.4633)
(0.3050, 0.4661)
(0.3100, 0.4689)
(0.3150, 0.4717)
(0.3200, 0.4746)
(0.3250, 0.4774)
(0.3300, 0.4803)
(0.3350, 0.4832)
(0.3400, 0.4861)
(0.3450, 0.4890)
(0.3500, 0.4920)
(0.3550, 0.4949)
(0.3600, 0.4979)
(0.3650, 0.5008)
(0.3700, 0.5038)
(0.3750, 0.5068)
(0.3800, 0.5098)
(0.3850, 0.5129)
(0.3900, 0.5159)
(0.3950, 0.5189)
(0.4000, 0.5220)
(0.4050, 0.5251)
(0.4100, 0.5282)
(0.4150, 0.5313)
(0.4200, 0.5344)
(0.4250, 0.5375)
(0.4300, 0.5407)
(0.4350, 0.5438)
(0.4400, 0.5470)
(0.4450, 0.5502)
(0.4500, 0.5534)
(0.4550, 0.5566)
(0.4600, 0.5598)
(0.4650, 0.5631)
(0.4700, 0.5663)
(0.4750, 0.5696)
(0.4800, 0.5729)
(0.4850, 0.5762)
(0.4900, 0.5795)
(0.4950, 0.5828)
(0.5000, 0.5862)
(0.5050, 0.5895)
(0.5100, 0.5929)
(0.5150, 0.5963)
(0.5200, 0.5997)
(0.5250, 0.6031)
(0.5300, 0.6065)
(0.5350, 0.6100)
(0.5400, 0.6134)
(0.5450, 0.6169)
(0.5500, 0.6204)
(0.5550, 0.6239)
(0.5600, 0.6274)
(0.5650, 0.6310)
(0.5700, 0.6345)
(0.5750, 0.6381)
(0.5800, 0.6416)
(0.5850, 0.6452)
(0.5900, 0.6488)
(0.5950, 0.6525)
(0.6000, 0.6561)
(0.6050, 0.6598)
(0.6100, 0.6634)
(0.6150, 0.6671)
(0.6200, 0.6708)
(0.6250, 0.6745)
(0.6300, 0.6782)
(0.6350, 0.6820)
(0.6400, 0.6857)
(0.6450, 0.6895)
(0.6500, 0.6933)
(0.6550, 0.6971)
(0.6600, 0.7009)
(0.6650, 0.7048)
(0.6700, 0.7086)
(0.6750, 0.7125)
(0.6800, 0.7164)
(0.6850, 0.7203)
(0.6900, 0.7242)
(0.6950, 0.7281)
(0.7000, 0.7321)
(0.7050, 0.7361)
(0.7100, 0.7400)
(0.7150, 0.7440)
(0.7200, 0.7481)
(0.7250, 0.7521)
(0.7300, 0.7561)
(0.7350, 0.7602)
(0.7400, 0.7643)
(0.7450, 0.7684)
(0.7500, 0.7725)
(0.7550, 0.7766)
(0.7600, 0.7807)
(0.7650, 0.7849)
(0.7700, 0.7891)
(0.7750, 0.7933)
(0.7800, 0.7975)
(0.7850, 0.8017)
(0.7900, 0.8060)
(0.7950, 0.8102)
(0.8000, 0.8145)
(0.8050, 0.8188)
(0.8100, 0.8231)
(0.8150, 0.8274)
(0.8200, 0.8318)
(0.8250, 0.8362)
(0.8300, 0.8405)
(0.8350, 0.8449)
(0.8400, 0.8493)
(0.8450, 0.8538)
(0.8500, 0.8582)
(0.8550, 0.8627)
(0.8600, 0.8672)
(0.8650, 0.8717)
(0.8700, 0.8762)
(0.8750, 0.8807)
(0.8800, 0.8853)
(0.8850, 0.8899)
(0.8900, 0.8945)
(0.8950, 0.8991)
(0.9000, 0.9037)
(0.9050, 0.9083)
(0.9100, 0.9130)
(0.9150, 0.9177)
(0.9200, 0.9224)
(0.9250, 0.9271)
(0.9300, 0.9318)
(0.9350, 0.9366)
(0.9400, 0.9413)
(0.9450, 0.9461)
(0.9500, 0.9509)
(0.9550, 0.9558)
(0.9600, 0.9606)
(0.9650, 0.9655)
(0.9700, 0.9703)
(0.9750, 0.9752)
(0.9800, 0.9801)
(0.9850, 0.9851)
(0.9900, 0.9900)
(0.9950, 0.9950)
(1.0000, 1.0000)
};
\addplot [draw=black, line width=0.4mm, densely dashed] coordinates {
(0.0050, 0.3697)
(0.0100, 0.3716)
(0.0150, 0.3734)
(0.0200, 0.3753)
(0.0250, 0.3772)
(0.0300, 0.3791)
(0.0350, 0.3810)
(0.0400, 0.3829)
(0.0450, 0.3848)
(0.0500, 0.3867)
(0.0550, 0.3887)
(0.0600, 0.3906)
(0.0650, 0.3926)
(0.0700, 0.3946)
(0.0750, 0.3965)
(0.0800, 0.3985)
(0.0850, 0.4005)
(0.0900, 0.4025)
(0.0950, 0.4045)
(0.1000, 0.4066)
(0.1050, 0.4086)
(0.1100, 0.4107)
(0.1150, 0.4127)
(0.1200, 0.4148)
(0.1250, 0.4169)
(0.1300, 0.4190)
(0.1350, 0.4211)
(0.1400, 0.4232)
(0.1450, 0.4253)
(0.1500, 0.4274)
(0.1550, 0.4296)
(0.1600, 0.4317)
(0.1650, 0.4339)
(0.1700, 0.4360)
(0.1750, 0.4382)
(0.1800, 0.4404)
(0.1850, 0.4426)
(0.1900, 0.4449)
(0.1950, 0.4471)
(0.2000, 0.4493)
(0.2050, 0.4516)
(0.2100, 0.4538)
(0.2150, 0.4561)
(0.2200, 0.4584)
(0.2250, 0.4607)
(0.2300, 0.4630)
(0.2350, 0.4653)
(0.2400, 0.4677)
(0.2450, 0.4700)
(0.2500, 0.4724)
(0.2550, 0.4747)
(0.2600, 0.4771)
(0.2650, 0.4795)
(0.2700, 0.4819)
(0.2750, 0.4843)
(0.2800, 0.4868)
(0.2850, 0.4892)
(0.2900, 0.4916)
(0.2950, 0.4941)
(0.3000, 0.4966)
(0.3050, 0.4991)
(0.3100, 0.5016)
(0.3150, 0.5041)
(0.3200, 0.5066)
(0.3250, 0.5092)
(0.3300, 0.5117)
(0.3350, 0.5143)
(0.3400, 0.5169)
(0.3450, 0.5194)
(0.3500, 0.5220)
(0.3550, 0.5247)
(0.3600, 0.5273)
(0.3650, 0.5299)
(0.3700, 0.5326)
(0.3750, 0.5353)
(0.3800, 0.5379)
(0.3850, 0.5406)
(0.3900, 0.5434)
(0.3950, 0.5461)
(0.4000, 0.5488)
(0.4050, 0.5516)
(0.4100, 0.5543)
(0.4150, 0.5571)
(0.4200, 0.5599)
(0.4250, 0.5627)
(0.4300, 0.5655)
(0.4350, 0.5684)
(0.4400, 0.5712)
(0.4450, 0.5741)
(0.4500, 0.5769)
(0.4550, 0.5798)
(0.4600, 0.5827)
(0.4650, 0.5857)
(0.4700, 0.5886)
(0.4750, 0.5916)
(0.4800, 0.5945)
(0.4850, 0.5975)
(0.4900, 0.6005)
(0.4950, 0.6035)
(0.5000, 0.6065)
(0.5050, 0.6096)
(0.5100, 0.6126)
(0.5150, 0.6157)
(0.5200, 0.6188)
(0.5250, 0.6219)
(0.5300, 0.6250)
(0.5350, 0.6281)
(0.5400, 0.6313)
(0.5450, 0.6344)
(0.5500, 0.6376)
(0.5550, 0.6408)
(0.5600, 0.6440)
(0.5650, 0.6473)
(0.5700, 0.6505)
(0.5750, 0.6538)
(0.5800, 0.6570)
(0.5850, 0.6603)
(0.5900, 0.6637)
(0.5950, 0.6670)
(0.6000, 0.6703)
(0.6050, 0.6737)
(0.6100, 0.6771)
(0.6150, 0.6805)
(0.6200, 0.6839)
(0.6250, 0.6873)
(0.6300, 0.6907)
(0.6350, 0.6942)
(0.6400, 0.6977)
(0.6450, 0.7012)
(0.6500, 0.7047)
(0.6550, 0.7082)
(0.6600, 0.7118)
(0.6650, 0.7153)
(0.6700, 0.7189)
(0.6750, 0.7225)
(0.6800, 0.7261)
(0.6850, 0.7298)
(0.6900, 0.7334)
(0.6950, 0.7371)
(0.7000, 0.7408)
(0.7050, 0.7445)
(0.7100, 0.7483)
(0.7150, 0.7520)
(0.7200, 0.7558)
(0.7250, 0.7596)
(0.7300, 0.7634)
(0.7350, 0.7672)
(0.7400, 0.7711)
(0.7450, 0.7749)
(0.7500, 0.7788)
(0.7550, 0.7827)
(0.7600, 0.7866)
(0.7650, 0.7906)
(0.7700, 0.7945)
(0.7750, 0.7985)
(0.7800, 0.8025)
(0.7850, 0.8065)
(0.7900, 0.8106)
(0.7950, 0.8146)
(0.8000, 0.8187)
(0.8050, 0.8228)
(0.8100, 0.8270)
(0.8150, 0.8311)
(0.8200, 0.8353)
(0.8250, 0.8395)
(0.8300, 0.8437)
(0.8350, 0.8479)
(0.8400, 0.8521)
(0.8450, 0.8564)
(0.8500, 0.8607)
(0.8550, 0.8650)
(0.8600, 0.8694)
(0.8650, 0.8737)
(0.8700, 0.8781)
(0.8750, 0.8825)
(0.8800, 0.8869)
(0.8850, 0.8914)
(0.8900, 0.8958)
(0.8950, 0.9003)
(0.9000, 0.9048)
(0.9050, 0.9094)
(0.9100, 0.9139)
(0.9150, 0.9185)
(0.9200, 0.9231)
(0.9250, 0.9277)
(0.9300, 0.9324)
(0.9350, 0.9371)
(0.9400, 0.9418)
(0.9450, 0.9465)
(0.9500, 0.9512)
(0.9550, 0.9560)
(0.9600, 0.9608)
(0.9650, 0.9656)
(0.9700, 0.9704)
(0.9750, 0.9753)
(0.9800, 0.9802)
(0.9850, 0.9851)
(0.9900, 0.9900)
(0.9950, 0.9950)
(1.0000, 1.0000)
};
\end{axis}\end{tikzpicture}
	}
	\subfloat[
    $k = K - 5$
	]
	{
	\begin{tikzpicture}\begin{axis}[height=4.5cm,width=4.5cm,xmin=0,xmax=1,ymin=0,ymax=1,xtick={0,1},xticklabels={0,1},ytick={0,1},yticklabels={0,1},]
\addplot [draw=black, line width=0.5mm] coordinates {
(0.0050, 0.3297)
(0.0100, 0.3318)
(0.0150, 0.3339)
(0.0200, 0.3360)
(0.0250, 0.3380)
(0.0300, 0.3402)
(0.0350, 0.3423)
(0.0400, 0.3444)
(0.0450, 0.3465)
(0.0500, 0.3487)
(0.0550, 0.3508)
(0.0600, 0.3530)
(0.0650, 0.3552)
(0.0700, 0.3574)
(0.0750, 0.3596)
(0.0800, 0.3618)
(0.0850, 0.3640)
(0.0900, 0.3662)
(0.0950, 0.3685)
(0.1000, 0.3707)
(0.1050, 0.3730)
(0.1100, 0.3753)
(0.1150, 0.3776)
(0.1200, 0.3799)
(0.1250, 0.3822)
(0.1300, 0.3845)
(0.1350, 0.3868)
(0.1400, 0.3892)
(0.1450, 0.3915)
(0.1500, 0.3939)
(0.1550, 0.3963)
(0.1600, 0.3987)
(0.1650, 0.4011)
(0.1700, 0.4035)
(0.1750, 0.4059)
(0.1800, 0.4083)
(0.1850, 0.4108)
(0.1900, 0.4133)
(0.1950, 0.4157)
(0.2000, 0.4182)
(0.2050, 0.4207)
(0.2100, 0.4232)
(0.2150, 0.4257)
(0.2200, 0.4283)
(0.2250, 0.4308)
(0.2300, 0.4334)
(0.2350, 0.4359)
(0.2400, 0.4385)
(0.2450, 0.4411)
(0.2500, 0.4437)
(0.2550, 0.4463)
(0.2600, 0.4489)
(0.2650, 0.4516)
(0.2700, 0.4542)
(0.2750, 0.4569)
(0.2800, 0.4596)
(0.2850, 0.4623)
(0.2900, 0.4650)
(0.2950, 0.4677)
(0.3000, 0.4704)
(0.3050, 0.4732)
(0.3100, 0.4759)
(0.3150, 0.4787)
(0.3200, 0.4815)
(0.3250, 0.4843)
(0.3300, 0.4871)
(0.3350, 0.4899)
(0.3400, 0.4927)
(0.3450, 0.4956)
(0.3500, 0.4984)
(0.3550, 0.5013)
(0.3600, 0.5042)
(0.3650, 0.5071)
(0.3700, 0.5100)
(0.3750, 0.5129)
(0.3800, 0.5158)
(0.3850, 0.5188)
(0.3900, 0.5218)
(0.3950, 0.5247)
(0.4000, 0.5277)
(0.4050, 0.5307)
(0.4100, 0.5338)
(0.4150, 0.5368)
(0.4200, 0.5398)
(0.4250, 0.5429)
(0.4300, 0.5460)
(0.4350, 0.5491)
(0.4400, 0.5522)
(0.4450, 0.5553)
(0.4500, 0.5584)
(0.4550, 0.5616)
(0.4600, 0.5647)
(0.4650, 0.5679)
(0.4700, 0.5711)
(0.4750, 0.5743)
(0.4800, 0.5775)
(0.4850, 0.5807)
(0.4900, 0.5840)
(0.4950, 0.5872)
(0.5000, 0.5905)
(0.5050, 0.5938)
(0.5100, 0.5971)
(0.5150, 0.6004)
(0.5200, 0.6037)
(0.5250, 0.6071)
(0.5300, 0.6104)
(0.5350, 0.6138)
(0.5400, 0.6172)
(0.5450, 0.6206)
(0.5500, 0.6240)
(0.5550, 0.6275)
(0.5600, 0.6309)
(0.5650, 0.6344)
(0.5700, 0.6379)
(0.5750, 0.6414)
(0.5800, 0.6449)
(0.5850, 0.6484)
(0.5900, 0.6519)
(0.5950, 0.6555)
(0.6000, 0.6591)
(0.6050, 0.6627)
(0.6100, 0.6663)
(0.6150, 0.6699)
(0.6200, 0.6735)
(0.6250, 0.6772)
(0.6300, 0.6809)
(0.6350, 0.6845)
(0.6400, 0.6882)
(0.6450, 0.6920)
(0.6500, 0.6957)
(0.6550, 0.6994)
(0.6600, 0.7032)
(0.6650, 0.7070)
(0.6700, 0.7108)
(0.6750, 0.7146)
(0.6800, 0.7184)
(0.6850, 0.7223)
(0.6900, 0.7261)
(0.6950, 0.7300)
(0.7000, 0.7339)
(0.7050, 0.7378)
(0.7100, 0.7417)
(0.7150, 0.7457)
(0.7200, 0.7497)
(0.7250, 0.7536)
(0.7300, 0.7576)
(0.7350, 0.7616)
(0.7400, 0.7657)
(0.7450, 0.7697)
(0.7500, 0.7738)
(0.7550, 0.7779)
(0.7600, 0.7820)
(0.7650, 0.7861)
(0.7700, 0.7902)
(0.7750, 0.7944)
(0.7800, 0.7985)
(0.7850, 0.8027)
(0.7900, 0.8069)
(0.7950, 0.8111)
(0.8000, 0.8154)
(0.8050, 0.8196)
(0.8100, 0.8239)
(0.8150, 0.8282)
(0.8200, 0.8325)
(0.8250, 0.8368)
(0.8300, 0.8412)
(0.8350, 0.8455)
(0.8400, 0.8499)
(0.8450, 0.8543)
(0.8500, 0.8587)
(0.8550, 0.8632)
(0.8600, 0.8676)
(0.8650, 0.8721)
(0.8700, 0.8766)
(0.8750, 0.8811)
(0.8800, 0.8856)
(0.8850, 0.8902)
(0.8900, 0.8947)
(0.8950, 0.8993)
(0.9000, 0.9039)
(0.9050, 0.9085)
(0.9100, 0.9132)
(0.9150, 0.9178)
(0.9200, 0.9225)
(0.9250, 0.9272)
(0.9300, 0.9319)
(0.9350, 0.9367)
(0.9400, 0.9414)
(0.9450, 0.9462)
(0.9500, 0.9510)
(0.9550, 0.9558)
(0.9600, 0.9606)
(0.9650, 0.9655)
(0.9700, 0.9704)
(0.9750, 0.9752)
(0.9800, 0.9802)
(0.9850, 0.9851)
(0.9900, 0.9900)
(0.9950, 0.9950)
(1.0000, 1.0000)
};
\addplot [draw=black, line width=0.4mm, densely dashed] coordinates {
(0.0050, 0.3697)
(0.0100, 0.3716)
(0.0150, 0.3734)
(0.0200, 0.3753)
(0.0250, 0.3772)
(0.0300, 0.3791)
(0.0350, 0.3810)
(0.0400, 0.3829)
(0.0450, 0.3848)
(0.0500, 0.3867)
(0.0550, 0.3887)
(0.0600, 0.3906)
(0.0650, 0.3926)
(0.0700, 0.3946)
(0.0750, 0.3965)
(0.0800, 0.3985)
(0.0850, 0.4005)
(0.0900, 0.4025)
(0.0950, 0.4045)
(0.1000, 0.4066)
(0.1050, 0.4086)
(0.1100, 0.4107)
(0.1150, 0.4127)
(0.1200, 0.4148)
(0.1250, 0.4169)
(0.1300, 0.4190)
(0.1350, 0.4211)
(0.1400, 0.4232)
(0.1450, 0.4253)
(0.1500, 0.4274)
(0.1550, 0.4296)
(0.1600, 0.4317)
(0.1650, 0.4339)
(0.1700, 0.4360)
(0.1750, 0.4382)
(0.1800, 0.4404)
(0.1850, 0.4426)
(0.1900, 0.4449)
(0.1950, 0.4471)
(0.2000, 0.4493)
(0.2050, 0.4516)
(0.2100, 0.4538)
(0.2150, 0.4561)
(0.2200, 0.4584)
(0.2250, 0.4607)
(0.2300, 0.4630)
(0.2350, 0.4653)
(0.2400, 0.4677)
(0.2450, 0.4700)
(0.2500, 0.4724)
(0.2550, 0.4747)
(0.2600, 0.4771)
(0.2650, 0.4795)
(0.2700, 0.4819)
(0.2750, 0.4843)
(0.2800, 0.4868)
(0.2850, 0.4892)
(0.2900, 0.4916)
(0.2950, 0.4941)
(0.3000, 0.4966)
(0.3050, 0.4991)
(0.3100, 0.5016)
(0.3150, 0.5041)
(0.3200, 0.5066)
(0.3250, 0.5092)
(0.3300, 0.5117)
(0.3350, 0.5143)
(0.3400, 0.5169)
(0.3450, 0.5194)
(0.3500, 0.5220)
(0.3550, 0.5247)
(0.3600, 0.5273)
(0.3650, 0.5299)
(0.3700, 0.5326)
(0.3750, 0.5353)
(0.3800, 0.5379)
(0.3850, 0.5406)
(0.3900, 0.5434)
(0.3950, 0.5461)
(0.4000, 0.5488)
(0.4050, 0.5516)
(0.4100, 0.5543)
(0.4150, 0.5571)
(0.4200, 0.5599)
(0.4250, 0.5627)
(0.4300, 0.5655)
(0.4350, 0.5684)
(0.4400, 0.5712)
(0.4450, 0.5741)
(0.4500, 0.5769)
(0.4550, 0.5798)
(0.4600, 0.5827)
(0.4650, 0.5857)
(0.4700, 0.5886)
(0.4750, 0.5916)
(0.4800, 0.5945)
(0.4850, 0.5975)
(0.4900, 0.6005)
(0.4950, 0.6035)
(0.5000, 0.6065)
(0.5050, 0.6096)
(0.5100, 0.6126)
(0.5150, 0.6157)
(0.5200, 0.6188)
(0.5250, 0.6219)
(0.5300, 0.6250)
(0.5350, 0.6281)
(0.5400, 0.6313)
(0.5450, 0.6344)
(0.5500, 0.6376)
(0.5550, 0.6408)
(0.5600, 0.6440)
(0.5650, 0.6473)
(0.5700, 0.6505)
(0.5750, 0.6538)
(0.5800, 0.6570)
(0.5850, 0.6603)
(0.5900, 0.6637)
(0.5950, 0.6670)
(0.6000, 0.6703)
(0.6050, 0.6737)
(0.6100, 0.6771)
(0.6150, 0.6805)
(0.6200, 0.6839)
(0.6250, 0.6873)
(0.6300, 0.6907)
(0.6350, 0.6942)
(0.6400, 0.6977)
(0.6450, 0.7012)
(0.6500, 0.7047)
(0.6550, 0.7082)
(0.6600, 0.7118)
(0.6650, 0.7153)
(0.6700, 0.7189)
(0.6750, 0.7225)
(0.6800, 0.7261)
(0.6850, 0.7298)
(0.6900, 0.7334)
(0.6950, 0.7371)
(0.7000, 0.7408)
(0.7050, 0.7445)
(0.7100, 0.7483)
(0.7150, 0.7520)
(0.7200, 0.7558)
(0.7250, 0.7596)
(0.7300, 0.7634)
(0.7350, 0.7672)
(0.7400, 0.7711)
(0.7450, 0.7749)
(0.7500, 0.7788)
(0.7550, 0.7827)
(0.7600, 0.7866)
(0.7650, 0.7906)
(0.7700, 0.7945)
(0.7750, 0.7985)
(0.7800, 0.8025)
(0.7850, 0.8065)
(0.7900, 0.8106)
(0.7950, 0.8146)
(0.8000, 0.8187)
(0.8050, 0.8228)
(0.8100, 0.8270)
(0.8150, 0.8311)
(0.8200, 0.8353)
(0.8250, 0.8395)
(0.8300, 0.8437)
(0.8350, 0.8479)
(0.8400, 0.8521)
(0.8450, 0.8564)
(0.8500, 0.8607)
(0.8550, 0.8650)
(0.8600, 0.8694)
(0.8650, 0.8737)
(0.8700, 0.8781)
(0.8750, 0.8825)
(0.8800, 0.8869)
(0.8850, 0.8914)
(0.8900, 0.8958)
(0.8950, 0.9003)
(0.9000, 0.9048)
(0.9050, 0.9094)
(0.9100, 0.9139)
(0.9150, 0.9185)
(0.9200, 0.9231)
(0.9250, 0.9277)
(0.9300, 0.9324)
(0.9350, 0.9371)
(0.9400, 0.9418)
(0.9450, 0.9465)
(0.9500, 0.9512)
(0.9550, 0.9560)
(0.9600, 0.9608)
(0.9650, 0.9656)
(0.9700, 0.9704)
(0.9750, 0.9753)
(0.9800, 0.9802)
(0.9850, 0.9851)
(0.9900, 0.9900)
(0.9950, 0.9950)
(1.0000, 1.0000)
};
\end{axis}\end{tikzpicture}
	}
	\caption{
	Polynomial regularizer $f_k(x)$
	in \Cref{prop:poly}
	(solid line)
	versus 
	$e^{x-1}$ (dashed line).
    }
	\label{fig:poly}
\end{figure}
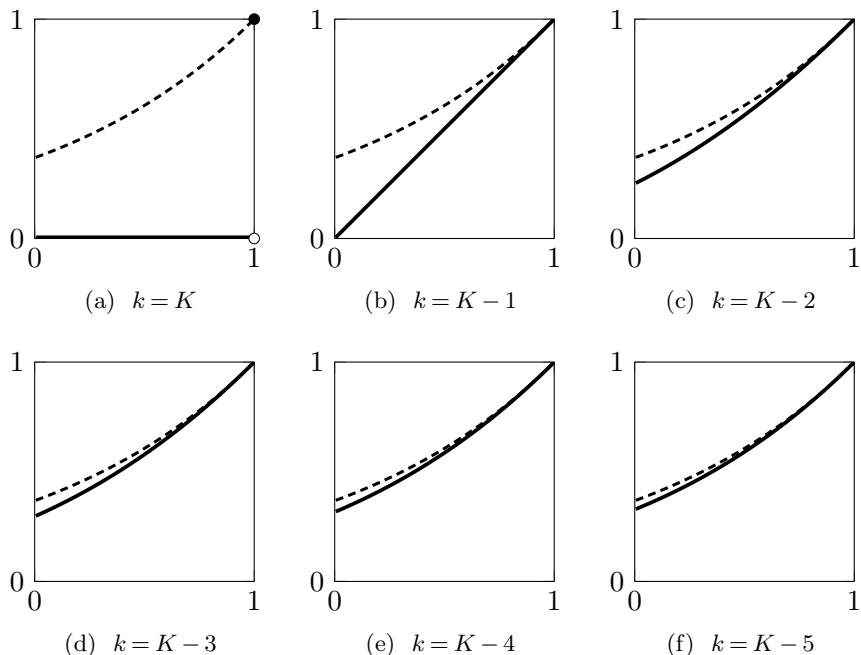

Our key technical ingredient that formalizes the above intuitive notion of regularization intensity is introducing a particular sequence $f_1,f_2,\ldots,f_K:[0,1]\rightarrow [0,1]$ of \emph{polynomials of decreasing degrees}, where $f_k$ is a polynomial of degree $(K-k)$ associated to stage $k\in[K]$.  See \Cref{fig:poly} and \Cref{prop:poly}. Here, the regularization intensity of each polynomial is essentially governed by its degree --- hence our sequences starts with the highest degree and goes in the descending order of polynomial degrees. We identify this sequence using a backward recursive equation (will be detailed later). We then show this sequence \emph{exactly} characterizes the regularizers of the optimal competitive multi-stage algorithm.
% More precisely, we use the primitive function $F_k(x)\triangleq\int_{0}^{x}f_k(y)dy$ to map the normalized fractional degree (or normalized allocation level) of each offline vertex to a penalty term in $[0,1]$ at stage $k$. We then subtract the weighted sum of these penalties from the total weight objective function to construct our regularized objective function; see convex programs \ref{eq:concave-k-bmatching} and \ref{eq:concave-adwords-k} for more details.
Interestingly, the first-stage polynomial $f_1$ converges to $f(x)= e^{x-1}$ as $K$ goes to infinity. Also, in the special case of batches of size one, replacing the polynomial $f_k$ at all stages with \emph{the same function} $f(x)=e^{x-1}$ recovers the $(1-1/e)$-competitive BALANCE algorithms in \cite{AGKM-11} for the online vertex weighted b-matching.

To analyze our polynomial regularized matching algorithms --- inspired by the technique introduced in \cite{feng2020two} for two-stage matching --- we use the strong duality of the convex programs, tied to a specific primal-dual analysis that is quite different from the known primal-dual analysis in the literature for various forms of online bipartite allocation problems. The sketch of our analysis is roughly the following: The Karush–Kuhn–Tucker (KKT) conditions at each stage $k$ offer a structural decomposition of the subgraph induced by the $k^{\text{th}}$ batch. The special unweighted case of this decomposition coincides with the well-studied \emph{matching skeleton} introduced by~\cite{GKK-12}, which itself is closely related to the Edmonds-Gallai decomposition~\citep{edm-65,gal-64} of bipartite graphs. Using this decomposition, and by carefully using our polynomials $f_1,\ldots,f_k$, we propose a novel \emph{recursive primal-dual framework} to bound the competitive ratio of our fractional matchings. The resulting dual constructions, together with our recursive formulation, shows how to improve the competitive ratios from $(1-1/e)$ to $\cratio{K}$ in the multi-stage setting with $K$ batches. In other words, this new algorithmic construct extends the dual constructions in previous work~\citep{MSVV-07,AGKM-11,BJN-07} in a novel fashion to the multi-stage setting with batch arrivals. As a side dish, our technique provides a convex-programming based interpretation for the existing primal-dual proofs in the special case of the online setting. 

\revcolor{
\smallskip
\noindent{\textbf{New Challenges for the Configuration Allocation~}}Switching to the multi-stage configuration allocation problem with free disposals, we aim to extend our earlier algorithm to this quite general model; nevertheless, as it turns out, our earlier algorithmic technique cannot be applied directly to the multi-stage configuration allocation problem because of three new fundamental technical challenges: (i)~\emph{Price-level treatment:} each advertiser pays for impressions at different price levels depending on the selected configuration, which is in a stark contrast to vertex-weighted matching where all impressions are essentially sold at the same price. Thus, it is not clear how to define a regularizer term to mimic the earlier approach,  (ii)~\emph{Lack of skeleton structure:} moving from matching to configuration allocation breaks several of the structural properties of the problem, most importantly the existence of a matching skeleton graph decomposition as mentioned earlier, which is crucial in the analysis of the multi-stage vertex-weighted matching, and (iii)~\emph{Managing preemption:} as the algorithm decides on the choice of the configuration at each time, it implicitly decides also on the number of impressions that each advertiser is responsible to pay at each price level at this point (and basically the rest are preempted); these price-level  dependent preemption or free-disposal decisions regarding previous impression-allocations are crucial to be able to obtain any bounded competitive algorithm, and a priori it is not  clear how to extend our previous approach to help with these preemption decisions.}

\revcolor{The first new key technical ingredient 
to circumvent the above challenges is the stage-dependent 
\emph{price-level regularized convex program}
\ref{eq:convex primal}
% stage-dependent convex programs
used by \Cref{alg:opt} in each stage $k\in[K]$. Similar to our previous convex programs, this new program maximizes the total revenue minus a regularization term by fractionally assigning a configuration to each user in batch $k$ and deciding which impressions should be preempted by each advertiser during stage $k$. This time, the regularization term is a summation of several terms, where we have one term for each advertiser
at \emph{each possible per-impression price level}. More precisely, we keep track of the price
distribution of top impressions allocated to each advertiser given her capacity
(i.e., the number of impressions at different per-impression price levels that the advertiser is willing to pay so far), instead of just her remaining capacity.
Having this distribution at the beginning of each stage $k$ as its input, the regularizer term corresponding to this advertiser is a separable function across all the price levels, where each term applies the primitive polynomial $F_k(x)$ in \Cref{prop:poly} to map the (normalized) number of allocated impressions at some price level to a penalty term in the objective. Intuitively speaking, this new form of regularization helps with gradually decreasing the hedging for each price level separately across stages.

The second new key technical ingredient is how to connect convex programs
\ref{eq:convex primal}
in each stage to an LP primal-dual analysis for the configuration allocation. As mentioned earlier, in contrast to the vertex-weighted matching, in general configuration allocation the resulting convex programs
\emph{do not admit} any matching skeleton or similar graph decompositions. The absence of this combinatorial structure, which has been the backbone of our previous primal-dual analysis, combined with price-level regularization and the need to incorporate preemption decisions in the analysis, make this new analysis more challenging. We circumvent this last obstacle by considering the Lagrangian dual program
\ref{eq:convex dual}
of the convex program
\ref{eq:convex primal}, 
and \emph{directly} using its optimal dual solution
to construct the dual assignment in the 
primal-dual analysis framework. Surprisingly, even with this different proposed dual assignment, we can connect the analysis of the competitive ratio to the same recursive functional equation across $K$ stages as before, and thus obtaining the $\cratio{K}$ competitive ratio by \Cref{alg:opt}. We should highlight that the connection between LP primal-dual analysis and the convex duality in our paper might be of independent interest in other online algorithm design problems. }

We finally show the competitive ratio $\cratio{K}$ is the best achievable by any fractional multi-stage algorithm, whether deterministic or randomized, by showing an upper-bound instance for the special case of the unweighted multi-stage matching. Notably, this instance relies on batches of size more than one, and hence cannot be used as an upper-bound instance for the special case of online arrival where batch sizes are all one. See \Cref{fig:bad-example} for more details.

% We finally turn our attention to the integral versions of the vertex weighted b-matching and AdWords problems, when the capacities are large (or equivalently, bid over budget ratios are small). By using the totally unimodularity of flow-type polytopes~\citep{sch-03}, we propose our deterministic rounding algorithm for the vertex weighted b-matching problem. The idea is finding integer lower and upper bounds on the fractional degrees of each offline vertex using the fractional solution of the convex program at each stage, and then searching for an integral matching satisfying these bounds.  For the AdWords problem, we use a straightforward variant of independent randomized rounding at each stage, after slightly reducing all the budgets. By applying standard concentration bounds for each offline vertex, we prove that maintaining feasibility of the resulting integral allocation only adds a small loss in expectation when the bid over budget ratios are small. 

\smallskip
\noindent\textbf{Managerial Insights~} Our study sheds insights on the role of batching in service operation of matching platforms, and provides algorithmic evidences on how they should adaptively tune their batched matching algorithms to achieve the right trade-off between \emph{``greedy-ness''} in the current decision and {``hedging''} for the sake of future decisions \revcolor{in a prior-free non-stationary environment}:\footnote{\revcolor{We would like to reiterate that in our paper we make no assumptions on the number of nodes in each batch.}}
\begin{itemize}
    \item In settings that allow for latency---and hence allow for batching the demand arrivals---batched matching algorithms (versus fully online algorithms) can be used by the algorithm designer to obtain performance improvements, \revcolor{even when the environment is highly non-stationary}.
    \item There is an inherent trade-off between the gain from batching (which is captured by the competitive ratio in our competitive analysis framework) and the degree of batching (which is captured by the inverse number of batches in our framework) \revcolor{in prior-free non-stationary dynamic allocations}: with less number of batches/stages---which accounts for a higher degree of batching---a well-designed multi-stage matching algorithm can obtain a higher performance; moreover, it gets closer to the offline optimal matching as the degree of batching increases.\footnote{Of course, there is no free lunch here and the actual cost of batching is the induced latency in decision making.}
    \item When designing batched matching algorithms, our result suggests that pushing the matching to be more robust to future uncertainty early on (i.e., to be more balanced) and then pushing it to be more greedy later on will help with the performance in a finite horizon problem. In particular, we could formally capture this design aspect using our decreasing degree polynomial regularizers; However, in general, decreasing the degree of hedging induced by the matching algorithm over time to obtain improved performance is the main operational takeaway message of our work, and there might be other ways of performing this task in practice.
\end{itemize}

\revcolor{\smallskip
\noindent\textbf{Organization~} In \Cref{sec:prelim}, we start by formalizing both our baseline model (\Cref{sec:base-model}) and extension model (\Cref{sec:extension-model}).
In \Cref{sec:vertex-weight}, we introduce our first algorithm  by starting with some illustrative examples (\Cref{sec:example}). We then present this optimal competitive multi-stage algorithm for the vertex weighted bipartite matching problem (\Cref{sec:upper-bound}), and analyze its competitive ratio lower bound (\Cref{sec:primal-dual}). In \Cref{sec:extensions}, we switch to our extension model. We present our second algorithm, which is an optimal competitive multi-stage algorithm for the configuration allocation problem (\Cref{sec:alg-convex}), and characterize some of its properties (\Cref{sec:convex program}). We then analyze its competitive ratio lower bound (\Cref{sec:main proof}).
We finally show that our competitive ratio lower-bounds \emph{exactly} match the upper-bound on the competitive ratio of \emph{any} multi-stage algorithm (\Cref{sec:lowerbound}) and numerically evaluate the performance of our multi-stage algorithms on synthetic instances (Appendix~\ref{apx:numerical}).}

\subsection{Further Related Literature}
\label{sec:related}
Here is a summary of the other work in the literature directly related to our work.

\vspace{2mm}
\noindent\emph{Multi-stage matching.} \revcolor{Our work is closely related to \cite{feng2020two}, which studies the optimal competitive two-stage vertex weighted matching and pricing under both adversarial and stochastic settings. We extend their result in the adversarial case to multi stages and settings beyond vertex weighted matching (in particular to the more general settings of multi-stage edge-weighted matching and configuration allocation with free disposal). In \cite{feng2020two} basically there is a large class of convex programs that can be used in the first stage to introduce hedging. In our paper, we diverge by considering a \emph{different} form of convex-programming based matching where we use regularizers, and we show by changing the regularizer at each stage the optimal competitive ratio can be achieved for any $K$. Notably, both papers use a primal-dual analysis to analyze the underlying algorithm, but our dual construction in the analysis follows a different logic and has a different algebraic form (even for the special case of the two-stage problem).} Our result for the unweighted multi-stage matching is closely related to the beautiful work of \cite{LS-17}, which also uses the matching skeleton of \cite{GKK-12}, but for a different model of arrivals and online problem.

% In their model, a batch of \emph{new edges} arrive in each stage. Their proposed algorithm first finds a maximum fractional matching using the matching skeleton in each stage, and then removes some of the edges sampled from this matching through a randomized procedure. This approach obtains an optimal $2/3$ competitive ratio for two stages and a non-optimal $1/2+O(2^{-K})$ competitive ratio for $K$ stages. 
\vspace{2mm}

\noindent{\emph{Two-stage stochastic programming.}}
Two-stage stochastic combinatorial
optimization problems have been studied extensively in the 
literature
\citep[e.g.,][]{BL-11, CCP-05,SS-04,SS-07,IKMM-04, hanasusanto2015k,hanasusanto2018conic}.
Also, see the excellent book of \cite{kuhn2006generalized} for more context regarding convex multi-stage stochastic programs.  For two-stage stochastic matching, various models with different objectives
have been studied in \citet{KS-06, EGMS-10, KKU-08} and more recently in \cite{feng2020two}. Another work that is conceptually related to us is the recent work of \cite{housni2020matching}, which is inspired by the idea of robust multi-stage optimization~\citep[e.g.,][]{bertsimas2010optimality,bertsimas2011theory}. They consider a two-stage robust optimization for cost-minimization of the matching, where future demand uncertainty is modeled using a set of demand scenarios (specified explicitly or implicitly). Our model and results diverge drastically from this work.

\vspace{2mm}
\noindent\emph{Online bipartite allocations.}
Our results are comparable to the literature on online bipartite matching (with vertex arrival) and its extensions. Besides the work mentioned earlier, related to us is \cite{birnbaum2008line}, that gives a simple non primal-dual proof of RANKING algorithm of \cite{KVV-90} for online bipartite matching. An elegant primal-dual analysis is discovered later in \cite{DJK-13}. Also, at high-level, our primal-dual analysis resembles some aspects of ~\cite{BJN-07,FKMMP-09,devanur2009adwords,devanur2012online,esfandiari2015online,HKTWZZ-18,FNS-19}. Other related models are online bipartite stochastic matching~ \citep[e.g.,][]{feldman2009online,bahmani2010improved,MGS-12,mehta2014online,JL-14} and online bipartite matching with stochastic rewards~\citep[e.g.,][]{mehta2012online,goyal2020online,huang2020online}. We refer interested readers to a \revcolor{comprehensive} survey by \cite{mehta2013online} for further related work. Finally, related to us is also the literature on stochastic online packing and convex programming with large budgets~\citep[e.g.,][]{FMMS-10,DJSW-11,AD-14}. In particular, the role of concave regularizers have been studied in these settings for different purposes, e.g., fairness~\citep{agrawal2018proportional,balseiro2021regularized}, robustness to adversarial corruptions~\citep{balseiro2020best}, and obtaining blackbox reductions from Bayesian mechanism design to algorithm design~\citep{dughmi2021bernoulli}. 

\vspace{2mm}
\noindent\emph{Applications in revenue management and service operations.} From both methodological and application points of view, our work is related to several streams of work that focus to design online allocations and matching algorithms for problems in revenue management and service operations. These work pertain to practical applications in online retail, ride-sharing and other online platforms. Some examples are online assortment optimization (e.g., 
\citealp{GNR-14,MS-19}), online assortment of reusable resources  (e.g.,
\citealp{gong2019online,FNS-19,goyal2020onlineb}), Bayesian online assortment with heterogeneous customers (e.g.,
\citealp{rusmevichientong2020dynamic,feng2020near}), 
online assortment with replenishment~\citep{feng2021online}, dynamic pricing with static calendar \citep{MSZ-18}, volunteer crowd-sourcing and nudging~\citep{manshadi2020online}, two-sided assortment optimization~\citep{aouad2020online}, online allocation strategies in two-sided media platforms~\citep{alaei2020revenue}, \revcolor{online allocation of patient demand requests~\citep{golrezaei2021online}}, and finally the line of work on dynamic windowed supply-demand matching/queuing---with applications in ride-sharing (e.g., \citealp{ashlagi2019edge,aouad2020dynamic,ozkan2020dynamic,ata2020dynamic,aveklouris2021matching,arnosti2021managing}). \revcolor{Another recent work related to us is \cite{xie2022benefits} which studies the effect of delay (or equivalently look-ahead) in large-traffic stochastic stationary settings such as the multi-secretary problem~\citep{vera2021bayesian}. Our paper diverges from this work by considering non-stationary and adversarial arrival under batching instead of delay.}

% \subsection{from old intro}

% A canonical formulation to capture gist of some existing tensions in the above dynamic decision making scenarios is the (adversarial) online bipartite matching problem, originated from the seminal work of Karp, Vazirani, and Vazriani~\citep{KVV-90}. In this model, the online and offline vertices simply play the role of the arriving demand requests and the available supply, respectively. A rich body of literature has extended this simple formulation to more practical two-sided online allocation scenarios such as the online budgeted allocation/AdWords~\citep{MSVV-07} and the vertex weighted budgeted-matching~\citep{AGKM-11}. Other extensions and variants tailored to applications such as display advertising, ride-sharing, and assortment optimization are also studied~\citep{FKMMP-09, mehta2012online,MGS-12,GNR-14,HKTWZZ-18}. See \cite{mehta2013online} for a survey. On the algorithmic side, a sequence of past work have developed an elegant primal-dual framework to design and analyze online competitive algorithms in these problems~\citep{BJN-07,DJK-13,devanur2012online}. See \cite{buchbinder2009design} for a survey on the primal-dual technique in online algorithms, and specifically for designing and analyzing online bipartite allocation algorithms. 

\vspace{2mm}
\noindent\emph{Video-ads allocation.}
The popularity of in-stream video platforms has ignited 
a new research area related to video-ads, see 
a recent survey by \citet{FOG-21} for a detailed discussion.
There is a limited literature on
the algorithmic study of video-ads allocation.
Comparing to more traditional applications such as display ads,
video-ads (especially in-stream ads) 
have a distinct feature, which is the
need to be shown to the user during the video session.
\citet{DGNR-09} 
introduce the ``online story scheduling'' problem in which
the task is to schedule in-stream ads with 
different time lengths and per-unit prices for a single user in an online 
fashion, and
provide a constant competitive algorithm. 
Further improvements and 
studying other variants of this problem 
have been the focus in the literature, e.g., \citet{AP-19,LMQTYZ-16}.
\citet{BKK-09} study an alternative media scheduling problem 
for multiple users in an offline fashion.
\citet{SKFFC-17}
consider a variant of AdWords problem \citep{MSVV-07},
where each offline node (which is an in-stream ad in their model) is associated with 
a time length and every online node (i.e., a user) can be matched with multiple offline nodes as long as the total time length is below some threshold. 

\section{Preliminaries}
\smallskip
\label{sec:prelim}
%\subsection{Baseline Model: Multi-stage Edge-weighted Matching}
\subsection{Baseline Model: Multi-stage Vertex Weighted Matching}
\label{sec:base-model}
We study \emph{vertex weighted matching} as our baseline bipartite allocation model for matching arriving demands requests to available supply.
An instance of our model consists of a bipartite graph $\mathcal{G}=(\Rider,\Driver,E)$, where vertices 
in $\Driver$ (supply) are present offline 
and vertices in $\Rider$ (demand) arrive over time. 
Specifically, we consider a \emph{multi-stage arrival} model, 
in which vertices in $\Rider$ arrive in $K\in\mathbb{N}$ stages. 
We assume the number of stages $K$ is fixed and known by the algorithm. At each stage $k\in[K]$, a new subset of online vertices $\Rider_k\subseteq \Rider$ arrive, which we often call an arriving \emph{batch}, and reveal their incident edges in $E$. \revcolor{We assume $\lvert\Rider_k\rvert \geq 1$ and we make no further assumptions on $\Rider_k$'s.} We highlight that $\Rider=\cup_{k\in[K]}\Rider_k $ and $\Rider_k\cap\Rider_{k'}=\emptyset$ for $k\neq k'$. Each offline vertex $j$ has a total capacity $n_j$ for allocations, as well as a non-negative weight $w_j$ for each unit of allocation. 
After arrival of batch $k$, the multi-stage allocation algorithm makes an irrevocable allocation decision by allocating each vertex $i\in\Rider_K$ (which corresponds to one unit of demand) to the set of neighboring offline vertices $j\in\Driver, (i,j)\in E$ that have non-zero remaining capacity. Let $E_k$ denote the set of edges in $E$ incident to $\Rider_k$. 
We denote the allocation of stage $k$ by the vector $\xbf_{k}\in\mathbb{R}^{E_k}$, where $x_{k,ij}\in[0,1]$ is the amount of allocation of vertex $i$ to $j$. 
This allocation can be fractional or integral. We mostly focus on fractional allocations and assume $n_j=1$ for all the offline vertices $j\in\Driver$ without the loss of generality.\footnote{\revcolor{We are essentially considering fractional algorithms and adversarial arrivals; in such a setting, without the loss of generality, we can focus on adversaries that choose the input instance adaptively.}}  The goal of the multi-stage vertex weighted matching algorithm is to maximize the total weight of offline vertices in the (fractional) allocation, while ensuring  feasibility, i.e., each online (resp. offline) vertex uses no more than one unit of demand (resp. capacity).

\revcolor{\subsection{Extension Model: Multi-stage Configuration Allocation}
\label{sec:extension-model}
Motivated by the application of displaying video-ads in video streaming platforms, we also study the extension of our baseline model to \emph{multi-stage configuration allocation}. An instance of this model consists of a set of offline advertisers $\Driver$ who are willing to pay the platform to display their ads to its users and a set of arriving users $\Rider$ who are planning to watch videos on this platform. Specifically, 
the platform needs to choose
an advertising \emph{configuration} $\config\in\Config$
for each user (or equivalently, for each video session watched by a user), where $\Config$ is the set of feasible configurations.\footnote{Our model and results allow a different feasible set for each user, but we omit that for simplicity.} Fixing a user $\rider\in \Rider$,
choosing each configuration $\config\in\Config$ 
allocates a certain number of reserved impressions 
$\alloc_{\threeindex}$
at a non-negative per-impression price $\weight_{\threeindex}$
to each advertiser $\driver\in\Driver$.
% advertiser $\driver$'s impression 
% have been displayed to user $\rider$
% with per-unit price $\weight_{\threeindex}$.
Each advertiser $\driver$
demands $n_\driver$ number of impressions in total. We assume free-disposal, which means that the total allocated impressions to advertiser $j$ can exceed $n_j$, but she is eventually only paying for the top $n_j$ most expensive impressions. Similar to our baseline model in \Cref{sec:base-model}, users 
arrive to the platform over time
in $K\in\N$ stages. At each stage $k\in [K]$
a new batch of online users $\Rider_k\subseteq\Rider$
arrives. Upon the arrival of this batch, the video platform observes 
$\allocs\ked = \{\alloci\}_{\rider\in\Rider_k,\config\in\Config,\driver\in \Driver}$
and 
$\weightbf\ked = \{\weight_{\threeindex}\}_{\rider\in\Rider_k,\config\in\Config,\driver\in \Driver}$. Then the multi-stage allocation algorithm makes an irrevocable decision by
assigning a configuration to each user $\rider\in\Rider_k$. These configuration allocations can be integral or fractional. 
In this paper we focus on the fractional allocations~\footnote{\label{footnote:integral}Using standard randomized rounding techniques, all of our results for the fractional allocation problem, including the algorithm and its competitive analysis, can easily be extended to the setting with integral randomized allocations under the \emph{large budget assumptions}, i.e, in the asymptotic regime when $\underset{i,c,j}{\min}\frac{n_j}{\alloci}$ is large.} and assume $n_j=1$ for all advertisers $j\in\Driver$ without the loss of generality.\footnote{Since we allow fractional solutions, it is without
loss of generality to normalize the parameters after a simple change of variables, so that we can assume $n_\driver = 1$ for all $\driver\in \Driver$ without loss of generality. We omit details for brevity.}
 The goal of the multi-stage configuration allocation algorithm is to choose configurations for the users to maximize the total collected payments from allocating impressions to the advertisers at the end, which we also refer to as the revenue of the platform.}

\revcolor{
\smallskip
\noindent\textbf{LP Formulation and Offline Benchmark}~~To help with understanding our extension model, we formulate the
optimum offline solution in
the configuration allocation
problem as a simple linear program, denoted by  
\ref{eq:LP primal}:
    \begin{equation}
\tag{\textsc{PRIMAL-MCA}}
\label{eq:LP primal}
\arraycolsep=1.4pt\def\arraystretch{1}
\begin{array}{lllll}
\max\limits_{\xbf,\zbf\geq \mathbf{0}}
\quad\quad\quad&
% \mathop{\displaystyle\sum
\displaystyle\sum_{\rider\in\Rider}
\displaystyle\sum_{\config\in\Config}
\displaystyle\sum_{\driver\in\Driver}
\weight_{\threeindex}\cdot
% }
% \limits_{(\threeindex)\in E}
 x_{\threeindex}
&~~\text{s.t.}
% \min\limits_{\alphabf,\betabf,\gammabf}
% \quad\quad\quad 
% &
% \displaystyle\sum_{\rider\in \Rider }{\alpha_\rider}+
% \displaystyle\sum_{\driver\in \Driver}{\beta_\driver}
% &~~\text{s.t.} 
\\[1.4em]
 &
%  \mathop{\displaystyle\sum
\displaystyle\sum_{\rider\in\Rider}
\displaystyle\sum_{\config\in\Config}
 x_{\threeindex} 
%  }
%  \limits_{\rider,\config:(\threeindex)\in E}
 \leq
 n_{\driver} 
 &~~\driver\in\Driver~,
% \gamma_{\threeindex} 
% +
% \beta_\driver \geq \weight_{\threeindex}
% & (\threeindex)\in E~,
\\[1.4em]
 &
 \displaystyle\sum\limits_{\config\in\Config}
 z_{\rider,\config}
 \leq 1 
 &~~
 \rider\in \Rider~, 
% \alpha_\rider 
% \geq 
% \displaystyle\sum\limits_{\driver\in \Driver}
% \alloci\cdot \gamma_{\threeindex} &i\in \Rider, \config\in \Config~, 
\\[1.4em]
 &
 x_{\threeindex} \leq 
 \alloci \cdot z_{\rider,\config}
 &~~
\rider\in\Rider,\config\in\Config,\driver\in\Driver~. 
% \alpha_\rider 
% \geq 0 &\rider \in \Rider~, 
% \\[1.4em]
%  &x_{\threeindex} \geq 0 &(\threeindex)\in E~, &
% & &
% \beta_\driver\geq 0  &\driver\in \Driver~,
% \\[1.4em]
%  &z_{\rider,\config} \geq 0 &\rider\in\Rider, \config\in \Config~. &
% & &
% \gamma_{\threeindex} \geq 0  &(\threeindex)\in E~.
\end{array}
\end{equation}
In the above formulation, $z_{\rider,\config}$
is the fraction of configuration $\config$ assigned
to user $\rider$,
and $x_{\threeindex}$
is the (fractional) number of impressions
created by user $\rider$ for advertiser $\driver$ based on configuration $\config$ that the advertiser is finally willing to pay. Note that the number $x_{\threeindex}$ of impressions that the advertiser is willing to pay is no more than number of allocated impressions $\alloci \cdot z_{\rider,\config}$ from user $\rider$ under configuration $\config$, and the first quantity can be strictly smaller than the second (as advertiser $\driver$ should only pay for at most $n_\driver$ impressions overall). If an impression is allocated but the advertiser is not responsible to pay for it, we refer to that impression as being \emph{preempted} or \emph{freely disposed}.

We finally note that the multi-stage configuration allocation problem
generalizes the multi-stage vertex weighted matching problem. Also, although we allow for free-disposal in the extension model as allocations of an advertiser's capacity can value differently (i.e., be at a different price level) for different users and configurations, for the special case of vertex weighted matching there is no point in free-disposal by any multi-stage algorithm, as all allocations of an offline node's capacity value the same. See the formal details in Appendix~\ref{apx:connection}. This distinction is the key that requires novel extensions of our techniques for the baseline model to capture different price levels.}

\subsection{Competitive Ratio Analysis}

We measure the performance of our multi-stage algorithms by their \emph{competitive ratio}, which is the worst-case ratio between the expected total weight \revcolor{(or total revenue)} of the algorithm and the optimum offline benchmark given the entire problem instance (formally defined below).
\begin{definition}[\textbf{Competitive Ratio}]
	\label{def:competitive ratio}
	Fix the number of stages $K\in\mathbb{N}$. For $\alpha\in [0,1]$, a multi-stage allocation algorithm $\ALG$ is \emph{$\alpha$-competitive} against the optimum offline solution in the multi-stage vertex weighted matching \revcolor{(resp. multi-stage configuration allocation)} problem, if we have:
	\begin{align*}
		\inf\limits_{I \in\mathcal{I}^{(K)}}~~
		\frac{\ALG(I)}{\texttt{OPT}(I)}~
		\geq \alpha~,
	\end{align*}
	where $\mathcal{I}^{(K)}$ is the set of instances of the multi-stage vertex weighted matching  \revcolor{(resp. multi-stage configuration allocation)} problem with $K$ stages, 
	and $\ALG(I)$ and $\texttt{OPT}(I)$ are the total weights \revcolor{(resp. total revenue)} of the multi-stage algorithm $\ALG$ and optimum offline solution on an instance $I$, respectively. 
	For randomized multi-stage algorithms, we abuse the notation and $\ALG(I)$ denotes the expected total weight \revcolor{(resp. expected total revenue)} of the algorithm on an instance $I$.
\end{definition}

%  As an important note, we are mainly interested in the competitive ratio of fractional $K$-stage algorithms.

%  We also consider the asymptotic competitive ratio of integral $K$-stage algorithms (possibly randomized) under the ``large budgets assumption'' in the vertex weighted b-matching problem (i.e., under the assumption that $\min_j B_j=\Omega(1)$), or equivalently under the small bid over budget assumption in the budgeted allocation problem (i.e., under the assumption that $\max_{i,j} \frac{b_{ij}}{B_j}=o(1)$ or equivalently $\min_j B_j=\Omega(1)$ when $b_{ij}\in[0,1]$). 

\subsection{A Recursive Sequence of Polynomials}
\label{sec:poly}
As a novel technical ingredient of our proposed multi-stage allocation algorithms, we identify a particular sequence of polynomials of decreasing degrees. Our construction of these polynomials is recursive, which gives us a polynomial of degree $K-k$ for each stage $k=1,2,\ldots,K-1$. Here, we introduce these polynomials, describe this underlying recursive equation, and elaborate a bit more on some of their properties (\Cref{prop:poly}; see the proof in Appendxi~\ref{apx:poly}). Later, we first show in \Cref{sec:vertex-weight} how these polynomials help us to design an optimal competitive multi-stage algorithm for the vertex weighted matching and how to develop a primal-dual analysis framework that incorporates this recursive equation. \revcolor{Then we show how to extend these ideas and use our polynomials for the general multi-stage configuration allocation problem in \Cref{sec:extensions}, where we require treating allocations at different price levels differently --- as will be clear later.}

\begin{proposition}[{\textbf{{Sequence of Polynomial Regularizers}}}]
\label{prop:poly}
Consider a sequence of functions $f_1,f_2,\ldots,f_{K}:[0,1]\rightarrow[0,1]$
satisfying the following backward recursive equation:
\begin{itemize}
    \item $\forall x\in[0,1]:~f_{K}(x)=\mathbb{I}\{{x=1\}}$,
    where $\mathbb{I}(\cdot)$ is 
    the indicator function,
    \item $\forall x\in[0,1], k\in[K-1]:~f_{k}(x)=\underset{y\in[0,1-x]}{\max}~(1-y)f_{k+1}(x+y)$.
\end{itemize}
Then, the following statements hold:
\begin{enumerate}[label=(\roman*)]
    \item The function sequence $\{f_k\}_{k=1}^{K-1}$ is a sequence of $K-1$ polynomials with decreasing degrees, where $f_k(x)=(1-\frac{1-x}{K-k})^{K-k}$ for $k\in[K-1]$.\footnote{As a convention, we refer to $f_K(x)=\mathbb{I}\{x=1\}$ as a polynomial of degree $0$ (although not mathematically rigorous).}
    \item $f_k(x)$ is monotone increasing and continuous over $[0,1]$ for $k\in[K-1]$. Hence, $F_k(x)\triangleq \int_{0}^x f_k(y)dy$ is differentiable and strictly convex over $[0,1]$ for $k\in[K-1]$.
    \item $\forall k\in[K]$, $f_k(x)\leq e^{x-1}$ for $x\in[0,1]$. Moreover, the first function $f_1(x)=(1-\frac{1-x}{K-1})^{K-1}$ converges point-wise from below to the function $e^{x-1}$ as $K\to+\infty$ (see \Cref{fig:poly}).
    \item $\underset{y\in[0,1]}{\min}1-(1-y)f_1(y)=\cratioexp\triangleq\cratio{K}$
\end{enumerate}
\end{proposition}

The aforementioned polynomials $\{f_k\}_{k\in[K-1]}$ are drawn in \Cref{fig:poly}. As it is clear from this figure, these polynomials provide careful point-wise estimations for the exponential function $f(x)=e^{x-1}$ at different levels of precision. When these polynomials are used to regularize the maximum weight matching (or total revenue) objective functions, they provide different levels of ``hedging'' intensity against the future uncertainty in the arriving input exactly because of these different levels of estimation precision. This point will be clear in the upcoming sections.

\section{Multi-stage Vertex Weighted Matching}
\label{sec:vertex-weight}
In this section we show the result
of the baseline model, 
which is an optimal competitive algorithm for the $K$-stage (fractional) vertex weighted matching problem. We start by a few simple examples to provide intuitions behind some of the main technical ideas used in the design of our algorithm. Then, in \Cref{sec:upper-bound}, we formally describe this algorithm and all of its ingredients. We then provide details of our new convex-programming based primal-dual framework in \Cref{sec:primal-dual}, and show how the recursive sequence of polynomials introduced in \Cref{prop:poly} help with analyzing the competitive ratio. 

% Note that in this section, for ease of exposition, 
% we focus on the the problem of designing fractional multi-stage allocations. 
% We extend our study to the integral vertex weighted b-matching
% and integral budgeted allocation, both  
% under the large budgets assumption, in \Cref{sec:extensions}.

\subsection{A Few Illustrative Examples}
\label{sec:example}

\begin{figure}[hbt!]
	\centering
	\subfloat[
	\label{fig:example-a}
    Ignoring the batch arrival
	]{
\includegraphics[trim=9cm 3cm 10.5cm 1cm,clip,width=0.45\textwidth]{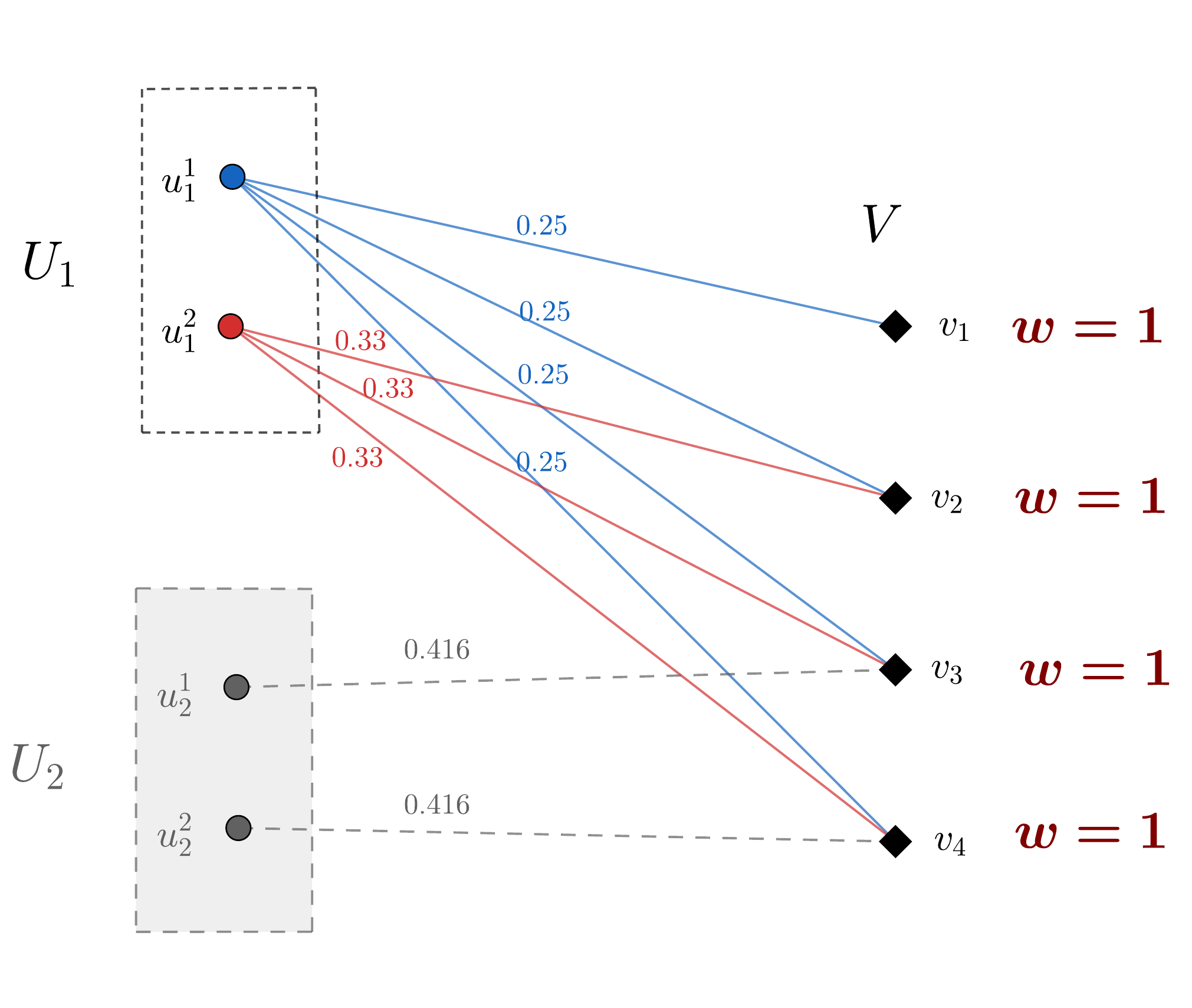}
	}
	\subfloat[
	\label{fig:example-b}
   Considering the batch arrival
	]{
\includegraphics[trim=7.5cm 3.5cm 12cm 0cm,clip,width=0.45\textwidth]{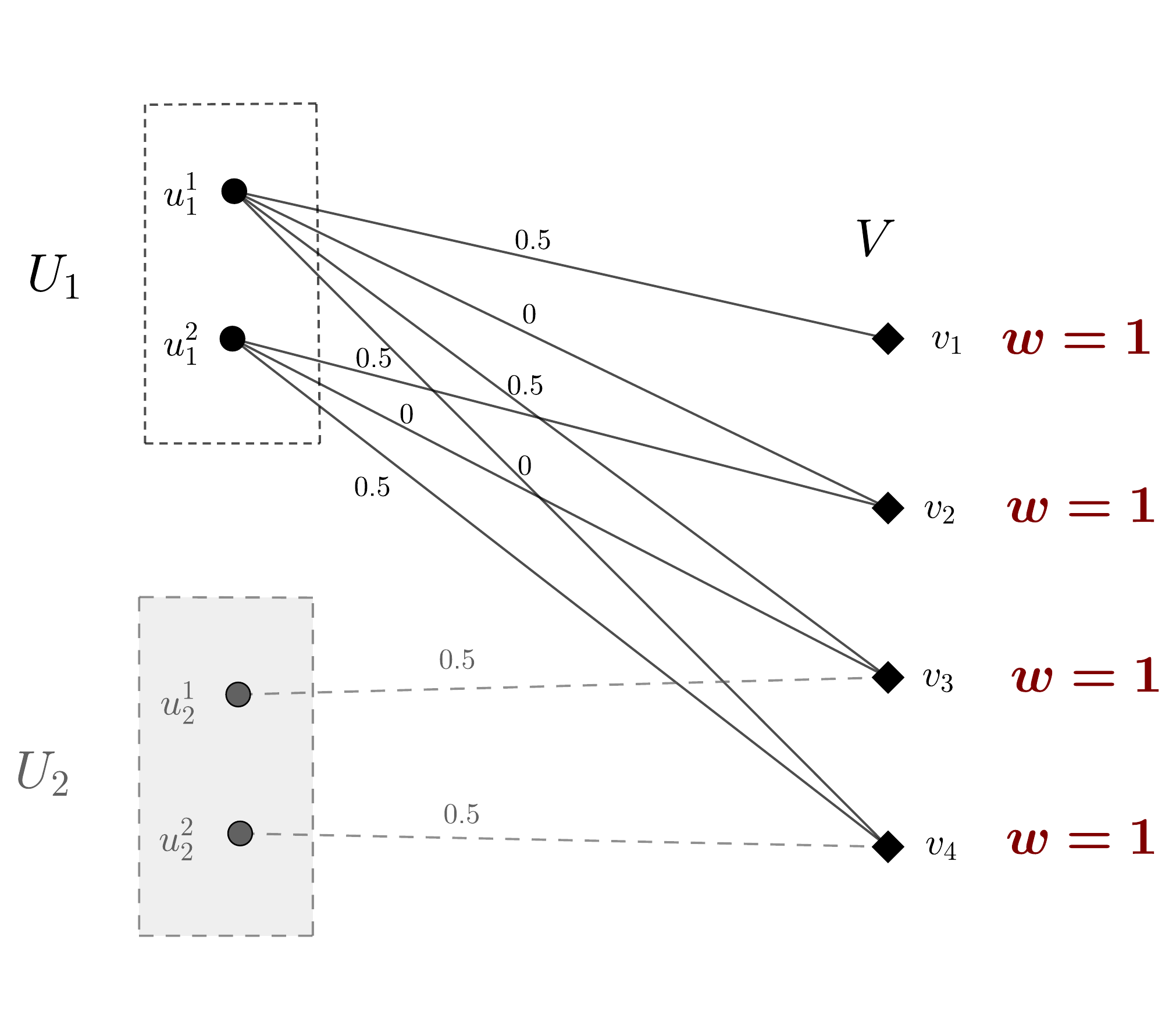}
	}
	\\
	\centering
	\subfloat[
	\label{fig:example-c}
    Ignoring the weights
	]
	{
\includegraphics[trim=9.5cm 3cm 11cm 2cm,clip,width=0.45\textwidth]{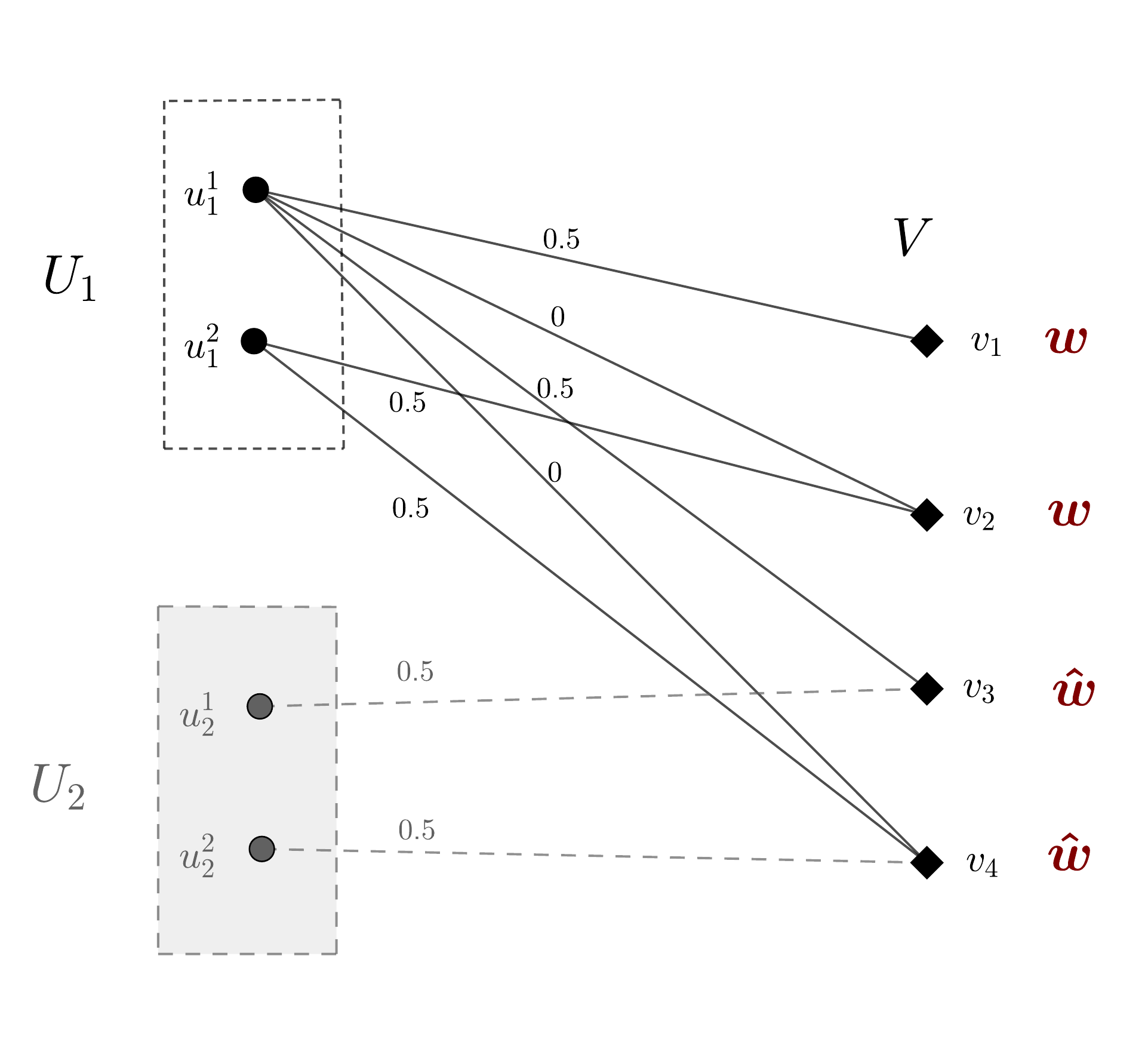}
	}
	\subfloat[
	\label{fig:example-d}
    Considering the weights
	]{
\includegraphics[trim=6.8cm 0cm 8.8cm 0cm,clip,width=0.45\textwidth]{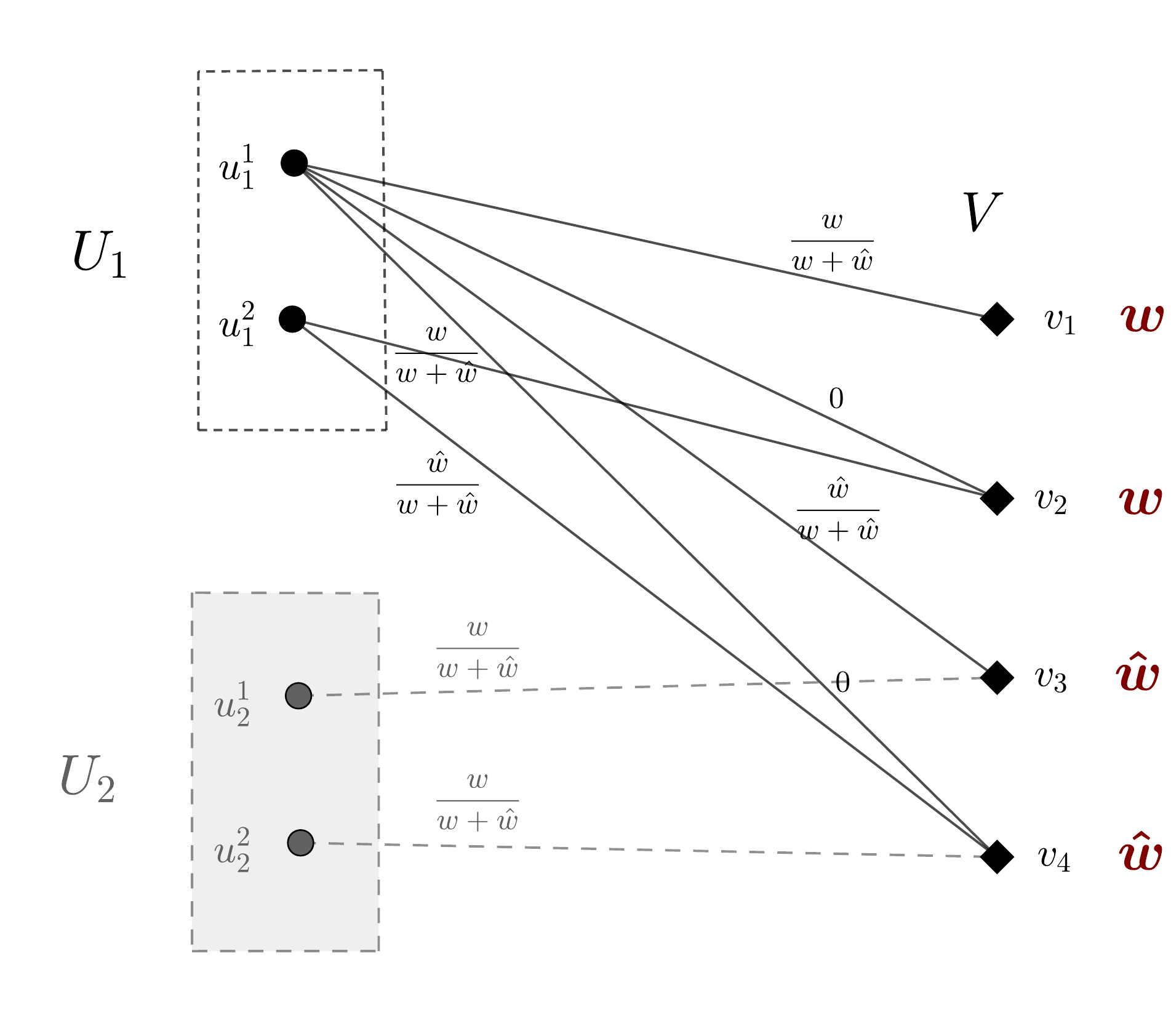}
	}
	\smallskip
	\caption{
Four simple examples of multi-stage vertex-weighted matching; We consider two different instances of this problem (one instance for (a) and (b), and one for (c) and (d)); Both problem instances have two stages, where $\boldsymbol{U_1=\{u_1^2,u_1^2\}$, $\boldsymbol{U_2=\{u_2^2,u_2^2\}}}$ and $\boldsymbol{V=\{v_1,v_2,v_3,v_4\}}$; edges are as drawn in the figures above; in (a) and (b) all the weights $\boldsymbol{w_1, w_2 ,w_3}$ and $\boldsymbol{w_4}$ are equal to $\boldsymbol{1}$; in (c) and (d), $\boldsymbol{w_1=w_2=w}$ and $\boldsymbol{w_3=w_4=\hat{w}}$; We run online water-filling, batched water-filling, batched water-filling (ignoring weights), and convex-programming based batched matching (considering the weights) algorithms in (a), (b), (c) and (d), respectively. 
    }
	
\end{figure}

\smallskip
\noindent{\textbf{Online Allocation vs.\ Batched Allocation}~} Consider a simple two-stage matching instance as in \Cref{fig:example-a}, where all the weights $w$ are equal to $1$. Now, consider the simple fractional allocation algorithm BALANCE, also known as \emph{water-filling}, that is commonly used in the literature of online bipartite matching~\citep{MSVV-07,AGKM-11}. Upon arrival of batch $\Rider_1$, BALANCE goes over demand nodes in $\Rider_1$ in some order and allocate each $\rider\in \Rider_1$ continuously and fractionally to the supply node in $\Driver$ that has the largest remaining capacity. As a result, no matter what ordering it picks, the algorithm produces the allocation vector $(0.25,0.25,0.25,0.25)$ for $u_1^1$ and $(1/3,1/3,1/3)$ for $u_1^2$, as it can be seen in \Cref{fig:example-a}. Fixing the first stage allocation, it is easy to check that in the worst-case (for the competitive ratio) the adversary will add two edges in the second stage to two of the supply nodes with the smallest remaining capacity, namely $(u_2^1,v_3)$ and $(u_2^2,v_4)$. In this case, the algorithm obtains a total objective value of $2+2\times(1-0.25-1/3)\approx 2.83$, while the optimum offline has an objective value of $4$; hence a ratio of $2.83/4<0.75$. Alternatively, consider a \emph{batched water-filling} allocation, where we try to fractionally allocate demand nodes in $\Rider_1$ simultaneously to supply nodes $\Driver$, so that the remaining capacities of supply nodes will be as balanced as possible. This algorithm produces the allocation vector $(0.5,0,0.5,0)$ for $u_1^1$ and $(0.5,0,0.5)$ for $u_1^2$, as it can be seen in \Cref{fig:example-b}. The same second stage would be the worst-case for competitive ratio of this algorithm, under which the algorithm obtains a total objective value of $2+1=3$, while the optimum offline obtains $4$; hence a ratio of $0.75$. As it can be seen from this example, the batched allocation could obtain the (optimal) ratio of $0.75$, while the algorithm that mimicked the online allocation could only obtain a ratio strictly smaller than $0.75$. 

\smallskip
\noindent\textbf{Naive vs.\ Weight-dependent Batched Allocations~} Consider a slightly different instance as in \Cref{fig:example-c}, in which we remove the edge $(u_1^2,v_3)$ in the first stage and supply nodes either have weight $w$ or $\hat{w}$, and $w\gg \hat{w}$.  A naive batched water-filling allocation as before will obtain an objective value of $0.5\times(2w+2\hat{w})+\hat{w}=w+2\hat{w}$, while the optimum offline has an objective value of $2w+2\hat{w}$; hence a ratio of $\frac{w+2\hat{w}}{2w+2\hat{w}}\approx 0.5$. Now consider a different batched allocation algorithm that does take weights into account: it finds a fractional allocation that maximizes $\sum_{j}w_j y_j-\frac{1}{2}\sum_{j}w_j y_j^2$, where $y_j$ is the final allocation amount of supply node $j$ after the first stage. Such a fractional matching can be found by solving a concave program, and it is easy to check that it produces the allocation vector  $(\frac{w}{w+\hat{w}},0,\frac{\hat{w}}{w+\hat{w}},0)$ for $u_1^1$ and $(\frac{w}{w+\hat{w}},\frac{\hat{w}}{w+\hat{w}})$ for $u_1^2$, as it can be seen in \Cref{fig:example-d}. The same second stage graph would also be the worst-case for competitive ratio of this algorithm. In that case, the algorithm obtains an objective value of 
$$
2w\times \frac{w}{w+\hat{w}}+2\hat{w}=\frac{2w^2+2\hat{w}^2+2w\hat{w}}{w+\hat{w}}~,
$$
while the optimum offline obtains an objective value of $2w+2\hat{w}$. Considering the ratio, we have
$$
\frac{\frac{2w^2+2\hat{w}^2+2w\hat{w}}{w+\hat{w}}}{2w+2\hat{w}}=\frac{w^2+\hat{w}^2+w\hat{w}}{w^2+\hat{w}^2+2w\hat{w}}\geq \frac{3}{4},
$$
simply because $4w^2+4\hat{w}^2+4w\hat{w}-3w^2-3\hat{w}^2-6w\hat{w}=(w-\hat{w})^2\geq 0$. Note that equality holds if and only if $w=\hat{w}$.

The above examples highlight that (i) in a multi-stage bipartite allocation problem, batched allocation can potentially help with improving the competitive ratio, and (ii) a batched allocation algorithm can benefit from using weights in a structured fashion. 
% However, it probably needs less aggressive hedging as it gets closer to the last stage; in particular, there is no need to output any balanced allocation in the last stage and an optimal competitive multi-stage allocation algorithm should always select the maximum weight allocation during that stage. 
Finally, the second example suggests that using regularizers for the maximum weight allocation problem is a promising approach to induce more balanced-ness to our allocation algorithm at each stage. This investigation leaves us with the following question: \emph{(a) how to use regularizers in a systematic fashion, to capture the information about the weights and the revealed subgraph at each stage?} \revcolor{Also, intuitively speaking, one might expect that the batched allocation algorithm benefits from having a more balanced allocation early on, to hedge against the uncertainty in future stages, and being more greedy later. Now we can ask:} \emph{(b) how to gradually change the balanced-ness of the allocation across different stages, so that we hedge more aggressively early on and less aggressively towards later stages?} As we will see in the next subsection, we address both of these questions by using the sequences of polynomials introduced in \Cref{sec:poly} to regularize the stage-wise maximum weight allocation linear programs.

\newcommand{\PRMWM}{\texttt{PR-MWM}}
\subsection{Polynomial Regularized Maximum Weight Matching}
\label{sec:upper-bound}

Recall the sequence of polynomials $f_1,f_2,\ldots,f_K$ defined in \Cref{sec:prelim} (see \Cref{prop:poly} and \Cref{fig:poly} for details). We are now ready to describe our main algorithm, which is an optimal competitive multi-stage algorithm for vertex weighted bipartite matching.

\smallskip
\noindent{\textbf{Overview of the Algorithm~}} Given the sequence of polynomials in \Cref{prop:poly}, our algorithm solves a different convex program
at each stage $k\in[K]$, and returns its unique optimal solution as the fractional matching of that stage. 
The program of stage $k$ is essentially a regularized maximum vertex weighted matching considering the current remaining capacity of each offline vertex. In this convex program, the regularization is done by a weighted combination of polynomials $F_k(x)=\int_0^{x}f_k(y)dy$ over offline vertices, each evaluated at the fractional degree of the offline vertex at the end of the allocation. \revcolor{Notably, the convex program at stage $k$ depends on the current state of the allocations, which is captured by the vector of fractional degrees of each offline vertex $j$ at the beginning of stage $k$ denoted by $\mathbf{y}^{(k)}$.} 
See \Cref{alg:b-matching} for a formal description.

\begin{algorithm}[htb]
 	\caption{Polynomial Regularized Max Weight Matching (\PRMWM)}
 	\label{alg:b-matching}
 	\KwIn{Number of batches/stages $K$}
 	    \vspace{2mm}
 	    
    $\forall j\in\Driver:~y^{(0)}_j\gets 0$.

 	\vspace{1mm}
 	\For{stage $k=1$ to $K$}
 	{
 	\vspace{2mm}
 	\tcc{$k^{\textrm{th}}$ batch of vertices in $\Rider$, denoted by $\Rider_k$, arrives. Let $E_k$ denote the set of edges incident to vertices in $\Rider_k$.}
 	\vspace{2mm}
 	
   Given $\mathbf{y}^{(k-1)}$ and subgraph $G_k=(\Rider_k,\Driver,E_k)$, solve the convex program~\ref{eq:concave-k-bmatching}:
   \vspace{-1mm}
 		\begin{align}
 	\label{eq:concave-k-bmatching}\tag{$\mathcal{P}^{\text{VWM}}_k$}
    \begin{array}{lll}
       \xbf^*_k= \underset{\xbf\geq \mathbf{0}}{\argmax} & 
        \displaystyle\sum_{(i,j) \in E_k}w_j x_{ij} - \sum_{j\in \Driver}w_j F_k\big(y^{(k-1)}_j+
        \sum_{i : (i,j)\in E_k}x_{ij}\big)
        & \text{s.t.}\\
         &\displaystyle
         \sum_{j: (i,j)\in E_k} x_{ij} \leq 1
         & i\in \Rider_k\\
         & 
         \displaystyle\sum_{i: (i,j)\in E_k} x_{ij} \leq 1-y^{(k-1)}_j
         & j\in \Driver
    \end{array}
\end{align}
\tcc{Recall that the $k^{\text{th}}$ regularizer is defined as $F_k(x)\triangleq\displaystyle\int_0^{x}f_k(y)dy$.}

$\forall j\in \Driver: y^{(k)}_j\gets y^{(k-1)}_j+\displaystyle\sum_{i\in\Rider_k}x^*_{k,ij}$.

Return $\xbf^*_k$ as the fractional matching of stage $k$.
 	}

%  	\vspace{2mm}
%  	\emph{\underline{Pre-processing:}} Compute the optimal assignment $\{\curalloctypeopt\}$ of $\EAR{\typedistributionsequence}$ by invoking the offline assortment oracle  (\Cref{asp:oracle})
%  	\vspace{2mm}
 	
%  	\For{$t=1$ to $T$}{
%  	\tcc{consumer $t$ with type $\type_t\sim\typedistribution_t$ arrives} 
%  	\vspace{1mm}
 	
%  	\emph{\underline{Simulation:}} Upon realizing consumer type $\type_t$, sample $\hat{\assortment}_t\sim \{\curalloctypeopt\}_{\assortment\in\assortmentspace}$
 	
%  	\vspace{2mm}
%  	\emph{\underline{Discarding:}} Initialize $\bar\assortment_t\gets\hat\assortment_t$
 	
%  	\For{each product $i\in\hat\assortment_t$}{
%  	\vspace{1mm}
%  	Flip an independent coin and remove $i$ from $\bar\assortment_t$ with probability $\gamma$
 	
%  	\If {there is no available unit of product $i$}{
%  	Remove $i$ from $\bar\assortment_t$}}
 	
%  	\vspace{2mm}
%  	\emph{\underline{Post-processing:}} Let $\tilde{S}_t\gets \textsc{Sub-assortment Sampling} \left(\choice^{\type_t}, \bar\assortment_t,\{\choice(\hat\assortment_t, i)\}_{i\in \bar\assortment_t}\right)$
 	
%  	\tcc{Send a query call to Procedure~\ref{alg:sample assortment} with appropriate input arguments}
%  	\vspace{2mm}
 	
%  	Offer assortment $\tilde{\assortment}_t$ to consumer $t$
 	
%  	}
\end{algorithm}
We are now ready to describe the main result of this section.

\begin{theorem}[\textbf{Optimal Multi-Stage Algorithm for Weighted Bipartite Matching}]
\label{thm:b-matching}
In the $K$-stage (fractional) vertex weighted matching problem,
 \PRMWM~(\Cref{alg:b-matching}) is $\cratio{K}$-competitive, where $\cratio{K}=\cratioexp$.
\end{theorem}

\smallskip
\noindent{\textbf{High-level Proof Overview}~} In a nutshell, we analyze our convex programming based algorithm by following two steps. First, we apply the theory of convex duality to the convex programs corresponding to each stage, which guides us to build a convex program specific graph decomposition for each stage's subgraph. Intuitively speaking, the graph decomposition in each stage partitions the subgraph of that stage into groups of demand and supply nodes, so that in each group there exists a fully ``weighted balanced'' fractional matching (measured by some definition related to the convex program). Second, we try to find an assignment to the dual linear program of our optimum offline LP that is built stage by stage, and in each stage uses the graph decomposition of that stage. We suggest a particular graph-decomposition based dual assignment, where we use the same assignment for all supply nodes in the same group and the same assignment for all demand nodes in the same group. Further, we show the dual objective value is equal to the primal algorithm's objective value from our construction. We finally show this dual assignment is approximately feasible, where the approximation factor is $\cratio{K}$. It turns out that the approximate feasibility of our dual assignment boils down to the requirement that functions $f_1,f_2,\ldots,f_K$ satisfy the specific recursive equation in the second bullet of \Cref{prop:poly} for $k=1,2,\ldots,K-1$. Hence, our \PRMWM~algorithm (\Cref{alg:b-matching}) obtains the desired competitive ratio of $\cratio{K}$, as it uses the sequence of  polynomials of decreasing degrees in place of $f_1,\ldots,f_K$ that are the unique solution to mentioned recursive functional equation. See the details of this proof overview in \Cref{sec:primal-dual}.

\begin{remark}[\textbf{Batched Water-filling}]
\label{rem:batched-water}
For the special case of unweighted matching, it is not hard to see that the solution of the convex program \ref{eq:concave-k-bmatching} does \emph{not} depend on the choice of the regularizer function $F_k$ (it only needs to be a strictly convex function). The algorithm for this special case is indeed the batched water-filling algorithm that we investigated in our first example in \Cref{sec:example}. \revcolor{This algorithm is oblivious to the choice of the regularizer and does not even need to know $K$.} Note that while in that simple example the batched water-filling ended up with fully balancing the remaining capacities of all the offline vertices in the first stage graph, this is not going to be necessarily the case in an arbitrary graph at any stage. We should also highlight that this algorithm generalizes the classic water-filling/BALANCE algorithm for the online bipartite fractional matching~\citep{MSVV-07,DJK-13} to the setting where we have batches. Importantly, even though our polynomials are not \emph{explicitly} used by the algorithm, they still help with its competitive ratio analysis and showing that it is $\cratio{K}$-competitive. 
\end{remark}
\begin{remark}[\textbf{Vertex-weighted Online Bipartite Matching}]
Another important special case is when batches have size one, which is basically the online arrival problem. As mentioned earlier, by replacing $f_k(x)$ with $e^{x-1}$ in \emph{every stage}, the final allocation of our algorithm will be the same as the classic BALANCE algorithm for the fractional vertex weighted online matching \citep{AGKM-11, DJK-13}. In other words, the special case of our algorithm provides a convex programming based interpretations of the BALANCE algorithm.\footnote{To provide more details, BALANCE can be viewed as running a particular first-order iterative method for solving this convex program (basically running a slight variant of the Frank-Wolfe algorithm\citep{frank1956algorithm}).}
\end{remark}
\begin{remark}[\textbf{Unweighted Online Bipartite Matching}]
Our result implies the following simple corollary for the very special case of the classic online bipartite matching problem: the water-filling/BALANCE algorithm is $\cratio{n}$ competitive for the (unweighted) online bipartite matching problem with a fixed number of $n$ online vertices. Interestingly, this corollary was also observed in \cite{goel2008online}, but for the RANKING algorithm~\citep{KVV-90} instead of water-filling, and through a more refined combinatorial analysis of RANKING using a factor revealing LP. As we will see later in \Cref{sec:lowerbound} and \Cref{rem:online-matching-with-finite-nodes}, this bound is not necessarily tight for the online bipartite matching problem with $n$ online nodes---despite the fact that the competitive ratio $\cratio{K}$ is indeed optimal in multi-stage matching with $K$ stages (\Cref{thm:tightness}).
\end{remark}

\subsection{Recursive Primal-dual Analysis of \texorpdfstring{\Cref{alg:b-matching}}{Lg}}
\label{sec:primal-dual}
We prove \Cref{thm:b-matching} by 
a primal-dual argument that carefully uses our polynomials.  
Consider the offline primal LP of the maximum vertex weighted matching in graph $G=(\Rider,\Driver,E)$, and its dual:
\begin{equation}
\tag{\textsc{Primal-Dual-MWM}}
\label{eq:LP-max-weight}
\arraycolsep=1.4pt\def\arraystretch{1}
\begin{array}{llllllll}
\underset{\xbf\geq\mathbf{0}}{\max}\quad\quad\quad&\displaystyle\sum_{(i,j)\in E}w_jx_{ij} &~~\text{s.t.}&
&\quad\quad\underset{\boldsymbol{\alpha},\boldsymbol{\beta}\geq\mathbf{0}}{\text{min}}\quad\quad\quad &\displaystyle\sum_{i\in \Rider }{\alpha_i}+\displaystyle\sum_{j\in \Driver}{\beta_j}&~~\text{s.t.} \\[1.4em]
 &\displaystyle\sum_{j:(i,j)\in E}{x_{ij}}\leq1 &~~i\in\Rider~,& 
& &\alpha_i+\beta_j\geq w_j& (i,j)\in E~,\\[1.4em]
 &\displaystyle\sum_{i:(i,j)\in E}{x_{ij}}\leq 1 &~~j\in \Driver~, &
& & \\
\end{array}
\end{equation}
Given the feasible primal assignment $\{\xbf^*_k\}_{k\in[K]}$ produced by \Cref{alg:b-matching}, we construct a dual assignment so that,
\begin{itemize}
    \item Primal and dual have the same objective values: $\displaystyle\sum_{k\in[K]}\displaystyle\sum_{(i,j)\in 
    E_k}w_jx^*_{k,ij}=\sum_{i\in\Rider}\alpha_i+\sum_{j\in\Driver}\beta_j$~,
    \vspace{1mm}
    \item Dual is approximately feasible: $\forall (i,j)\in E: \alpha_i+\beta_j\geq \cratio{K}\cdot w_j$~.
\end{itemize}
Existence of such a dual assignment will guarantee that primal is $\cratio{K}$-competitive.

\revcolor{Our dual assignment is constructed stage-by-stage. Moreover, our construction relies on a structural decomposition of the subgraph in each stage $k$, which itself is tied to the solution of the convex program~\ref{eq:concave-k-bmatching} of stage $k$.  
We highlight that this decomposition is a slightly modified version of a simpler decomposition first introduced in \cite{feng2020two} (for only the first stage in a two-stage problem and for a slightly different convex program) which actually \emph{does not depend} on the choice of the convex function they used. It also extends the well-known matching skeleton for unweighted bipartite graphs~\citep{GKK-12}. Notably, in contrast to both of of these decompositions, this new decomposition actually \emph{depends} on the choice of the convex function used in its definition. This is a key technical property used in the analysis of our algorithm. We first define this decomposition in \Cref{lemma:structure} and show how it can be constructed through convex programming duality (KKT conditions). 
% See the proof of this lemma in \Cref{apx:skeleton}.
We then show how it helps us with our dual construction.}

\begin{proposition}[\revcolor{Stage-wise Structural Decomposition}]
\label{lemma:structure}
Fix any $k \in [K - 1]$.
Let $\xbf^*_k$ be the optimal solution of the convex program~\ref{eq:concave-k-bmatching}. Consider the subgraph $G'_k=(\Rider_k,\Driver,E'_k)$ of $G_k$, where $E'_k=\{(i,j)\in E_k :x^*_{k,ij}>0\}$. Let $\Driver^{(0)}_k$ be the set of offline vertices in $\Driver$ fully matched by $\xbf^*_k$ given the current $\ybf$, and 
$\Rider^{(0)}_k$ be the set of online vertices who are neighbors of $\Driver^{(0)}_k$ in $G_k'$, that is, 
$$\Driver^{(0)}_k\triangleq \{j\in \Driver: y_j+\sum_{i:(i,j)\in E_k}x^*_{k,ij}=1\}~~\textrm{and}~~\Rider^{(0)}_k\triangleq \{i\in \Rider_k:\exists j\in \Driver^{(0)}_k, x^*_{k,ij}>0\}~.$$
Moreover, let the pairs 
$\left\{\left(\Rider_k^{(l)},\Driver_k^{(l)}\right)\right\}_{l=1}^{L_k}$ 
identify the $L_k$ connected components of the induced subgraph $G_k'[\Rider_k\setminus\Rider_k^{(0)},\Driver\setminus\Driver_k^{(0)}]$ of $G_k'$. Then:
\begin{enumerate}%[label=\roman*.]
    \item \underline{Uniformity}: ~$\forall~ l\in[0:L_k],j,j'\in \Driveri_k:$ $$w_j\left(1-
    f_k\left(y_j + \sum_{i:(i,j)\in E_k}x^*_{k,ij}\right)
    \right)=
    w_{j'}\left(1-
    f_k\left(y_{j'} + \sum_{i:(i,j')\in E_k}x^*_{k,ij'}\right)
    \right)\triangleq c_k^{(l)}$$
    \item \underline{Monotonicity}:  $\forall~ l,l'\in[0:L_k]$: there exists no edge in $E_k$ between $\Rider_k^{(l)}$ and $\Driver_k^{(l')}$ if $c_k^{(l)}<c_k^{(l')}$. 
    \item \underline{Saturation}: $\forall~ l\in[L_k]$: all online vertices $i\in\Rider_k^{(l)}$ are fully matched by $\xbf_k^*$, i.e., $\displaystyle\sum_{j:(i,j)\in E_k}x^*_{k,ij}=1$.
\end{enumerate}
\end{proposition}

We prove the above graph-based structural decomposition by investigating the program~\ref{eq:concave-k-bmatching} through convex strong duality. See the details in Appendix~\ref{apx:kkt}. \revcolor{At a high-level, the above graph decomposition partitions the subgraph revealed at each stage $k$ based on a (weighted) notion of \emph{``match-ability''} of the offline nodes (i.e., $c_k^{(l)}$ values). Having such a partitioning helps us to identify a charging scheme, so that we can charge the primal contributions of our algorithm between the dual variables corresponding to offline and online nodes appropriately.} We now have all the ingredients to prove our main theorem of this section using a primal-dual analysis. 
% Note that 
% $0=c_k^{(0)}\leq c_k^{(1)}\leq \ldots\leq c_k^{(L_k)}$.

% \begin{definition}[batch-water-filling (BWF)]
% In any stage $k < K$, let 
% $\{y_{k, j}\}_{j\in D}$ be the capacity 
% of drivers which has been used so far,
% batch-water-filling (BWF) selects
% the fractional matching as the solution of 
% program \eqref{eq:concave program k} 
% on $(\Rider_k, D, E_k)$ 
% with initial capacity $\{y_{k, j}\}_{j\in D}$ and $K - k$.
% In the last stage $k = K$, 
% BWF selects the maximum weighted fractional matching.
% \end{definition}

% \begin{proof}
\ifMS \proof{\textsl{Proof of \Cref{thm:b-matching}.}}
\else \begin{proof}[Proof of \Cref{thm:b-matching}]\fi
For each stage $k\in[K-1]$, let $\{(\Rider^{(0)}_k, \Driver^{(0)}_k),\dots, (\Rider^{(L_k)}_k, \Driver^{(L_k)}_k)\}$ be the 
structural decomposition of $(\Rider_k, \Driver, E_k)$ as in \Cref{lemma:structure},
and $\{x_{k, ij}^*\}_{(i,j)\in E_k}$
be the optimal solution of the program~\ref{eq:concave-k-bmatching}. Note that if $c_k^{(l)}=c_k^{(l')}$ for $l\neq l'$ and some $k\in[K-1]$, we can simply merge the two pairs $\left(\Rider_k^{(l)},\Driver_k^{(l)}\right)$ and $\left(\Rider_k^{(l')},\Driver_k^{(l')}\right)$ to $\left(\Rider_k^{(l)}\cup \Rider_k^{(l')},\Driver_k^{(l)}\cup \Driver_k^{(l')}\right)$, and still our decomposition satisfies the three properties of \Cref{lemma:structure}; these properties are all we need for our technical arguments. Therefore, without loss of generality, we can assume $c_k^{(l)}$'s are non-identical. Moreover, $0=c_k^{(0)}<c_k^{(1)}<\ldots<c_k^{(L_k)}$.

First, we define the following notations which will be useful in this proof:
\begin{align*}
\forall j\in \Driver,~k\in[K]:\quad\quad\quad &y^*_{k,j}\triangleq\sum_{s=1}^{k-1}\sum_{i:(i,j)\in E_s}x^*_{s,ij}
% ,\\
%  &y^*_{0,j}=0.
\end{align*}
Now, consider the offline primal-dual linear programs in \ref{eq:LP-max-weight}. Given the feasible primal assignment $\{\xbf^*_{k}\}_{k\in[K]}$, construct a dual assignment as follows. First, set $\alpha_i\gets 0$ and $\beta_j\gets 0$ for all $i\in\Rider,j\in \Driver$. At each stage $k\in[K-1]$, update the dual variables as follows:
\begin{align*}
   &\forall i \in \Rider^{(l)}_k,~l\in[0:L_k]: &&\alpha_i\gets c^{(l)}_k
   \\
    &\forall j \in \Driver^{(l)}_k,~l\in[0:L_k]: &&\beta_{j}\gets\beta_{j}+\Delta\beta_{k,j}\\
    &&&\Delta\beta_{k,j}\triangleq \left(
\sum_{i:(i,j)\in E_k}x_{k,ij}^* \right)\cdot \left(w_j - c_k^{(l)}\right)\overset{(a)}{=}w_j(y^*_{k+1,j}-y^*_{k,j})f_k(y^*_{k+1,j})
\end{align*}
where in eq.\ (a) above 
% \YF{label eq.~(1) is not defined in our paper}
we use the uniformity property and the definition of $c^{(l)}_k$ in \Cref{lemma:structure}. For the last stage $K$, note that $F_K(x)=0$ for all $x\in[0,1]$. Therefore, \Cref{alg:b-matching} essentially solves the maximum vertex weighted matching in the last stage graph $(\Rider_K, \Driver, E_K)$ where each $j\in\Driver$ has initial capacity $1-y^*_{K, j}$. Let $\{\hat{\alpha}_{K,i}, \hat{\beta}_{K,j}\}_{i\in\Rider_K,j\in\Driver}$ 
be the \emph{optimal} dual solution of this LP. Then update the  dual variables as follows:
\begin{align*}
&\forall i\in\Rider_K:~~\alpha_i \gets \hat{\alpha}_{K,i} \\
&\forall j\in \Driver~:~~\beta_j\gets \beta_j+\Delta \beta_{K,j},~~\Delta\beta_{K, j} \triangleq (1 - y^*_{K, j})\hat{\beta}_{K,j}
\end{align*}

% Consider the following construction of dual solution.
% At stage $k < K$, let 
% $\bar f_k(x) \triangleq f_{K - k}(x) = \left(
% \frac{K - k - 1 + x}{K-k}
% \right)^{K-k}$, and 
% $\{(\Rider^{(0)}_k, \Driver^{(0)}_k),\dots, (\Rider^{(L)}_k, \Driver^{(L)}_k)\}$ be the 
% partition of $(\Rider_k, D, E_k)$ in \Cref{lemma:structure},
% and 
% $\{x_{k, ij}^*\}_{(i,j)\in E_k}$
% be the optimal solution  
% from the concave program~\eqref{eq:concave program k}
% solved by BWF.
% Construct the dual solution as follows,
% for any $k < K$ 
% \begin{align*}
%     \alpha_i &= c^{(l)}_k
%     \qquad \forall i \in \Rider^{(l)}_k \\
%     \Delta \beta_{k,j} &= \left(
%     % 1 - 
%     \sum_{i\in N_k(j)}x_{k,ij}^*
%     % - y_{k,j}
%     \right)\cdot 
%     \left(w_j - c_k^{(l)}\right)
%     \qquad \forall j \in \Driver^{(l)}
% \end{align*}
% where $N_k(j)$ denotes all neighbor vertices of item $j$
% in graph $(\Rider_k, D, E_k)$ at stage $k$.
The rest of the proof is done in two steps:
\vspace{1mm}

\noindent[\textit{Step \rom{1}}]~\emph{Comparing objective values in primal and dual.}
we first show that 
the objective value of the above dual assignment
is equal to the objective value of the primal assignment of \Cref{alg:b-matching}.
To show this, we consider 
each stage $k$ separately.
For the last stage $K$, 
this holds by construction.
For each stage $k \in [K-1]$, note that the primal objective value is decomposed across pairs $(\Rider_k^{(l)}, \Driver_k^{(l)})$, 
$l\in[0: L_k]$, 
simply because $x^*_{k,ij}>0$ only if $i$ and $j$ belong to the same pair (\Cref{lemma:structure}). As these pairs partition sets of vertices $\Rider$ and $\Driver$, it is enough to compare the changes in the primal and dual objectives in each pair $(\Rider_k^{(l)}, \Driver_k^{(l)})$ separately. Notice that if $l = 0$ then $c^{(l)}_k = 0$, and thus,
\begin{align*}
    \Delta(\textrm{Dual})^{(0)}\triangleq\sum_{i \in \Rider^{(0)}_k} \alpha_i
    +
    \sum_{j \in \Driver^{(0)}_k} \Delta\beta_{k,j}
    &= 
    \sum_{j \in \Driver^{(0)}_k}\sum_{i:(i,j)\in E_0} w_jx_{k, ij}^*\triangleq\Delta(\textrm{Primal})^{(0)}. 
\end{align*}
Similarly, for $l \in [L_k]$, 
\begin{align*}
     \Delta(\textrm{Dual})^{(l)}&\triangleq\sum_{i \in \Rider^{(l)}_k} \alpha_i
    +
    \sum_{j \in \Driver^{(l)}_k} \Delta\beta_{k,j}
    =
    |\Rider^{(l)}_k|\cdot c^{(l)}_k
    +
    \sum_{j \in \Driver^{(l)}_k}\sum_ {i:(i,j)\in E_k} 
    x_{k, ij}^*\left(w_j - c^{(l)}_k\right)\\
    &= 
     |\Rider^{(l)}_k|\cdot c^{(l)}_k+\sum_{j\in\Driver_k^{(l)}}\sum_{i:(i,j)\in E_k} 
    w_jx_{k, ij}^*-c^{(l)}_k\sum_{i\in\Rider_k}\left(\sum_{j:(i,j)\in E_k}x^*_{k,ij}\right)\\
    &=\sum_{j\in\Driver_k^{(l)}}\sum_{i:(i,j)\in E_k} 
    w_jx_{k, ij}^*\triangleq \Delta(\textrm{Primal})^{(l)},
\end{align*}
where the last equality uses the saturation property in \Cref{lemma:structure} 
for $l\in[L_k]$.
\vspace{1mm}

\noindent[\textit{Step \rom{2}}]~\emph{Checking approximate feasibility of dual.} 
To show approximate dual feasibility, it is enough to show that for any stage $k\in[K]$ and any arbitrary edge $(i, j) \in E_k$,
\begin{align}
\label{eq:dual constraint feasibility}
    \alpha_i + \sum_{s = 1}^k\Delta\beta_{s, j}
    \geq 
    w_j\left(1-(1-1/K)^K\right)
\end{align}
We next show \eqref{eq:dual constraint feasibility}
holds for $k\in[K-1]$ by using the following technical lemma, which carefully uses our sequence of polynomial regularizers to establish a lower bound on the LHS of \eqref{eq:dual constraint feasibility}.

  \begin{lemma}
    \label{lem:polynomial inequality 2}
    For any $K \in \N,~k \in [K - 1]$, 
    $x \in [0, 1]$, and
    any $0 = y_1 \leq 
    y_2 \leq \dots \leq y_k \leq  1 - x$,
    \begin{align*}
     (1-x)f_{k}(y_k + x) -
    \sum_{s = 1}^{k - 1} 
    (y_{s + 1} - y_s)
    f_{s}(y_{s + 1})
     \leq 1-\cratio{K}=
    (1-1/K)^K.
    \end{align*}
    \end{lemma}
Now assume $i \in \Rider_k^{(l)}$ and $j \in \Driver_k^{(l')}$, which implies $c^{(l')}_k \leq c^{(l)}_k$ (\Cref{lemma:structure}). Then we have:
\begin{align*}
     \alpha_i + \sum_{s = 1}^{k}\Delta\beta_{s, j}&=c_k^{(l)} + 
    \left(\sum_{i':(i',j)\in E_k}x_{k, i'j}^*\right)
    \left(w_j - c_k^{(l')}\right)
    +
    \sum_{s= 1}^{k - 1} \Delta\beta_{s, j} \\
    &\geq c_k^{(l')} + 
    \left(\sum_{i':(i',j)\in E_k}x_{k, i'j}^*\right)
    \left(w_j - c_k^{(l')}\right)
    +
    \sum_{s= 1}^{k - 1} \Delta\beta_{s, j} \\
    &=w_j\left(1-f_k\left(y^*_{k,j}+\sum_{i':(i',j)\in E_k}x_{k, i'j}^*\right)+\left(\sum_{i':(i',j)\in E_k}x_{k, i'j}^*\right)f_k\left(y^*_{k,j}+\sum_{i':(i',j)\in E_k}x_{k, i'j}^*\right)\right)\\
    &+ \sum_{s= 1}^{k - 1} \Delta\beta_{s, j} \\
    &=w_j\left(1-\left(1-\sum_{i':(i',j)\in E_k}x_{k, i'j}^*\right)f_k\left(y^*_{k,j}+\sum_{i':(i',j)\in E_k}x_{k, i'j}^*\right)+\sum_{s=1}^{k-1}\left(y^*_{s+1,j}-y^*_{s,j}\right)f_k\left(y^*_{s+1,j}\right)\right)\\
    &\geq w_j \cratio{K}~,
\end{align*}
where the first inequality uses $c^{(l')}_k \leq c^{(l)}_k$, the second equality uses the expression for $c^{(l')}_k$ in \Cref{lemma:structure} due to uniformity, and the last inequality invokes \Cref{lem:polynomial inequality 2} for the sequence: 
$$
0=y^*_{1,j}\leq y^*_{2,j}\leq \ldots \leq y^*_{k,j}\leq 1-\sum_{i':(i',j)\in E_k}x_{k, i'j}^*~.
$$
% \begin{align*}
%      & c_k^{(l)} + 
%     \left(\sum_{i'\in N_k(j)}x_{k, i'j}^*\right)
%     \left(w_j - c_k^{(l)}\right)
%     +
%     \sum_{s' = 1}^{k - 1} \Delta\beta_{s', j} \\
%     &\leq 
%     c_k^{(l')} + 
%     \left(\sum_{i'\in N_k(j)}x_{k, i'j}^*\right)
%     \left(w_j - c_k^{(l)}\right)
%     +
%     \sum_{s' = 1}^{k - 1} \Delta\beta_{s', j} \\
%     &=
%     \alpha_i + \sum_{s' = 1}^{k}\Delta\beta_{s', j}
% \end{align*}

So far, we have shown \eqref{eq:dual constraint feasibility} holds for all stages $k < K$.
For the last stage $K$, and any edge $(i, j) \in E_K$,
notice that 
$\alpha_i+\Delta\beta_{K, j}=\hat{\alpha}_{K,i} + (1 - y^*_{K, j})\hat\beta_{K, j} \geq 
w_j(1 - y^*_{K, j})$ by construction.
 If $y^*_{K, j} = 0$, inequality \eqref{eq:dual constraint feasibility} holds. Otherwise, let $k' < K$ be the latest stage before $K$ where 
there exists $l^\dagger$ 
such that 
$j \in \Driver_{k'}^{(l^\dagger)}$ and 
$\Delta\beta_{k',j} > 0$. By definition, we have
$y^*_{k', j} + \displaystyle\sum_{i':(i',j)\in E_{k'}}x_{k', i'j}^* = y^*_{k'+1,j}= y^*_{K,j}$. Therefore:
\begin{align*}
    \alpha_i + \sum_{s = 1}^{K}\Delta\beta_{s, j}
    &\geq    
    w_j(1 - y^*_{K, j})
    + \sum_{s = 1}^{K - 1}\Delta\beta_{s, j} = 
    w_j(1 - y^*_{K, j}) + \sum_{s = 1}^{k'}\Delta\beta_{s, j} \\
    &\geq w_j\left(1 - f_{k'}(y^*_{k'+1, j})\right) + \sum_{s = 1}^{k'}\Delta\beta_{s, j} =c^{(l^\dagger)}_{k'} + \sum_{s = 1}^{k'}\Delta\beta_{s, j} \\
    &\geq w_j(1 - (1 - 1/K)^K)~,
\end{align*}
where the first equality uses the fact $\Delta\beta_{s, j} = 0$ for all $s$
between $k' + 1$ and $K - 1$; the second inequality uses the fact that 
that $f_{s}(x) \geq x$ for all $x\in [0, 1]$ and $s\in [K - 1]$
and $y^*_{k'+1,j}= y^*_{K,j}$,
the second equality uses the definition of $c^{(l^\dagger)}_{k'}$; and finally the last inequality holds by exactly the same argument as in the case of $k\in[K-1]$.\ifMS \hfill\Halmos \fi
\ifMS
\endproof
\else
\end{proof}
\fi

We finish the analysis by proving the technical lemma used in the proof of \Cref{thm:b-matching}. It will be further used in the analysis of our multi-stage configuration allocation algorithm in \Cref{sec:extensions}.

\ifMS \proof{\textsl{Proof of \Cref{lem:polynomial inequality 2}.}}
\else \begin{proof}{proof}[Proof of \Cref{lem:polynomial inequality 2}]\fi
    We show the statement by induction on $k$.
    
    \vspace{2mm}
    \noindent\textsl{Base case $(k = 1)$.} 
    The left hand side of inequality becomes
    \begin{align*}
        (1-x)f_{1}(x) 
        \leq 
        1-\cratio{K},
    \end{align*}
    which holds because of \Cref{prop:poly}, part (vi). 
    
    \vspace{2mm}
    \noindent\textsl{Inductive step $(1< k\leq K-1)$}.
    For any instance $(K, k, x, y_1, \dots, y_k)$,
    consider another instance 
    $(K, k - 1, x'\triangleq y_k - y_{k - 1}, y_1, \dots, y_{k-1})$.
    Note that
    \begin{align*}
         &(1 - x)f_{k}(y_k + x) - 
     \sum_{s = 1}^{k - 1} 
    (y_{s + 1} - y_s)
    f_{s}(y_{s + 1}) \overset{(a)}{\leq} f_{k-1}(y_k)
    -
     \sum_{s = 1}^{k - 1} 
    (y_{s + 1} - y_s)
    f_{s}(y_{s + 1})\\
   &= \left(1-(y_k-y_{k-1})\right)f_{k - 1}(y_{k - 1} + (y_k - y_{k - 1}))-
    \sum_{s = 1}^{k - 2} 
    (y_{s + 1} - y_s)
    f_{s}(y_{s + 1})\\
    &=\left(1-x'\right)f_{k - 1}(y_{k - 1} + x')-
    \sum_{s = 1}^{k - 2} 
    (y_{s + 1} - y_s)
    f_{s}(y_{s + 1}) \overset{(b)}{\leq} 1-\cratio{K}~,
    \end{align*}
    % \begin{align*}
    %     &
    % 1 - (1 - x)f_{K - k}(y_k + x) + 
    %  \sum_{s = 1}^{k - 1} 
    % (y_{s + 1} - y_s)
    % f_{K - s}(y_{s + 1})\\
    %     &1 - (1 - 1/K)^K  \\
    %     \leq &
    %     1 - f_{K - k + 1}(y_{k - 1} + (y_k - y_{k - 1})) 
    %     + 
    % \sum_{s = 1}^{k - 2} 
    % (y_{s + 1} - y_s)
    % f_{K - s}(y_{s + 1})
    % +
    % (y_k-y_{k-1}) f_{K - {k - 1}}(y_{k - 1} + (y_k - y_{k - 1}))
    % \\
    % =&
    % 1 - f_{K - k + 1}(y_k)
    % +
    %  \sum_{s = 1}^{k - 1} 
    % (y_{s + 1} - y_s)
    % f_{K - s}(y_{s + 1})
    % \end{align*}
    where ineq.\ (a) uses
    \Cref{prop:poly} and the recursive definition of $\{f_k(x)\}_{k\in[K]}$, and ineq.\ (b) uses the induction hypothesis on instance 
    $(K, k - 1, x', y_1, \dots, y_{k-1})$, where $x'=y_k - y_{k - 1}$.\ifMS
\hfill\Halmos\fi
\ifMS\endproof
\else
\end{proof}
\fi

\revcolor{\section{Extension: Multi-stage Configuration Allocation}
\label{sec:extensions}
In this section, we discuss how 
our approach can be generalized 
to the multi-stage configuration allocation problem. There are several key challenges towards this extension. First, in contrast to vertex weighted matching, here each unit of an advertiser's capacity can be allocated at \emph{different prices}. Thus, it is not clear how to use the regularization approach of the previous section. 
This important distinction also implies that a similar graph-based structural decomposition to analyze a potential convex programming based algorithm might not exist in this model, roughly because the degree of match-ability of an offline node might vary across different price levels. Finally, a configuration allocation algorithm should make disposal (preemption) decisions jointly with its allocation whenever needed. A priori, it is not clear how to incorporate preemption into our previous convex programs.\footnote{Note that edge-weighted online matching is a special case; hence disposals are definitely needed for having a bounded competitive ratio even when there is one offline node.} 

We start by introducing a novel algorithmic construct to address the above challenges. We present our new convex-programming based algorithm (\Cref{alg:opt}) based on this new technique in \Cref{sec:alg-convex}. Due to the lack of a structural decomposition, we study the key properties of this convex program directly in \Cref{sec:convex program} through convex duality. We then try to mimic the recursive primal-dual analysis of \Cref{sec:vertex-weight} in \Cref{sec:main proof}, with the major difference that we directly use the optimal Lagrangian dual solution of our convex programs in each stage to find an approximate dual certificate for our primal algorithm in the LP formulation of the offline problem.}

% that show the approximate optimality of our algorithm from our stage-wise convex program.s 

% We then set up the stage for our competitive analysis in \Cref{sec:convex program} by identifying a few key properties of the convex programs defining our algorithm in each stage $k\in[K]$ (by using convex duality). We then introduce our convex-programming based recursive primal-dual analysis in \Cref{sec:main proof}, in which we directly use the optimal Lagrangian dual solution of our convex programs in each stage to find an approximate dual certificate for our primal algorithm in the LP formulation of the offline problem.

\revcolor{\subsection{Price-level Regularized Convex Programs and Configuration Allocation}
\label{sec:alg-convex}
Before describing our algorithm, we make an
assumptions for simplicity (without loss of generality) that there exists a finite universe $0 = \weightHat_0 < \weightHat_1
< \dots < \weightHat_{\totalweight}$ of $T\in\mathbb{N}$ possible per-impression prices.
Namely, for each per-impression price $\weight_{\threeindex}$,
there exists an index $\wi\in[\totalweight]$ such that 
$\weight_{\threeindex} = \weightHat_\wi$.
As we will see soon, our algorithm does not need to know 
$\{\weightHat_\wi\}_{\wi\in[\totalweight]}$
ahead of time, and its competitive ratio will have no dependency on this finite set.\footnote{\label{footnote:weightHat}Notably, the discreetness assumption is mainly for ease of exposition, and our results can immediately be extended to the setting with continuous per-impression prices.}

\smallskip
\noindent\textbf{Overview of the Algorithm}~~~At each stage $k\in[K]$, our algorithm solves a stage-dependent convex program that is essentially a regularized version of the LP for a greedy allocation at that stage. Intuitively speaking, the main new ideas behind this convex program are:
\begin{enumerate}
    \item Regularizing the revenue function \emph{separately at different price levels} using a convex function in a structured fashion, in order to create a controlled hedging through maintaining balancedness in allocation separately at each price level, 
    \item Setting it up in a way that we can extract a balanced configuration allocation, together with the necessary amount of preemption at each price level, from the solution of the program.
\end{enumerate}
Formally, we solve the following convex program~\ref{eq:convex primal} in stage $k$ of our algorithm:
\begin{align}
\tag{$\mathcal{P}^{\textsc{MCA}}_k$}
\label{eq:convex primal}
&\arraycolsep=1.4pt\def\arraystretch{2.2}
\begin{array}{lll}
        \max\limits_{\xbf,\ybf,\zbf \geq \mathbf{0}}~~~~
        &
        \displaystyle\sum_{\rider\in\Rider_k}
        \displaystyle\sum_{\config\in\Config}
        \displaystyle\sum_{\driver\in\Driver}
        \weight_{\threeindex} 
        \cdot
        x_{\threeindex}
        - 
        \sum_{\driver\in\Driver}
        \sum_{\wi \in [\totalweight]}
        \weightHat_\wi \cdot 
        \left(
        \comsumptioni-
        y_{\twoindex}
        \right)&
        \\
        &
        \hspace{10mm}
        -
        \displaystyle\sum\limits_{\driver\in \Driver}
        \displaystyle\sum\limits_{\wi \in [\totalweight]}
        \left(
        \weightHat_\wi - \weightHat_{\wi - 1}
        \right)
        \cdot 
        F_k\left(
        \mathop{
        \displaystyle\sum
        x_{\threeindex}
        }
        \limits_{
        \substack{
        \rider\in\Rider_k,\config\in\Config:
        % (\threeindex)\in E_k,\\
        \\
        ~~\weight_{\threeindex}\geq 
        \weightHat_\wi}
        }
        % x_{\threeindex}
        +
        \displaystyle\sum\limits_{\wi' \geq \wi}
        y_{\driver,\wi'}
        \right)
        & \text{s.t.}
        \\
         &
        %  \mathop{
         \displaystyle\sum_{\rider\in\Rider_k}
         \displaystyle\sum_{\config\in\Config}
        %  x_{\threeindex}
        %  }
        %  \limits_{\rider,\config:(\threeindex)\in E_k}
         x_{\threeindex}
         +
         \displaystyle\sum_{\wi\in[\totalweight]}
         y_{\twoindex}
         \leq 1
         &
         \driver\in\Driver~,
         \\
         &
         \displaystyle\sum_{\config\in\Config} z_{\rider,\config}\leq 1
         &
         \rider\in\Rider_k~,
         \\
         &
         x_{\threeindex}\leq 
         \alloci\cdot z_{\rider,\config}
         &
        %  (\threeindex)\in E_k
        \rider\in\Rider_k,\config\in\Config,\driver\in\Driver~,
         \\
         &
         y_{\twoindex}\leq \comsumptioni
         &
         \driver\in \Driver, \wi\in[\totalweight]~.
        %  \\
        %  &x_{\threeindex}\geq 0,~z_{\rider,\config}\geq 0
        %  &\rider\in\Rider_k,\config\in\Config,\driver\in\Driver~, 
        %  \\
        %  &y_{\twoindex} \geq 0
        %  &\driver\in\Driver,\wi\in[\totalweight]~.
        %  \\
        \end{array}
\end{align}
This convex program depends on the set $\Rider_k$ of arriving users in stage $k$, the vector of impression numbers $\allocs\ked = \{\alloci\}_{\rider\in\Rider_k}$, and the vector of per-impression prices $\weightbf\ked = \{\weight_{\threeindex}\}_{\rider\in\Rider_k}$, all of which are revealed to the algorithm at the beginning of stage $k$. Furthermore, our algorithm keeps track of $\boldsymbol{\comsumption}\ked = \{\comsumption_{\twoindex}\ked\}$, where $\comsumption_{\twoindex}\ked$ is the fractional number of impressions that each advertiser $j$ is willing to pay for each per-impression price $\tau$ at the end of each stage $k$. The vector $\boldsymbol{\comsumption}^{(k-1)}$, which is basically the price distribution of top impressions allocated to advertiser $j$ with total capacity of $n_j=1$, captures the state of the allocation at the beginning of stage $k$. Notably, the above convex program \ref{eq:convex primal} depends on this state information  $\boldsymbol{\comsumption}\kminused$. As for the decision variables, variable $z_{\rider,\config}$
in the above program specifies the fractional allocation of
user $\rider\in\Rider_k$ with configuration $\config\in\Config$. Moreover, variable $x_{\threeindex}$
specifies the fractional number of impressions created by user $\rider$ form configuration $\config$ that advertiser $\driver$ is willing to pay for after the allocation of stage $k$,
and thus 
$\alloci\cdot z_{\rider,\config} - 
x_{\threeindex}$ impressions of user $\rider$ for advertiser $\driver$ are immediately preempted at the end of stage $k$.
Finally, variable $y_{\twoindex}$
specifies the total number of impressions of previous $k - 1$ stages that advertiser $\driver$ is still willing to pay for at 
per-impression price of 
$\weightHat_{\wi}$ \emph{after} allocation/preemption decisions of stage $k$. In other words,
$\comsumptioni - y_{\twoindex}$
number of impressions of advertiser $\driver$ which
were finalized in the previous $k-1$ stages at
per-impression price of $\weightHat_{\wi}$ will be preempted after stage $k$.  See \Cref{alg:opt} for a formal description.

\begin{algorithm}[htb]
 	\caption{Polynomial Regularized Max Configuration Allocation (\PRMCA)}
 	\label{alg:opt}
 	\KwIn{Number of stages $K$, convex functions $\{F_k:[0,1]\rightarrow[0,1]\}_{k\in[K]}$}
 	    \vspace{2mm}
 	    
    $\forall \driver\in\Driver,\wi\in[\totalweight]:
    ~\comsumption_{\twoindex}^{(0)}\gets 0$.

 	\vspace{1mm}
 	\For{stage $k=1$ to $K$}
 	{
 	\vspace{2mm}
 	\tcc{$k^{\textrm{th}}$ batch of users
 	in $\Rider$, denoted by $\Rider_k$, arrives. 
%  	Let $E_k$ denote the set of edges incident to vertices in $\Rider_k$.
 	}
 	\vspace{2mm}
 	
  Given $\boldsymbol{\comsumption}\kminused$,
  $\Rider_k$,
$\allocs\ked$ and $\weightbf\ked$, solve the convex program~\ref{eq:convex primal} to obtain $(\xbf\starred_k,\ybf\starred_k,\zbf\starred_k)$.
$\forall \driver\in \Driver,\wi\in[\totalweight]:
\comsumption_{\twoindex}\ked\gets
        \mathop{
        \displaystyle\sum
        x_{\threeindex}\starred
        }
        \limits_{
        \substack{
        \rider\in\Rider_k,\config\in\Config:
        % (\threeindex)\in E_k,\\
        \\
        ~~\weight_{\threeindex}= 
        \weightHat_\wi}
        }
        % x_{\threeindex}
        +
        y_{\driver,\wi}\starred$.

Return $(\xbf\starred_k,\ybf_k\starred,\zbf_k\starred)$
as the fractional solution of stage $k$.

\tcc{$\zbf_k\starred:$~configuration allocation of stage $k$, $\boldsymbol{\comsumption}\kminused-\ybf_k\starred:$~preemption decisions of stage $k$.}
 	}
\end{algorithm}

We are now ready to present the main result of this section. We prove this theorem in \Cref{sec:main proof}.

\begin{theorem}[\textbf{Optimal Multi-Stage Algorithm for Configuration Allocation}]
\label{thm:main result configuration allocation}
In the $K$-stage (fractional) configuration allocation problem,
 \PRMCA~(\Cref{alg:opt}) is $\cratio{K}$-competitive, where $\cratio{K}=\cratioexp$.
\end{theorem}

\begin{remark}
% [\textbf{Optimality of the $\boldsymbol{\cratio{K}}$ Competitive Ratio for Configuration Allocation Problem}]
We note that the multi-stage configuration allocation
problem generalizes the multi-stage edge-weighted matching with free-disposal~\citep{FKMMP-09}, budgeted allocation or AdWords with no disposal~\citep{MSVV-07}, and vertex weighted matching with no disposal~\citep{AGKM-11}. As we show later in \Cref{sec:lowerbound}, no integral or fractional multi-stage algorithm can obtain a competitive ratio better than $\cratio{K}$ in the multi-stage (unweighted) matching, and hence $\cratio{K}$ is the optimal competitive ratio for the multi-stage configuration allocation.
\end{remark}}

% The convex program \ref{eq:convex primal}
% plays a crucial role in our optimal 
% competitive algorithm.
% In \Cref{sec:program analysis}, we 
% provide detailed characterizations
% of 
% this convex program by considering its dual
% program.
% Then, 
% combining such characterizations with 
% a primal dual analysis framework,
% we proof the competitive ratio of 
% \Cref{alg:opt}
% in \Cref{sec:main proof}.

\revcolor{\subsection{Properties of Price-Level Regularized Convex Programs via Convex Duality}
\label{sec:convex program}

Fix any stage $k\in[K]$. We now aim to identify/characterize some useful properties of the convex program \ref{eq:convex primal}. To simplify our study, we define the following notation for any $(\xbf,\ybf)$:
$$\cumcomsumptioni(\xbf,\ybf)\triangleq\sum
% \limits
_{
        \substack{\rider\in\Rider_k,\config\in\Config:
        \\
        ~~\weight_{\threeindex}\geq 
        \weightHat_\wi}
        }
        x_{\threeindex}
        +
        \sum_{\wi' \geq \wi}
        y_{\driver,\tau'}$$
Note that if 
$(\xbf\ked, \ybf\ked,\zbf\ked)$ is the fractional allocation selected by \Cref{alg:opt} (i.e., optimal solution of \ref{eq:convex primal}), then 
by definition, $\cumcomsumptioni(\xbf\ked,\ybf\ked) = \sum_{\tau'\geq \tau} \comsumption_{\twoindex}\ked$. In other words, $1-\cumcomsumptioni(\xbf\ked,\ybf\ked)$ is essentially the cumulative per-impression price distribution of advertiser $j$ at the end of stage $k$. Note that in the first $K - 1$ stages, 
the objective function in program~\ref{eq:convex primal}
is strictly concave,
while the convex program
degenerates to a linear program in the last stage $K$. For this reason, we study the properties of stages $k \in [K - 1]$
and the last stage $K$ separately. 

\smallskip
\noindent\textbf{Properties of Stage $\boldsymbol{k\in [K - 1]}$.}
To characterize the properties of the optimal solution in 
\ref{eq:convex primal} for stage $k\in[K-1]$,
we consider its 
Lagrangian relaxation 
and KKT optimality conditions.
Specifically, 
the Lagrangian dual \ref{eq:convex dual}
of \ref{eq:convex primal}
can be written as follows,
\begin{align*}
\arraycolsep=1.4pt\def\arraystretch{2.2}
    \tag{$\mathcal{D}^{\textsc{MCA}}_k$}
\label{eq:convex dual}
\begin{array}{ll}
   \min\limits_{\substack{\lambdabf,\thetabf,\mubf,\pibf,\\
   \psibf,\phibf,\iotabf\geq 0}} 
   &
   \max\limits_{\xbf,\ybf,\zbf\geq \mathbf{0}} 
           \displaystyle\sum_{\rider\in\Rider}
           \displaystyle\sum_{\config\in\Config}
           \displaystyle\sum_{\driver\in\Driver}
        \weight_{\threeindex} 
        \cdot
        x_{\threeindex}
        - 
        \sum_{\driver\in\Driver}
        \sum_{\wi \in [\totalweight]}
        \weightHat_\wi \cdot 
        \left(
        \comsumptioni-
        y_{\twoindex}
        \right)
    \\
     & 
        \hspace{10mm}
        -
        \displaystyle\sum\limits_{\driver\in \Driver}
        \displaystyle\sum\limits_{\wi \in [\totalweight]}
        \left(
        \weightHat_\wi - \weightHat_{\wi - 1}
        \right)
        \cdot 
        F_k\left(
        % \displaystyle\sum\limits_{
        % \substack{\rider,\config:
        % (\threeindex)\in E_k,\\
        % ~~\weight_{\threeindex}\geq 
        % \weightHat_\wi}
        % }
        % x_{\threeindex}
        % +
        % \displaystyle\sum\limits_{\wi\primed \geq \wi}
        % y_{\driver,\tau\primed}
        \cumcomsumptioni(\xbf,\ybf)
        \right)\\
        &
        \hspace{10mm}
        - 
        \displaystyle\sum_{\driver\in \Driver}
        \thetai\cdot 
        \left(
        % \mathop{
        % \displaystyle\sum
        %  x_{\threeindex}
        %  }
        %  \limits_{\rider,\config:(\threeindex)\in E_k}
        \displaystyle\sum_{\rider\in\Rider_k}
        \displaystyle\sum_{\config\in\Config}
         x_{\threeindex}
         +
         \displaystyle\sum_{\wi\in[\totalweight]}
         y_{\twoindex}
         - 1
        \right)
        -
        \displaystyle\sum_{\rider\in\Rider_k}
        \lambdai\cdot 
                \left(
        \displaystyle\sum_{\config\in\Config} z_{\rider,\config}
         - 1
        \right)
        \\
        &
        \hspace{10mm}
        -
        % \mathop{\displaystyle\sum
        % \mui}
        % \limits_{(\threeindex)\in E_k}
           \displaystyle\sum_{\rider\in\Rider}
           \displaystyle\sum_{\config\in\Config}
           \displaystyle\sum_{\driver\in\Driver}
           \mui\cdot 
        \left(x_{\threeindex} - \alloci\cdot z_{\rider,\config}
        \right)
        -
        \displaystyle\sum_{\driver\in \Driver}
        \displaystyle\sum_{\wi\in[\totalweight]}
        \pi_{\twoindex}\left(
        y_{\twoindex} - \comsumptioni
        \right)
        \\
        &
        \hspace{10mm}
        +
        % \mathop{\displaystyle\sum
        % \psii\cdot
        % }\limits_{(\threeindex)\in E_k}
           \displaystyle\sum_{\rider\in\Rider}
           \displaystyle\sum_{\config\in\Config}
           \displaystyle\sum_{\driver\in\Driver}
        \psii\cdot 
        x_{\threeindex}
        +
        % \mathop{\displaystyle\sum
        %  \phii  \cdot z_{\rider,\config}}
        % \limits_{\rider,\config:\exists (\threeindex)\in E_k}
        \displaystyle\sum_{\rider\in\Rider}
           \displaystyle\sum_{\config\in\Config}
           \phii  \cdot z_{\rider,\config}
        +
        \displaystyle\sum
        \limits_{\driver\in \Driver}
        \displaystyle\sum_{\wi\in[\totalweight]}
        % \cdot
        \iota_{\twoindex}\cdot
        y_{\twoindex}
\end{array}
\end{align*}
We start with the following lemma (proved in Appendix~\ref{apx:kkt-CA}) which builds 
a connection between the optimal solutions of
 \ref{eq:convex primal} and \ref{eq:convex dual}.
Briefly speaking, this lemma can be thought as
a collection of equalities extracted from
KKT conditions, which will later be used in the
competitive ratio analysis of \Cref{alg:opt}.
\begin{lemma}[Duality-based Connection]
\label{lem:KKT restate}
Fix any $k\in[K - 1]$.
Suppose $(\xbf\starred,\ybf\starred,\zbf\starred)$
is the optimal solution of the 
convex program \ref{eq:convex primal}.
There exists non-negative ($\lambdabf\starred, \thetabf\starred,
\mubf\starred, \psibf\starred,\phibf\starred$)
such that for all $(\threeindex)\in E_k$
\begin{align}
    \tag{complementary slackness}
    \label{eq:cs}
    \begin{split}
        % \thetai\starred
        % \left(
        % \displaystyle\sum_{\rider,\config:(\threeindex)\in E_k}
        %  x_{\threeindex}\starred
        %  +
        %  \displaystyle\sum_{\wi\in[\totalweight]}
        %  y_{\twoindex}\starred
        %  - 1
        % \right)
        % = 0 \\
        \lambdai\starred
                \left(
        \displaystyle\sum_{\config\in\Config} z_{\rider,\config}\starred
         - 1
        \right)
        =0
        \hspace{10mm}
        \mui\starred
        \left(x_{\threeindex}\starred
        - \alloci\cdot z_{\rider,\config}\starred
        \right) = 0
        \\
        \thetai\starred=0 \hspace{10mm}\psii\starred\cdot x_{\threeindex}\starred
        =0
        \hspace{10mm}
        \phii\starred\cdot z_{\rider,\config}\starred
        =0
    \end{split}
\end{align}
\begin{align}
    \label{eq:stationary z}
\tag{stationarity-1}
    \sum_{\driver\in \Driver}\alloci \cdot \mui\starred - \lambdai\starred + \phii\starred = 0
    % &
    % \hspace{10mm}
    % \forall \rider \in \Rider_k
    \\
\tag{stationarity-2}
\label{eq:stationary x}
    \weight_{\threeindex}
    -
    \sum\limits_{\wi\in[\totalweight]:
    \weightHat(\wi) \leq \weight_{\threeindex}}
    \left(\weightHat_\wi - \weightHat_{\wi-1}\right)
    \cdot 
    f_{k}(\cumcomsumptioni(\xbf\starred, \ybf\starred))
    % -
    % \thetai\starred 
    - \mui\starred + \psii\starred = 0
    % &
    % \hspace{10mm}
    % \forall (\threeindex)\in E_k 
\end{align}
\end{lemma}

Besides the above lemma, we intuitively expects that the optimal solution 
$(\xbf\starred,\ybf\starred,\zbf\starred)$
of \ref{eq:convex primal}
satisfies the following property:
whenever the entire capacity of a given advertiser is exhausted
and preemption is needed, 
it is optimal to dispose impressions which are allocated with the smallest price. 
% for any advertiser $\driver$,
% if there exists $x_{\threeindex}\starred > 0$
% and $\comsumptioni - y_{\twoindex} >0$,
% then we have $\weight_{\threeindex} \geq 
% \weightHat_{\wi}$,
% i.e., \Cref{alg:opt} only preempt  
% units of a advertiser from lower revenue level 
% to higher revenue level.
This intuition can be formalized below
and proved by convex duality (see Appendix~\ref{apx:kkt-CA-smallest}).
\begin{lemma}[Disposing Smallest Price]
\label{lem:preempt monotone}
Fix $k\in[K - 1]$.
Suppose $(\xbf\starred,\ybf\starred,\zbf\starred)$
is the optimal solution of \ref{eq:convex primal}.
For any advertiser $\driver\in \Driver$, and $\wi\in [\totalweight]$,
if $\comsumptioni - y_{\twoindex}^* > 0$,
then $\cumcomsumptioni(\xbf\starred,\ybf\starred) = 1$.
\end{lemma}

\Cref{lem:preempt monotone} gives us the following
corollary (proved in Appendix~\ref{apx:kkt-CA-cor1}), which again will later help with our competitive ratio analysis of \Cref{alg:opt}.

\begin{corollary}
Fix $k\in[K - 1]$.
\label{coro:cumulative comsumption monoton} 
Suppose $(\xbf\starred,\ybf\starred,\zbf\starred)$
is the optimal solution of \ref{eq:convex primal}. For any advertiser $\driver\in \Driver$, and $\wi\in [\totalweight]$, we have:
\begin{itemize}
    \item (\textbf{Distribution Monotonicity}) $\displaystyle\cumcomsumptioni(\xbf\starred,\ybf\starred) \geq 
\sum_{\wi'\geq \wi}\comsumptioni$.
\item (\textbf{Only Disposing Smallest}) $\displaystyle\sum_{\wi'\geq \wi}\left(\comsumption_{\driver,\wi'}\kminused - y_{\driver,\wi'}^*\right)
=
\sum_{\wi'\geq \wi}\left(\comsumption_{\driver,\wi'}\kminused - y_{\driver,\wi'}^*\right)
\cdot
f_k\left(\cumcomsumptioni(\xbf\starred,\ybf\starred)\right)$.
\end{itemize}
\end{corollary}

% \begin{corollary}
% Fix $k\in[K - 1]$.
% \label{coro:preempt complementary slackness}
% Suppose $(\xbf\starred,\ybf\starred,\zbf\starred)$
% is the optimal solution of \ref{eq:convex primal}.
% For any advertiser $\driver\in \Driver$, and $\wi\in [\totalweight]$,
% $\sum_{\wi'\geq \wi}\left(\comsumption_{\driver,\wi'}\kminused - y_{\driver,\wi'}^*\right)
% =
% \sum_{\wi'\geq \wi}\left(\comsumption_{\driver,\wi'}\kminused - y_{\driver,\wi'}^*\right)
% \cdot
% f_k\left(\cumcomsumptioni(\xbf\starred,\ybf\starred)\right)$.
% \end{corollary}

\smallskip
\noindent\textbf{Properties of the Last Stage $K$.}
For
stage $K$, 
note that $F_K(a) = 0$ for all $a \in [0, 1]$
and thus convex program \ref{eq:convex primal} 
degenerates to a linear program,
which solves the optimal configuration allocation
for the last stage graph under $\boldsymbol{\comsumption}\KMINUSED$.
This linear program \ref{eq:convex primal}  
admits a dual as follows,
\begin{align}
\tag{LP-$\mathcal{D}_K$}
\label{eq:convex dual K}
\arraycolsep=1.4pt\def\arraystretch{2.2}
    \begin{array}{lll}
    \min\limits_{\lambdabf,\thetabf,\mubf,\pibf\geq 0} &
    \displaystyle\sum_{\rider\in \Rider_K} \lambdai 
    +
    \displaystyle\sum_{\driver\in \Driver} \thetai
    -
    \displaystyle\sum_{\driver\in\Driver}
    \displaystyle\sum_{\wi\in[\totalweight]}
    \comsumption_{\twoindex}\KMINUSED
    \cdot (\weightHat_{\wi}- \pi_{\twoindex})
    &\text{s.t.}
    \\
    &
    \mui + \thetai \geq \weight_{\threeindex}
    & \rider\in\Rider_K, \config\in\Config,
    \driver\in \Driver~,
    \\
    &
    \thetai + \pi_{\twoindex} \geq \weightHat_{\wi}
    &
    \driver\in\Driver,\wi\in[\totalweight]~,
    \\
    &
    \lambdai \geq \displaystyle\sum_{\driver\in\Driver}
    \alloci\cdot \mui
    &
    \rider\in \Rider_K,
    \config\in \Config~.
    \end{array}
\end{align}

Technically speaking, the optimal 
solution of the dual program \ref{eq:convex dual K} for the last stage
satisfies the same KKT-like equalities/inequalities
as the solution of the Lagrangian dual program
\ref{eq:convex dual} for the first $K-1$ stages.
The main difference is the value of 
the optimal
$\thetai\starred$.
As we showed in \Cref{lem:KKT restate},
$\thetai\starred = 0$ in the first $K - 1$ stages.
This follows from 
the construction of function $F_k(\cdot)$, and in particular, the property that $\frac{d}{dx}F_k(1) = f_k(1) = 1$.
In contrast, $F_K\equiv 0$ in the last stage $K$,
and the value of the optimal $\thetai\starred$
can be strictly positive. It turns out this makes the competitive ratio analysis of \Cref{alg:opt} in the last stage a bit different from the previous stages. We introduce the following auxiliary lemma
to characterize the required properties of dual program \ref{eq:convex dual K} to help with this different part of the analysis. The proof is deferred to Appendix~\ref{apx:last-stage-lemma}.
\begin{lemma}
\label{lem:dual restate theta K}
Suppose 
$(\lambdabf\starred,\thetabf\starred,\mubf\starred,\pibf\starred)$
is the optimal solution of the dual program \ref{eq:convex dual K}.
For any advertiser $\driver\in\Driver$,
$\wi\in[\totalweight]$,
\begin{itemize}
    \item(\textbf{Low Prices of Last Stage}) \mbox{if $\weightHat_\wi\leq \thetai\starred $}:~~~~$\comsumption_{\twoindex}\KMINUSED\cdot \pi_{\twoindex}\starred = 0$
    \item(\textbf{High Prices of Last Stage}) \mbox{if $\weightHat_\wi>\thetai\starred $}:~~~~$\comsumption_{\twoindex}\KMINUSED
     \cdot 
     (
     \weightHat_{\wi} - \pi_{\twoindex}\starred)
    = 
     \comsumption_{\twoindex}\KMINUSED\cdot \thetai\starred$
\end{itemize}
% \begin{align*}
%     \comsumption_{\twoindex}\KMINUSED\cdot \pi_{\twoindex}\starred = 0
%     \;\;\mbox{if $\weightHat_\wi \leq \thetai\starred$; and}
%     \;\;
%      \comsumption_{\twoindex}\KMINUSED
%      \cdot 
%      (
%      \weightHat_{\wi} - \pi_{\twoindex}\starred)
%     = 
%      \comsumption_{\twoindex}\KMINUSED\cdot \thetai\starred
%     \;\;\mbox{if $\weightHat_\wi \geq \thetai\starred$}
% \end{align*}

% if $\weightHat_\wi \leq \thetai\starred$,
% then
% % \begin{align*}
%     $\comsumption_{\twoindex}\KMINUSED\cdot \pi_{\twoindex}\starred = 0$;
% % \end{align*}
% % \intertext{
% and if $\weightHat_\wi \geq \thetai\starred$,
% then
% % }
% % \begin{align*}
%     $\weightHat_{\wi} - \pi_{\twoindex}\starred 
%     = 
%     \thetai\starred$
% % \end{align*}
\end{lemma}

% First, we present the following lemma 
% which characterizes the optimal solution
% in program \ref{eq:convex primal}
% for stage $K$
% and its dual program 
% \ref{eq:convex dual K}.

% \begin{lemma}
% \label{lem:LP dual restated}
% Suppose $(\xbf\starred,\ybf\starred,\zbf\starred)$
% is 
% the optimal solution of the linear
% program \ref{eq:convex primal}
% for the last stage $K$ and
% $(\lambdabf\starred,\thetabf\starred,\mubf\starred,\pibf\starred)$
% is the optimal solution of the dual program \ref{eq:convex dual K}.
% Then,
% \begin{align}
%     \theta_{j}\starred =  
%      \weightHat_{\wi} - \pi_{\twoindex}\starred
%      \text{ if $y_{\twoindex}\starred > 0$}
% \end{align}
% \end{lemma}

\subsection{Competitive Ratio Analysis of \Cref{alg:opt}}
\label{sec:main proof}
In this section we show the competitive ratio of $\cratio{K}$  for \Cref{alg:opt} as stated in \Cref{thm:main result configuration allocation}. 
Remember the optimum offline 
LP \ref{eq:LP primal} in \Cref{sec:prelim}. We start by considering its dual:
\vspace{1mm}
\begin{equation}
\tag{\textsc{OPT-PRIMAL-DUAL}}
\label{eq:LP}
\arraycolsep=1.4pt\def\arraystretch{1}
\begin{array}{llllllll}
\max\limits_{\xbf,\zbf\geq \boldsymbol{0}}
\quad~~&
% \mathop{\displaystyle\sum
% \weight_{\threeindex}\cdot}
% \limits_{(\threeindex)\in E}
\displaystyle\sum_{\rider\in\Rider}
\displaystyle\sum_{\config\in\Config}
\displaystyle\sum_{\driver\in\Driver}
 \weight_{\threeindex}\cdot x_{\threeindex}
&~\text{s.t.}&
&
\min\limits_{\alphabf,\betabf,\gammabf\geq \boldsymbol{0}}
\quad~~
&
\displaystyle\sum_{\rider\in \Rider }{\alpha_\rider}+
\displaystyle\sum_{\driver\in \Driver}{\beta_\driver}
&~~\text{s.t.} \\[1.4em]
 &
%  \mathop{\displaystyle\sum
%  x_{\threeindex} }
%  \limits_{\rider,\config:(\threeindex)\in E}
\displaystyle\sum_{\rider\in\Rider}
\displaystyle\sum_{\config\in\Config}
x_{\threeindex}
 \leq1 &
 \driver\in\Driver~,& 
& 
&\gamma_{\threeindex} 
+
\beta_\driver \geq \weight_{\threeindex}
& 
\rider\in \Rider,
\config\in\Config,
\driver\in\Driver~,
\\[1.4em]
 &
 \displaystyle\sum\limits_{\config\in\Config}
 z_{\rider,\config}
 \leq 1 
 &
 \rider\in \Rider~, &
& &\alpha_\rider 
\geq 
\displaystyle\sum\limits_{\driver\in \Driver}
\alloci\cdot \gamma_{\threeindex} &i\in \Rider, \config\in \Config~, \\[1.4em]
 &
 x_{\threeindex} \leq 
 \alloci \cdot z_{\rider,\config}
 &
\rider\in \Rider,
\config\in\Config,
\driver\in\Driver~,
 &
& &
 &
\end{array}
\end{equation}

Similar to \Cref{sec:primal-dual}, given the 
feasible primal assignment $\{\xbf\ked,\ybf\ked,\zbf\ked\}_{k\in[K]}$ produced by \Cref{alg:opt} (which leads to a final assignment of $\hat{\xbf}$ when considering the impressions disposals at the end), we construct a non-negative dual assignment so that,
\begin{itemize}
    \item Primal and dual have the same objective values: $$\displaystyle\sum_{\rider\in\Rider}
    \displaystyle\sum_{\config\in\Config}
    \displaystyle\sum_{\driver\in\Driver}\weight_{\threeindex}\cdot \hat{x}_{\threeindex}=\displaystyle\sum_{k\in[K]}\left(
    % \displaystyle\sum_{(\threeindex)\in 
    % E_k}
    \displaystyle\sum_{\rider\in\Rider_k}
    \displaystyle\sum_{\config\in\Config}
    \displaystyle\sum_{\driver\in\Driver}
    \weight_{\threeindex}\cdot x_{\threeindex}\ked - \sum_{\driver\in \Driver}
    \sum_{\wi\in[\totalweight]}
    \weightHat_{\wi}\cdot
    \left(\comsumptioni - y_{\twoindex}\ked
    \right)\right)
    =
    \displaystyle\sum_{\rider\in\Rider}\alpha_\rider
    +
    \displaystyle\sum_{\driver\in\Driver}\beta_\driver$$
    
     \vspace{1mm}
    \item Dual is approximately feasible: 
    % in particular,
    \begin{align*}
    \forall \rider\in\Rider,\config\in\Config,\driver\in\Driver:
    ~~
    &\gamma_{\threeindex}
    +
    \beta_\driver \geq \cratio{K}\cdot \weight_{\threeindex}
    \intertext{and}
    \forall \rider\in \Rider,\config\in\Config: 
    ~~
    &\alpha_{\rider} \geq 
    \displaystyle\sum_{\driver\in \Driver}
    \alloci \cdot \gamma_{\threeindex}
    \end{align*}
\end{itemize}
Existence of such a dual assignment will guarantee that 
\Cref{alg:opt}
is $\cratio{K}$-competitive.

Our dual assignment is constructed stage-by-stage.
First, set $\alpha_i\gets 0$, $\beta_j\gets 0$,
and $\gamma_{\threeindex} \gets 0$
for all $\rider\in\Rider,\config\in\Config,\driver\in\Driver$. 
At each stage $k\in[K-1]$, 
let $(\lambdabf\ked,\thetabf\ked,\mubf\ked,\psibf\ked,\phibf\ked)$
to denote 
$(\lambdabf\starred,\thetabf\starred,\mubf\starred,\psibf\starred,\phibf\starred)$ in \Cref{lem:KKT restate},
and $\cumcomsumptioni\ked$
to denote $\cumcomsumptioni(\xbf\ked,\ybf\ked)$.
Consider the following assignment corresponding to stages $k\in[K-1]$ for the dual program in \ref{eq:LP}:
\begin{align*}
   &\forall \rider \in \Rider_k: 
   &
   &\alpha_\rider\gets 
%   \sum_{\config \in \Config}
%   z_{\rider,\config}\ked\cdot 
   \lambda_\rider\ked
   \\
   &\forall \rider\in\Rider_k,
   \config\in\Config,
   \driver\in\Driver:
%   \in E_k:
   &
   &
   \gamma_{\threeindex} \gets
   \mu_{\threeindex}\ked 
   \\
    &\forall \driver \in \Driver: &&\beta_{\driver}\gets\beta_{\driver}+\Delta\beta_{\driver}\ked\\
    &&&
    \Delta\beta_{\driver}\ked
    \triangleq 
    \sum_{\rider\in\Rider_k}
    \sum_{\config\in\Config}
    % \sum_{\rider,\config:(\threeindex)\in E_k}
    x_{\threeindex}\ked
    \left(
    % \theta_{\driver}\ked + 
    \sum_{\wi\in[\totalweight]:
    \weightHat(\wi) \leq \weight_{\threeindex}}
    \left(\weightHat_\wi - \weightHat_{\wi-1}\right)
    \cdot 
    % F_{\twoindex}(\xbf\ked, \ybf\ked)
    f_k\left(\cumcomsumptioni\ked\right)
    \right)
    \\
    &&&
    \hspace{25mm}
    -
    \sum_{\wi\in[\totalweight]}
    \weightHat_{\wi}\cdot
    \left(\comsumptioni - y_{\twoindex}\ked
    \right)
\end{align*}
For the last stage $K$,
let $(\lambdabf\KED,\thetabf\KED,\mubf\KED,\pibf\KED)$
be the optimal solution of \ref{eq:convex dual K}.
We consider the following assignment corresponding to stage $K$ for the dual program in \ref{eq:LP}:
\begin{align*}
   &\forall \rider \in \Rider_k: 
   &
   &\alpha_\rider\gets 
   \lambda_\rider\KED
   \\
   &\forall (\threeindex) \in E_k:
   &
   &
   \gamma_{\threeindex} \gets
   \mu_{\threeindex}\KED 
   \\
    &\forall \driver \in \Driver: &&\beta_{\driver}\gets\beta_{\driver}+\Delta\beta_{\driver}\KED\\
    &&&
    \Delta\beta_{\driver}\KED
    \triangleq 
    \thetai\KED 
    -
    \sum_{\driver\in\Driver,\wi\in[\totalweight]}
    \comsumption_{\twoindex}\KMINUSED
    \cdot (\weightHat_{\wi} - \pi_{\twoindex}\KED)
\end{align*}

The rest of the proof is quite technical and we postpone to Appendix~\ref{apx:proof-primal-dual-CA}. Interestingly, the proof heavily relies on all the properties that we identified in \Cref{sec:convex program}; in particular, we use the \emph{``Connection''} (\Cref{lem:KKT restate}) combined with the \emph{``Disposing Smallest Price''} (\Cref{lem:preempt monotone}) and its corollary (\Cref{coro:cumulative comsumption monoton}) to show for all stages $k\in[K-1]$ we have an equal change in the dual and primal objectives. For the last stage $K$, this property holds by construction. In order to show the approximate dual feasibility for constraints corresponding to stages $k\in[K-1]$, again we rely on the \emph{``Connection''} (\Cref{lem:KKT restate}). However, showing dual approximate feasibility for the constraints corresponding to last stage is a bit different. There, we heavily use \emph{``Low Prices of Last Stage''} (first part of \Cref{lem:dual restate theta K}) and \emph{``High Prices of Last Stage''} (second part of \Cref{lem:dual restate theta K}) to do the analysis.

}
\section{Competitive Ratio Upper-bound}
\label{sec:lowerbound}
In this section, we show a tight upper-bound of $\cratio{K}=\cratioexp$ on the worst-case competitive ratio of any multi-stage fractional matching algorithm, even for the special case of unweighted matching. This upper-bound clearly applies to the case of integral algorithms and the case of vertex weighted matching problem, and shows our earlier lower-bounds are the best achievable competitive ratios.

\begin{theorem}[\textbf{Tight Upper-bound for Multi-stage Matching}]
\label{thm:tightness}
No multi-stage fractional matching algorithm
can achieve a competitive ratio 
better than $\cratio{K}=\cratioexp$
in the $K$-stage (unweighted)
matching problem.
\end{theorem}
\smallskip
\noindent{\textbf{High-level Proof Overview}~} In order to show the upper-bound of $\cratio{K}$, we propose a distribution over deterministic instances of the multi-stage weighted fractional matching problem (i.e., a randomized instance), and show no fractional multi-stage algorithm obtains a competitive ratio better than $\cratio{K}$ over this randomized instance. To do so, we incorporate a simple factor revealing linear program whose optimal objective value provides an upper-bound on the competitive ratio of any feasible multi-stage matching algorithm. We then use LP primal-dual and by finding a feasible solution for its dual program we obtain the $\cratio{K}$ upper-bound. Interestingly, our upper-bound instance is a distribution over unweighted instances. Hence, the same upper-bound holds for the special case of multi-stage fractional matching in which all weights are equal.  

\begin{remark}[\textbf{Upper-bound for the Online Fractional Matching}]
\label{rem:online-matching-with-finite-nodes}
Notably, we need batches of size more than one in this instance, and hence our upper-bound \emph{does not} apply to the online matching problem with $K$ arrivals and batches of size $1$. So, while the competitive ratio $\cratio{K}$ is optimal for the multi-stage batch arrival problem, the competitive ratio of $\cratio{n}$ might not be optimal for the special case of online matching with $n$ vertex arrivals. To the best of our knowledge, the best known (almost) non-asymptotic competitive ratio upper-bound for the online fractional matching with $n$ arrivals is due to \cite{feige2019tighter}, where a bound of $\Theta(n)=\left(1-\frac{1}{e}\right)\left(1+\frac{1}{2n}\right)+O\left(\frac{1}{n^2}\right)$ is established. \revcolor{Note that a simple asymptotic expansion shows that $\Gamma(n)=\left(1-\frac{1}{e}\right)+\frac{1}{2en}+O\left(\frac{1}{n^2}\right) < \Theta(n)$, and hence the two bounds have a gap. Establishing the tight upper-bound for this problem is still open.}
 
\end{remark}
 
\subsection{Proof of \Cref{thm:tightness}}

Let $\M = K^K$.
Consider a bipartite graph 
$G=(\Rider,\Driver,E)$ with online vertices $\Rider$
and offline vertices $\Driver$,
where $\Rider = \Driver =
\{1,\dots, K\cdot \M\}= [K\M]$.
Online vertices $\Rider$
are partitioned into $K$ batches
$\{\Rider_1,\dots, \Rider_K\}$
where
the size of $\Rider_k$ is  $\left(\frac{K-1}{K}\right)^{k-1}\M$
for $k \in [K -1]$,
and $K\left(\frac{K-1}{K}\right)^{K-1}\M$
for $k = K$.
As a sanity check, notice that $\sum_{k = 1}^{K-1}
\left(\frac{K-1}{K}\right)^{k-1}\M +
K\left(\frac{K-1}{K}\right)^{K-1}\M = K\M$.

\begin{figure}[hbt!]
	\centering
	
\includegraphics[trim=6.7cm 0cm 8.2cm 0cm,clip,width=0.8\textwidth]{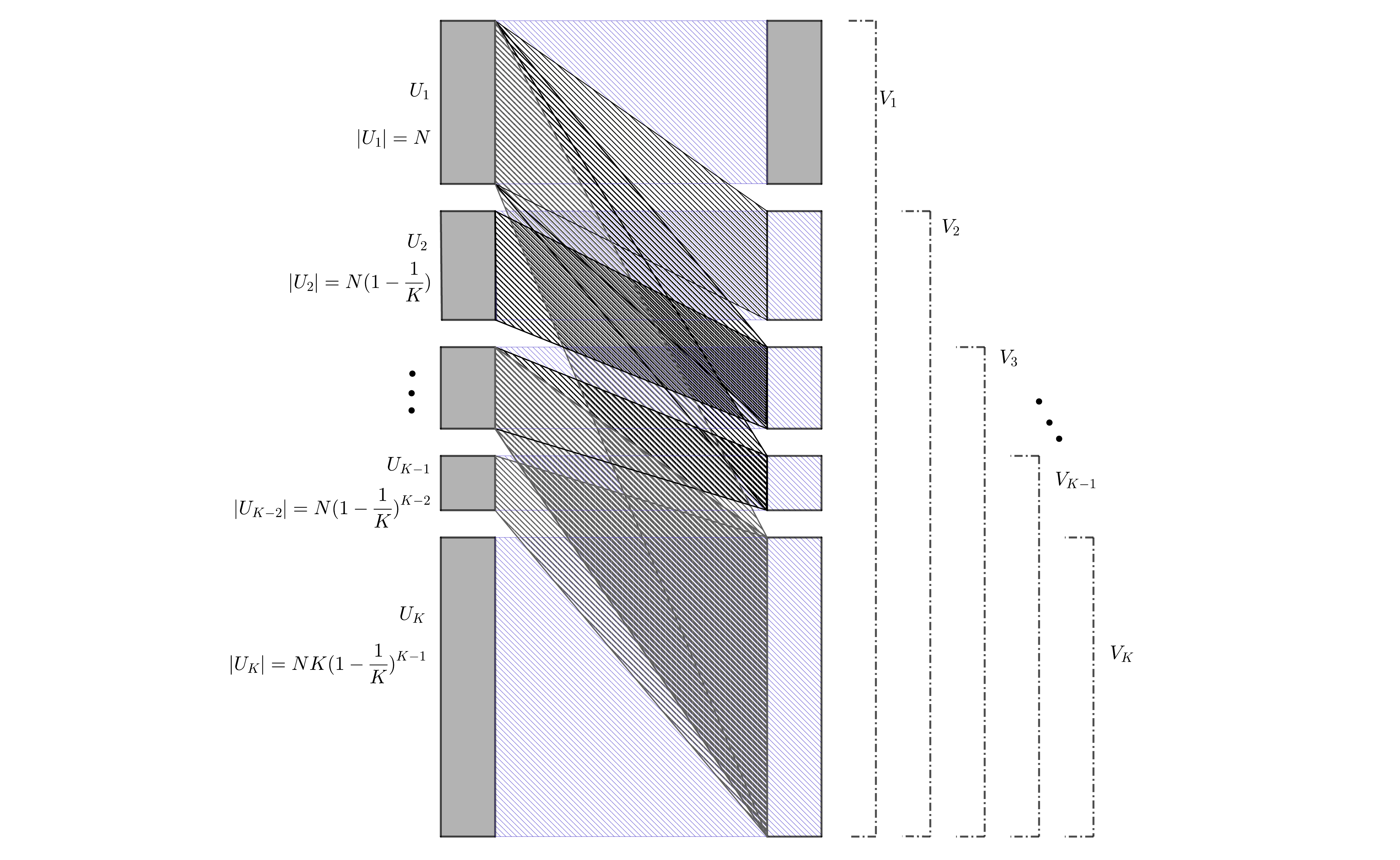}
\caption{\label{fig:bad-example}The upper-bound instance when $\boldsymbol{\pi(i)=NK-i+1, i\in[NK]}$}
\end{figure}

Let $\permu:[K\M]\rightarrow[K\M]$ be a uniform random permutation over $[K\M]$. Define the edge set $E$ of the graph $G$ as follows:
for each $k \in [K - 1]$,
each online vertex $i\in \Rider_k$
and offline vertex $j \in \Driver$,
there is an edge $(i, j)$ if $\permu(j) \leq K\,|\Rider_k|$;
for each online vertex $i\in\Rider_K$
and offline vertex $j\in \Driver$,
there is an edge $(i, j)$ if $\permu(j) \leq |\Rider_K|$.
Namely, at each stage $k$, we can partition offline vertices $\Driver$ into $\Driver_k$ and $\bar\Driver_k$
where $\Rider_k$ and $\Driver_k$
form a complete bipartite subgraph while 
there is no edge between $\Rider_k$
and $\bar\Driver_k$. The construction also ensures
that $\Driver_K \subseteq \Driver_{K - 1} \subseteq 
\dots \subseteq \Driver_1$. See \Cref{fig:bad-example} for an illustration of this randomized problem instance when permutation $\pi(i)=NK-i+1, i\in[NK]$ is realized.

First, we claim that there exists a perfect matching:
by backward induction on $k$ starting from $k=K$,
note that matching
all online vertices in
$\Rider_k$ with 
all offline vertices in
$
% \left\{
% j\in 
\Driver_k \backslash \Driver_{k+1} 
% \sum_{k' = k+1}^K |\Rider_{k'}| <
% \permu(j)
% \leq 
% \sum_{k' = k}^K |\Rider_{k'}|
% \right\}
$
for $k=1,2,\ldots,K$ produces a perfect matching between $\Rider$ and $\Driver$.

Next, we show that the expected size of the matching
produced by any given (possibly randomized) multi-stage fractional matching algorithm 
is at most $\cratio{K}\cdot K\M$,
where the expectation is taken over the randomness of the multi-stage algorithm
and the permutation $\permu$.
Let $\lusage_{k, j}^{\permu}$ be the expected degree
of offline vertex $j$ matched by the algorithm at stage $k$
under permutation $\permu$.
% The symmetry of our graph construction
% ensures that 
Note that:
$$\expect[\permu]{\lusage_{k, j}^{\permu} \condition j \not\in \Driver_k} = 0~~~~~\textrm{and}~~~~~\expect[\permu]{\lusage_{k, j}^{\permu} \condition j \in \Driver_k} 
= 
\expect[\permu]{\lusage_{k, j'}^{\permu}\condition j' \in \Driver_k}~$$ 
for all stages $k\in[K]$ and offline vertices $j, j'\in \Driver$.
The latter fact holds due to
the symmetry of our graph construction, i.e., conditioned on the event 
$[j\in\Driver_k~\&~j'\in \Driver_k]$,
it is an automorphism that exchanges $j$ and $j'$.
Thus, we can denote 
$\expect[\permu]{\lusage_{k, j}^{\permu} \condition j \in \Driver_k}$
by $\lusage_k$,
and write the following linear program and its dual 
to 
upper bound the expected size of the fractional matching
produced by any given (randomized or deterministic, integral or fractional) multi-stage algorithm:
\begin{equation*}
% \tag{\textsc{Primal-Dual-Upperbound}}
% \label{eq:LP-lower-bound}
\arraycolsep=1.4pt\def\arraystretch{1}
\begin{array}{llllllll}
\max\quad\quad\quad&
\displaystyle\sum_{k\in[K]}
|\Driver_k|\lusage_k
\qquad\qquad&\text{s.t.}&
&\quad\quad\text{min}\quad\quad\quad &
\mu + 
\displaystyle\sum_{k\in[K] }
{|\Rider_k|\zeta_k}
\qquad\qquad&\text{s.t.} \\[1.4em]
 &\displaystyle\sum_{k\in [K]}{\lusage_k}\leq 1~, & 
 & 
& &
\mu +|\Driver_k|\zeta_k
\geq |\Driver_k| & k\in[K]~,\\[1.4em]
 &
 |\Driver_k|\lusage_k
 \leq 
 |\Rider_k|&k\in [K]~, &
& &\mu\geq 0~, & \\
 &\lusage_k \geq 0 &k\in[K]~. &
& &\zeta_k \geq 0  &k\in [K]~. 
\end{array}
\end{equation*}
\ydrevcolor{Here, the variables are $\{\lusage_k\}_{k\in[K]}$
in the primal program and $\{\mu, \zeta_k\}_{k\in[K]}$ in the dual program. 
The primal objective function is the expected size of 
fractional matching induced by $\{\lusage_k\}_{k\in[K]}$,
the first primal constraint ensures that the total
expected allocation (over all $K$ stages) 
is at most one for each offline node,
and the second primal constraint
ensures that the expected 
fractional allocation on the offline side
is no more than the size of online nodes for each stage $k\in[K]$.}

Consider the following feasible dual solution 
with objective value $\cratio{K}\cdot K\M$
in the dual program:
$$\mu = |\Driver_K|~~~~,~~~~\zeta_k = \frac{|\Driver_k| - |\Driver_K|}{|\Driver_k|},~\forall k\in[K].$$
Invoking the weak duality between the above primal-dual linear
programs finishes 
the proof.\ifMS \hfill\Halmos \fi

\section{Conclusion and Discussions}
\label{sec:conclusion}
Motivated by the trade-off between batching and inefficiency in two-sided matching platforms, we studied the 
$K$-stage vertex weighted matching and its 
extension to the general model of configuration allocation with free-disposal. We designed optimal $\left(1-(1-1/K)^K\right)$-competitive 
fractional algorithms for these problems.
A key ingredient of our algorithms 
and analysis is 
a sequence of polynomials of decreasing degrees, used as regularizers of maximum weight matching. We also presented the idea of price-level regularization to handle weight heterogeneity of the allocation on the offline side, and showed our our approach can incorporate this form of regularization. Our work also identifies deep connections between convex duality and primal-dual analysis of multi-stage resource allocation algorithms.
There are several  open questions and future research directions stemming from this work. See Appendix~\ref{apx:open} for more discussions. 

% We also presented a deterministic rounding scheme
% with the competitive ratio $1-(1-1/K)^K 
% - O\left(1/\minbudget\right)$
% for the $K$-stage vertex weighted integral b-matching problem and 
% a randomized rounding scheme 
% with the competitive ratio $1-(1-1/K)^K 
% - O(\sqrt{\log(\minbudget)/\minbudget})$
% for the $K$-stage AdWords problem.

% Third, when batches have size one 
% (i.e., the fully online setting),
% identify the $K$-stage optimal algorithm and its 
% competitive ratio. For this last open problem, we conjecture the algorithms in this paper are optimal competitive.

% \section*{Acknowledgement}
% The authors would like to thank Yeganeh Alimohammadi for participating in the early stages of this project, and to Amin Saberi for helpful discussions regarding the presentation of this work. We are also grateful to Rene Caldentey, Ozan Candogan, and Niusha Navidi for helpful comments and feedback on an earlier version of this manuscript. 

\newcommand{\newblock}{}
\setlength{\bibsep}{0.0pt}
\bibliographystyle{plainnat}
\OneAndAHalfSpacedXI
{\footnotesize
\bibliography{refs}}

% \newcommand{\newblock}{}
% \bibliographystyle{apalike}
% % \bibliographystyle{plainnat}
% \bibliography{refs}

\newpage
\renewcommand{\theHchapter}{A\arabic{chapter}}
\renewcommand{\theHsection}{A\arabic{section}}

\begin{APPENDIX}{}

\revcolor{\section{Connection Between Baseline and Extension Models}
\label{apx:connection}
We formally show that
the multi-stage configuration allocation 
generalizes the multi-stage vertex weighted matching.
To see this, we consider a reduction in which every instance in the vertex weighted matching 
can be thought as an
instance in the configuration allocation as follows:
Each advertiser $\driver$ has a unit demand 
($n_\driver = 1$).
The feasible configuration set encodes 
all edges $E$ in the baseline model
($\Config = E$).
Moreover, for every 
edge $(\rider\primed,\driver\primed)\in E$,
the corresponding
configuration $\config$ 
encodes the 
number of reserved impression $\{\alloci\}_{\rider,\driver}$
as 
$\alloci =\indicator{\rider = \rider\primed,
\driver = \driver\primed}$.
The 
per-impression prices are identical for each advertiser $j$ and only depend on the weight $w_j$ of the offline vertex $j$
($\weight_{\threeindex} =w_j \indicator{(i,j)\in E}$).
Finally, note that
since per-impression price are identical for each advertiser
and number of reserved impression $\{\alloci\}$
is encoded as the aforementioned indicator function,
without loss of generality, 
we can restrict attention to online algorithms 
whose allocations never exceed demand $\{n_\driver\}$
of advertisers, and thus
no
preemption or free disposal is needed.}

\section{Missing Proofs}
\label{sec:apx}
 \subsection{Proof of \Cref{prop:poly}}
 \label{apx:poly}

We first prove (i) by backward induction on $k$:
\begin{itemize}
    \item \textsl{Base case $(k = K-1)$:} The maximum of function $(1-y)f_{K}(x+y)=(1-y)\mathbb{I}\{x+y=1\}$ over $y\in[0,1-x]$ happens at $y^*=1-x$, and therefore $f_{K-1}(x)=x$. 
    \item \textsl{Inductive step $(k\in [K-2])$.}
Suppose $f_{k+1}(x)=(1-\frac{1-x}{K-k-1})^{K-k-1}$. Fix $x\in[0,1]$ and let $y^*=\underset{y\in[0,1-x]}{\argmax}~(1-y)f_{k+1}(x+y)$. By writing the first-order condition due to optimality of $y^*$, we have:
\begin{equation}
\label{eq:poly}
  \left(1-\frac{1-x-y^*}{K-k-1}\right)^{K-k-2}-y^*\left(1-\frac{1-x-y^*}{K-k-1}\right)^{K-k-2}
    -\left(1-\frac{1-x-y^*}{K-k-1}\right)^{K-k-1}= 0~,
\end{equation}
which implies that $y^*=\frac{1-x}{K-k}$. Note that $y^*\in[0,1-x]$. To check whether $y^*$ is the maximum point in $[0,1-x]$, we consider the second-order condition with respect to $y^*$. After algebraic simplifications, we have (details are omitted for brevity):
$$ \frac{\partial^2}{\partial^2 y}\left[(1-y)f_{k+1}(x+y)\right]_{y=y^*}=(x-2)(1-\frac{1-x-y^*}{K-k-1})^{K-k-3}\leq 0
$$
Plugging back $y^*=\frac{1-x}{K-k}$ into the objective, we have:
$$
f_k(x)=(1-\frac{1-x}{K-k})f_{k+1}(x+\frac{1-x}{K-k})=(1-\frac{1-x}{K-k})(1-\frac{1-x}{K-k})^{K-k-1}=(1-\frac{1-x}{K-k})^{K-k}~,
$$
as desired. 
\end{itemize}
Given (i), proof of (ii) is immediate as the polynomial $f_k(x)=(1-\frac{1-x}{K-k})^{K-k}$ is monotone increasing for $k\in[K-1]$. Also, (iii) simply holds as $(1-x)\leq e^{-x}$ when $x\in[0,1]$, and that $\lim_{N\to+\infty}(1-\frac{x}{N})^N=e^{-x}$ for $x\in[0,1]$. Finally, to prove (iv), consider defining $f_0(x)=\underset{y\in[0,1-x]}{\max}(1-y)f_1(x+y)$. Following the backward induction in the proof of (i), $f_0(x)\triangleq(1-(1-x)/K)^K$. Therefore, 
$$\underset{y\in[0,1]}{\min}1-(1-y)f_1(y)=1-\underset{y\in[0,1]}{\max}(1-y)f_1(y)=1-f_0(0)=\cratio{K}.
$$

\revcolor{\subsection{Proof of \Cref{lemma:structure}}
 \label{apx:kkt}
 
 We first define the Lagrangian dual function $\Lag_k(\cdot)$ and then re-state the KKT optimality conditions of $\xbf^*_k$ in our specific convex program.
\begin{definition}
\label{def:lagrange}
The \emph{Lagrangian dual} $\Lag_k(.)$ for the concave program~\ref{eq:concave-k-bmatching} is defined as:
\begin{align*}
    \Lag_k(\xbf,\lambdabf,\thetabf,\gammabf)\triangleq& \displaystyle\sum_{j\in \Driver}w_j\left(
        \sum_{i:(i,j)\in E_k}x_{ij} - 
        F_k\left(y_j+
        \sum_{i: (i,j)\in E_k}x_{ij}\right)
        \right) \\
    &-\sum_{i\in \Rider_k}\lambda_i\left(\sum_{j:(i,j)\in E_k}x_{ij}-1\right)
    -\sum_{j\in \Driver}\theta_j\left(\sum_{i:(i,j)\in E_k}x_{ij}+y_j-1\right)
    \\
    &+\sum_{(i,j)\in E_k}\gamma_{ij}x_{ij}~~,
\end{align*}
where $\{\lambda_i\}$, $\{\theta_j\}$, and $\{\gamma_{ij}\}$ are the \emph{Lagrange multipliers}, also known as the \emph{dual variables}, 
corresponding to different constraints 
in program~\ref{eq:concave-k-bmatching}.
\end{definition}
\begin{proposition}[KKT Conditions for Multi-stage Polynomial Regularized Matching]
\label{prop:kkt}
Suppose $\xbf_k^*$ is the optimal solution of the concave program \ref{eq:concave-k-bmatching}. 
Consider the Lagrangian dual $\Lag_k$, as in \Cref{def:lagrange}. Then:
\begin{equation*}
    \underset{\xbf}{\max}~\underset{\lambdabf,\thetabf,\gammabf}{\min}~\Lag_k(\xbf,\lambdabf,\thetabf,\gammabf)= \underset{\lambdabf,\thetabf,\gammabf}{\min}~\underset{\xbf}{\max}~\Lag_k(\xbf,\lambdabf,\thetabf,\gammabf)=\underset{\lambdabf,\thetabf,\gammabf}{\min}~\Lag_k(\xbf_k^*,\lambdabf,\thetabf,\gammabf)~~.
\end{equation*}
Moreover, if $\lambdabf^*$, $\thetabf^*$, and $\gammabf^*$ (together with $\xbf_k^*)$ are the solutions of the above minmax/maxmin, then:
\begin{itemize}
    \item (Stationarity)~~$\forall (i,j)\in E_k:~~\frac{\partial \Lag_k}
    {\partial x_{ij}}(\xbf_k^*,\lambdabf^*,\thetabf^*,\gammabf^*)=0,$
    \item (Dual feasibility)~~$\forall i\in \Rider_k,j\in \Driver,(i,j)\in E_k:~\lambda^*_i\geq 0,~\theta^*_j\geq 0$, and $\gamma^*_{ij}\geq 0$~,
    \item (Complementary slackness)~~$\forall i\in \Rider_k,j\in \Driver,(i,j)\in E_k:$ 
    \begin{itemize}
        \item  $\gamma^*_{ij}=0$ or $\left(\gamma^*_{ij}>0~\textrm{and}~ x^*_{k,ij}=0\right)~,$
        \item $\lambda^*_{i}=0$ or $\left(\lambda^*_{i}>0~\textrm{and}~ \sum_{j: (i,j)\in E_k}x^*_{k,ij}=1\right)~,$
        \item  $\theta^*_{j}=0$ or $\left(\theta^*_{j}>0~\textrm{and}~ y_j+\sum_{i: (i,j)\in E_k}x^*_{k,ij} =1\right)~.$
    \end{itemize}
\end{itemize}
\end{proposition}
With the above ingredients, we prove the structural lemma.

\ifMS \proof{\textsl{Proof of \Cref{lemma:structure}.}}
\else \begin{proof}[Proof of \Cref{lemma:structure}]\fi
We first show the uniformity property. 
For $j\in \Driver_k^{(0)}$,
note that $$w_j\left(1-
    f_k\left(y_j + \sum_{i:(i,j)\in E_k}x^*_{k,ij}\right)
    \right)
    =w_j(1 - f_k(1))=0=c^{(0)}.$$ 
    For vertices in $\Driver_k^{(l)}$ 
    with $l\in [L_k]$,
    since
    they are in the same connected component of $G_k'[\Rider_k\setminus\Rider_k^{(0)},\Driver\setminus \Driver_k^{(0)}]$,
    it is sufficient to restrict attention to $j, j'\in\Driver_k^{(l)}$
    such that
    there exists a vertex $i\in\Rider_k^{(l)}$ such that $x^*_{k,ij}>0$ and $x^*_{k,ij'}>0$ (and therefore, $\gamma^*_{ij}=\gamma^*_{ij'}=0$, because of the complementary slackness in \Cref{prop:kkt}). Also, $\theta^*_j=\theta^*_{j'}=0$, again because of complementary slackness and the fact that $y_j+\sum_{i:(i,j)\in E_k}x^*_{k,ij}<1$ and $y_{j'}+\sum_{i: (i,j')\in E_k}x^*_{k,ij'}<1$.
Now, applying the stationarity condition in \Cref{prop:kkt}, we have:
$$
w_j\left(1-
    f_k\left(y_j + \sum_{i:(i,j)\in E_k}x^*_{k,ij}\right)
    \right)
-\lambda^*_i-\theta^*_j+\gamma^*_{ij}
=0=
w_{j'}\left(1-
    f_k\left(y_{j'} + \sum_{i: (i,j')\in E_k}x^*_{k,ij'}\right)
    \right)
-\lambda^*_i-\theta^*_{j'}+\gamma^*_{ij'}~,
$$
and therefore 
$w_j\left(1-
    f_k\left(y_j + \sum_{i:(i,j)\in E_k}x^*_{k,ij}\right)
    \right)
    =
    w_{j'}\left(1-
    f_k\left(y_{j'} + \sum_{i: (i,j')\in E_k}x^*_{k,ij'}\right)
    \right)$, 
    finishing the proof of the uniformity property. 
    
    To show saturation property, 
    note that $\lambda^*_i=
    w_j\left(1-
    f_k\left(y_j + \sum_{i:(i,j)\in E_k}x^*_{k,ij}\right)
    \right)>0$ for any $i \in \Rider_k^{(l)}, j \in \Driver_k^{(l)}$ with $l \in[L_k]$ such that $x_{ij}^* > 0$, simply because $f_k(1)=1$ and $f_k(x)$ is (strictly) monotone increasing.
    Again, by using complementary slackness, we have $\sum_{j:(i,j)\in E_k}x^*_{k,ij}=1$, which proves the saturation property.

Now suppose there exists an edge $(i,j')\in E_k$, where $i\in \Rider_k^{(l)}$ and $j'\in \Driver_k^{(l')}$, for $l,l'\in[0:L_k], l\neq l'$. 
First note that there should 
exist a vertex $j\in\Driver_k^{(l)}$ so that $x^*_{k,ij}>0$; this holds because $i$ is either fully matched by $\xbf_k^*$ if $l\in[L_k]$ 
-- thanks to the saturation property --  
or $l=0$ and $i$ is connected to a vertex  $j\in\Driver_k^{(0)}$ with $x^*_{k,ij'}>0$ by definition. 
Because of complementary slackness $\gamma^*_{ij}=0$. 
Now, by writing the stationarity condition of \Cref{prop:kkt} for the edge $(i,j')$, and noting that 
$w_{j'}\left(1-
    f_k\left(y_{j'} + \sum_{i: (i,j')\in E_k}x^*_{k,ij'}\right)
    \right)=c^{(l')}$ 
because of the uniformity property, we have:
$$
c^{(l')}-\lambda^*_i-\theta^*_{j'}+\gamma^*_{ij'}=0~,
$$
By writing the same condition for the edge $(i,j)$ we have: 
$$
c^{(l)} - \lambda^*_i - \theta^*_{j}=0~,
$$
and therefore 
$$
c^{(l)}\geq c^{(l)}-\theta^*_{j}
=c^{(l')}-\theta^*_{j'}+\gamma^*_{ij'}
\geq 
c^{(l')}-\theta^*_{j'}
$$
If $l'=0$, then clearly $c^{(l)}\geq 0= c^{(l')}$. 
If $l'\in [L_k]$, then $\sum_{i': (i',j')\in E_k}x^*_{i'j'}+y_{j'}<1$ 
and $\theta^*_{j'}=0$ because of complementary slackness. Therefore, $c^{(l)}\geq c^{(l')}$, which finishes the proof of the monotonicity property. \ifMS \hfill\Halmos \fi
\ifMS
\endproof
\else
\end{proof}
\fi

\subsection{Proof of \Cref{lem:KKT restate}}
\label{apx:kkt-CA}
Let ($\lambdabf\starred, \thetabf\starred,
\mubf\starred, \pibf\starred,
\psibf\starred,\phibf\starred,\iotabf\starred$)
be the optimal solution in 
the Lagrangian dual program \ref{eq:convex dual}. The property
\ref{eq:cs} 
is satisfied by considering 
the complementary slackness
KKT conditions.
\ref{eq:stationary z}
is due to 
the
stationarity/first-order KKT condition
with respect to $z_{\rider,\config}$.
To show \ref{eq:stationary x}, note that 
the
stationarity/first-order KKT condition
with respect to $x_{\threeindex}$ implies that 
\begin{align*}
    \weight_{\threeindex}
    -
    \sum_{\wi\in[\totalweight]:
    \weightHat(\wi) \leq \weight_{\threeindex}}
    \left(\weightHat_\wi - \weightHat_{\wi-1}\right)
    \cdot 
    f_{k}(\cumcomsumptioni(\xbf\starred, \ybf\starred))
    -
    \thetai\starred 
    - \mui\starred + \psii\starred = 0
\end{align*}
Thus, it is sufficient to show that $\thetai\starred = 0$
for every advertiser $\driver$.
By complementary slackness in KKT conditions,
we know that $\thetai\starred = 0$ if 
$
\cumcomsumption_{\driver,0}(\xbf\starred, \ybf\starred)
~(\equiv
\sum_{\rider\in\Rider_k}
\sum_{\config\in\Config}
         x_{\threeindex}\starred
         +
         \sum_{\wi\in[\totalweight]}
         y_{\twoindex}\starred)~
         < 1$.

Now we consider any advertiser $\driver$ where 
$
\cumcomsumption_{\driver,0}(\xbf\starred, \ybf\starred)
% \sum_{\rider,\config:(\threeindex)\in E_k}
%          x_{\threeindex}\starred
%          +
%          \sum_{\wi\in[\totalweight]}
        %  y_{\twoindex}\starred 
         = 1$.
Let $\taup$ be the smallest index $\tau$ such that
(i) $y_{\driver,\taup}\starred > 0$ or 
(ii) there exists $x_{\threeindex}\starred > 0$ 
with $\weight_{\threeindex} = \weightHat_{\taup}$.
Note that by definition, for any $\wi \leq \taup$,
\begin{align}
\label{eq:full capacity F}
    f_{k}(\cumcomsumptioni(\xbf\starred,\ybf\starred)) &= 
% F\left(
%         \displaystyle\sum\limits_{
%         \substack{\rider,\config:
%         (\threeindex)\in E_k,\\
%         ~~\weight_{\threeindex}\geq 
%         \weightHat_\wi}
%         }
%         x_{\threeindex}
%         +
%         \displaystyle\sum\limits_{\wi' \geq \wi}
%         y_{\driver,\tau'}
%         \right)  
        % =
        f_k(1) = 1
\end{align}
We use similar arguments for both case (i) and case (ii).
\begin{itemize}
    \item Suppose $y_{\driver,\taup}\starred > 0$.   
We have that 
\begin{align*}
    0 
    &\overset{(a)}{=} 
    \weightHat_{\taup} - 
    \sum_{\wi\leq \taup}
    (\weightHat_{\wi} - \weightHat_{\wi - 1})
    \cdot 
    % F_{\twoindex}(\xbf\starred,\ybf\starred)
    f_{k}(\cumcomsumptioni(\xbf\starred,\ybf\starred))
    - \thetai\starred - \pi_{\driver,\taup}\starred 
    + \iota_{\driver,\taup}\starred  
    \\
    & \overset{(b)}{=} 
    \weightHat_{\taup} - 
    \sum_{\wi\leq \taup}
    (\weightHat_{\wi} - \weightHat_{\wi - 1})
    - \thetai\starred - \pi_{\driver,\taup}\starred 
    \overset{(c)}{\leq} 
    -\thetai\starred \overset{(d)}{\leq} 0
\end{align*}
where equation (a) is due to 
the stationarity/first-order KKT condition with respect to 
$y_{\driver,\taup}$;
equation (b) is due to the complementary slackness 
KKT condition (i.e., $\iota_{\driver,\taup}\starred\cdot y_{\driver,\taup}\starred = 0$) 
and equation~\eqref{eq:full capacity F};
and inequality (c) and (d) is due to the non-negativity of 
$\thetai^*$ and $\pi_{\driver,\taup}\starred$.

\item Similarly, suppose 
there exists $x_{\threeindex}\starred > 0$ 
with $\weight_{\threeindex} = \weightHat_{\taup}$. 
Applying the stationarity/first-order KKT condition with respect to 
$x_{\threeindex}\starred$ and invoking 
equation \eqref{eq:full capacity F}, we have
\begin{align*}
    0 
    &\overset{}{=} 
     \weight_{\threeindex}
    -
    \sum_{\wi\in[\totalweight]:
    \weightHat_\wi \leq \weight_{\threeindex}}
    \left(\weightHat_\wi - \weightHat_{\wi-1}\right)
    \cdot 
    % F_{\twoindex}(\xbf\starred, \ybf\starred)
    f_{k}(\cumcomsumptioni(\xbf\starred,\ybf\starred))
    -
    \thetai\starred 
    - \mui\starred + \psii\starred
    \\
    & \overset{}{=} 
    \weightHat_{\taup} - 
    \sum_{\wi\leq \taup}
    (\weightHat_{\wi} - \weightHat_{\wi - 1})
    - \thetai\starred - \mui\starred 
    \overset{}{\leq} 
    -\thetai\starred \overset{}{\leq} 0
\end{align*}
Thus, we conclude that $\thetai\starred = 0$ for all advertiser $\driver$,
and finish the proof of property~\ref{eq:stationary x}.
\halmos
\end{itemize}

\subsection{Proof of \Cref{lem:preempt monotone}}
\label{apx:kkt-CA-smallest}
Let ($\lambdabf\starred, \thetabf\starred,
\mubf\starred, \pibf\starred,
\psibf\starred,\phibf\starred,\iotabf\starred$)
be the optimal solution in 
the Lagrangian dual program \ref{eq:convex dual}.
Suppose $\comsumptioni - y_{\twoindex}^* > 0$
for $\driver\in \Driver$ and $\wi\in[\totalweight]$.
The stationarity/first-order KKT condition with respect to
$y_{\twoindex}$ implies that
\begin{align*}
    \weightHat_{\wi} - 
    \sum_{\wi'\leq \wi}
    (\weightHat_{\wi'} - \weightHat_{\wi' - 1})
    \cdot 
    % F_{\twoindex}(\xbf\starred,\ybf\starred)
    f_{k}(\cumcomsumption_{\driver,\wi'}(\xbf\starred,\ybf\starred))
    - \thetai\starred - \pi_{\driver,\wi}\starred 
    + \iota_{\driver,\wi}\starred  
    = 0
\end{align*}
Note that $\iota_{\twoindex} \geq 0$.
The complementary slackness in
KKT condition implies that 
$\pi_{\twoindex}\starred = 0$
since $\comsumptioni - y_{\twoindex}^* > 0$.
Additionally, as we already 
shown in \Cref{lem:KKT restate},
$\thetai\starred = 0$. Hence,
we have
\begin{align*}
    \weightHat_{\wi} - 
    \sum_{\wi'\leq \wi}
    (\weightHat_{\wi'} - \weightHat_{\wi' - 1})
    \cdot 
    % F_{\twoindex}(\xbf\starred,\ybf\starred)
    f_{k}(\cumcomsumption_{\driver,\wi'}(\xbf\starred,\ybf\starred)) \leq 0
\end{align*}
Since $f_k(a)\leq 1$ for all $a\in[0, 1]$
with equality hold when $a = 1$, we conclude that 
$f_{k}(\cumcomsumption_{\driver,\wi}(\xbf\starred,\ybf\starred)) = 1$ 
and $\cumcomsumption_{\driver,\wi}(\xbf\starred,\ybf\starred) = 1$.
\hfill\halmos

\subsection{Proof of \Cref{coro:cumulative comsumption monoton}}
\label{apx:kkt-CA-cor1}
To see the proof of the first item (\textbf{Distribution Monotonicity)}, by definition, $\cumcomsumptioni(\xbf\starred,\ybf\starred) \geq 
\sum_{\wi'\geq \wi} y_{\driver,\wi'}^*$. Thus
it is sufficient to show the statement when 
$\sum_{\wi'\geq \wi}\left(\comsumption_{\driver,\wi'}\kminused - y_{\driver,\wi'}^*\right) > 0$.
In this case, 
let index $\taup$ be the index such that 
$\taup \geq \wi$ and $\comsumption_{\driver,\wi\primed}\kminused - y_{\driver,\wi\primed}^*>0$.
By \Cref{lem:preempt monotone},
we have $ 
\cumcomsumption_{\driver,\taup}(\xbf\starred,\ybf\starred) = 1$.
Combining with the fact that 
$
\cumcomsumption_{\driver,\wi}(\xbf\starred,\ybf\starred)\geq 
\cumcomsumption_{\driver,\taup}(\xbf\starred,\ybf\starred)
$
and $\sum_{\wi'\geq \wi}\comsumption_{\driver,\wi'}\kminused \leq 1$ finishes the proof of the first item.

For the proof of the second item (\textbf{Only Disposing Smallest}), the argument is similar to \Cref{coro:cumulative comsumption monoton}.
Note that it is sufficient to show the statement 
when 
$\sum_{\wi'\geq \wi}\left(\comsumption_{\driver,\wi'}\kminused - y_{\driver,\wi'}^*\right) > 0$.
In this case, let index $\taup$ be the index such that 
$\taup \geq \wi$ and $\comsumption_{\driver,\wi\primed}\kminused - y_{\driver,\wi\primed}^*>0$.
By \Cref{lem:preempt monotone},
we have $
f_k(\cumcomsumption_{\driver,\taup}(\xbf\starred,\ybf\starred)) = 1$.
Combining with the fact that 
$
f_k(\cumcomsumption_{\driver,\taup}(\xbf\starred,\ybf\starred))
\leq
f_k(\cumcomsumption_{\driver,\wi}(\xbf\starred,\ybf\starred))\leq 1$,
we conclude that 
$f_k(\cumcomsumption_{\driver,\wi}(\xbf\starred,\ybf\starred)) = 1$ which finishes the proof.

\hfill\halmos

\subsection{Proof of \Cref{lem:dual restate theta K}}
\label{apx:last-stage-lemma}
Let $(\xbf\starred,\ybf\starred,\zbf\starred)$
be the optimal solution of the linear program
\ref{eq:convex primal} in the last stage.

Suppose $\weightHat_\wi < \thetai\starred$.
The complementary slackness 
(i.e., $y_{\twoindex}\cdot (\thetai\starred+\pi_{\twoindex\starred} - \weightHat_{\wi}) = 0$)
implies that $y_{\twoindex} = 0$.
Applying the complementary slackness (i.e., 
$\pi_{\twoindex}\starred\cdot(\comsumption_{\twoindex}\KMINUSED - y_{\twoindex}\starred) = 0$)
again, we conclude that $\comsumption_{\twoindex}\KMINUSED\cdot \pi_{\twoindex}\starred = 0$.

Suppose $\weightHat_\wi > \thetai\starred$.
We know that $\pi_{\twoindex}\starred 
\geq \weightHat_\wi - \thetai\starred > 0$.
Then the complementary slackness (i.e., 
$\pi_{\twoindex}\starred\cdot(\comsumption_{\twoindex}\KMINUSED - y_{\twoindex}\starred) = 0$)
suggests that $\comsumption_{\twoindex}\KMINUSED =
y_{\twoindex}\starred$.
Applying the complementary slackness 
(i.e., $y_{\twoindex}\cdot (\thetai\starred+\pi_{\twoindex\starred} - \weightHat_{\wi}) = 0$),
we conclude that 
$ \comsumption_{\twoindex}\KMINUSED
     \cdot 
     (
     \weightHat_{\wi} - \pi_{\twoindex}\starred)
    = 
     \comsumption_{\twoindex}\KMINUSED\cdot \thetai\starred$.
     
     Finally, 
     suppose $\weightHat_\wi = \thetai\starred$.
     As a sanity check, note that 
     equality $\comsumption_{\twoindex}\KMINUSED\cdot \pi_{\twoindex}\starred = 0$
     is equivalent to
     equality 
     $ \comsumption_{\twoindex}\KMINUSED
     \cdot 
     (
     \weightHat_{\wi} - \pi_{\twoindex}\starred)
    = 
     \comsumption_{\twoindex}\KMINUSED\cdot \thetai\starred$.
     If $y_{\twoindex}\starred = 0$,
     following the same argument as 
     the case $\weightHat_\wi < \thetai\starred$
     finishes the proof.
     Otherwise (i.e., $y_{\twoindex}\starred > 0$),
     following the same argument as 
     the case $\weightHat_\wi > \thetai\starred$
     finishes the proof.
\hfill\halmos

\subsection{Proof of \Cref{thm:main result configuration allocation}}
\label{apx:proof-primal-dual-CA}
Consider the suggested dual assignment in \Cref{sec:main proof}. We finish the proof in two steps.

\smallskip
\noindent[\textit{Step \rom{1}}]~\emph{Comparing objective values in primal and dual.}
we first show that 
the objective value of the above dual assignment
is equal to the objective value of the primal assignment of \Cref{alg:opt}.
To show this, we consider 
each stage $k$ separately.
For the last stage $K$, 
this holds by construction.
For each stage $k \in [K-1]$, note that
by \Cref{lem:preempt monotone}
the primal objective value is 
\begin{align*}
    \Delta(\textrm{Primal})\ked = 
    \sum\limits_{\rider\in\Rider_k}
    \sum_{\config\in\Config}
    \sum_{\driver\in\Driver}
    \weight_{\threeindex} \cdot x_{\threeindex}\ked
    -
    \sum_{\driver\in \Driver}
    \sum_{\wi\in[\totalweight]}
    \weightHat_{\wi}\cdot
    \left(\comsumptioni - y_{\twoindex}\ked
    \right)
\end{align*}
To consider the dual objective value, 
we start with the observation that 
\begin{align*}
    \alpha_\rider &= 
   \lambda_\rider\ked\\
   &\overset{(a)}{=} 
    \sum_{\config \in \Config}
   z_{\rider,\config}\ked\cdot 
   \lambda_\rider\ked
   \\
   &\overset{(b)}{=}
    \sum_{\config \in \Config}
   z_{\rider,\config}\ked\cdot 
   \left(
    \sum_{\driver\in \Driver}\alloci \cdot \mui\ked
    + \phii\ked 
    \right)
    \\
    &= 
    \sum_{\config \in \Config}
    \sum_{\driver\in \Driver}
    \alloci\cdot 
    z_{\rider,\config}\ked\cdot 
    \mui\ked
    +
    \sum_{\config \in \Config}
    z_{\rider,\config}\ked\cdot  \phii\ked
    \\
    &\overset{(c)}{=}
    \sum_{\config \in \Config}
    \sum_{\driver\in \Driver}
    x_{\threeindex}\ked\cdot
    \mui\ked   
    \\
    &\overset{(d)}{=}
    \sum_{\config \in \Config}
    \sum_{\driver\in \Driver}
    x_{\threeindex}\ked \cdot 
    \left(
    \weight_{\threeindex}
    -
    \sum_{\wi\in[\totalweight]:
    \weightHat(\wi) \leq \weight_{\threeindex}}
    \left(\weightHat_\wi - \weightHat_{\wi-1}\right)
    \cdot f_k\left(\cumcomsumptioni\ked\right)
    % -
    % \thetai\ked  
    + \psii\ked 
    \right) 
    \\
    &\overset{(e)}{=}
    \sum_{\config \in \Config}
    \sum_{\driver\in \Driver}
    x_{\threeindex}\ked \cdot 
    \left(
    \weight_{\threeindex}
    -
    \sum_{\wi\in[\totalweight]:
    \weightHat(\wi) \leq \weight_{\threeindex}}
    \left(\weightHat_\wi - \weightHat_{\wi-1}\right)
    \cdot f_k\left(\cumcomsumptioni\ked\right)
    % -
    % \thetai\ked  
    \right) 
\end{align*}
Here we heavily use \Cref{lem:KKT restate}:
\ref{eq:cs} for equality~(a)~(c)~(e),
\ref{eq:stationary z} for equality~(b),
and 
\ref{eq:stationary x} for equality~(d).
Thus, we can compute the dual objective as follows,
\begin{align*}
    \Delta(\textrm{Dual})\ked = 
    \sum_{\rider\in \Rider_k}
    \alpha_{\rider} 
    +
    \sum_{\driver\in \Driver}
    \Delta\beta_{\driver}\ked
    =
    \sum\limits_{\rider\in\Rider_k}
    \sum_{\config\in\Config}
    \sum_{\driver\in\Driver}
    \weight_{\threeindex} \cdot x_{\threeindex}\ked
    -
    \sum_{\driver\in \Driver}
    \sum_{\wi\in[\totalweight]}
    \weightHat_{\wi}\cdot
    \left(\comsumptioni - y_{\twoindex}\ked
    \right)
    \equiv
     \Delta(\textrm{Primal})\ked 
\end{align*}

\vspace{1mm}

\noindent[\textit{Step \rom{2}}]~\emph{Checking approximate feasibility of dual.} 
First, we show that all dual assignment are non-negative. 
Note that the non-negativity of $\alphabf$ and $\gammabf$
is guaranteed by construction. 
For $\betabf$, we show that $\Delta\beta_{\driver}\ked\geq 0$
for all $\driver\in \Driver$ and $k\in [K]$.
To see this, note that for any stage $k\in [K - 1]$
\begin{align}
    % \label{eq:beta reformulate}
% \begin{split}
\nonumber
    \Delta\beta_{\driver}\ked
    &=
    \sum_{\rider\in\Rider_k}
    \sum_{\config\in\Config}
    x_{\rider,\config,\driver}\ked
    \left(
    % \theta_{\driver}\ked + 
    \sum_{\wi\in[\totalweight]:
    \weightHat(\wi) \leq \weight_{\rider,\config,\driver}}
    \left(\weightHat_\wi - \weightHat_{\wi-1}\right)
    \cdot 
    f_k\left(\cumcomsumptioni\ked\right)
    \right)
    -
    \sum_{\wi\in[\totalweight]}
    \weightHat_{\wi}\cdot
    \left(\comsumption_{\twoindex}\kminused - y_{\twoindex}\ked
    \right)\\
\nonumber
    &
    \overset{(a)}{=}
    \sum_{\wi\in [\totalweight]}
    \left(\weightHat_\wi - \weightHat_{\wi-1}\right)
    \cdot
    \left(
    \displaystyle\sum\limits_{
        \substack{\rider\in\Rider_k,\config\in\Config:
    \\
        ~~\weight_{\rider,\config,\driver}\geq 
        \weightHat_\wi}
        }
        x_{\rider,\config,\driver}\ked
    \right)
    \cdot 
    % \left(
   f_k\left(\cumcomsumptioni\ked\right)
    % \right)
    -
    \sum_{\wi\in[\totalweight]}
    \left(\weightHat_{\wi} - \weightHat_{\wi-1}\right)\cdot
    % \left(
    \sum_{\wi'\geq \wi}\left(
    \comsumption_{\driver,\wi'}\kminused - y_{\driver,\wi'}\ked
    % \right)
    \right) 
    \\
\nonumber
    &
    \overset{(b)}{=}
    \sum_{\wi\in [\totalweight]}
    \left(\weightHat_\wi - \weightHat_{\wi-1}\right)
    \cdot
    \left(
    \displaystyle\sum\limits_{
        \substack{\rider\in\Rider_k,\config\in\Config:
        % (\rider,\config,\driver)\in E_s,
        \\
        ~~\weight_{\rider,\config,\driver}\geq 
        \weightHat_\wi}
        }
        x_{\rider,\config,\driver}\ked
    \right)
    \cdot 
    % \left(
   f_k\left(\cumcomsumptioni\ked\right)
    % \right)
    -
    \sum_{\wi\in[\totalweight]}
    \left(\weightHat_{\wi} - \weightHat_{\wi-1}\right)\cdot
    % \left(
    \sum_{\wi'\geq \wi}\left(
    \comsumption_{\driver,\wi'}\kminused - y_{\driver,\wi'}\ked
    % \right)
    \right) 
    \cdot 
    f_k\left(\cumcomsumptioni\ked\right)
    \\
    \label{eq:beta reformulate}
    &=
    \sum_{\wi\in [\totalweight]}
    \left(\weightHat_\wi - \weightHat_{\wi-1}\right)
    \cdot
    \left(
    \cumcomsumptioni\ked - \cumcomsumptioni\kminused
    \right)
    \cdot 
    % \left(
   f_k\left(\cumcomsumptioni\ked\right)
    \overset{(c)}{\geq} 0
    % \right)
\end{align}
where (a) is by changing the summation order,
(b) and (c) are due to
\Cref{coro:cumulative comsumption monoton}.
For the last stage $K$,
\begin{align*}
  \begin{split}
    \Delta\beta_{\driver}\KED
    &=
     \thetai\KED 
    -
    \displaystyle\sum_{\driver\in\Driver,\wi\in[\totalweight]}
    \comsumption_{\twoindex}\KMINUSED
    \cdot (\weightHat_{\wi} - \pi_{\twoindex}\KED)
    \overset{(a)}{\geq} 
     \thetai\KED 
    -
    \displaystyle\sum_{\driver\in\Driver,\wi\in[\totalweight]}
    \comsumption_{\twoindex}\KMINUSED
    \cdot \thetai\KED \geq 0
    \end{split}  
\end{align*}
where (a) uses the fact that $\thetai\KED + \pi_{\twoindex}\KED \geq \weightHat_{\wi}$ by definition.

Second, we show that $\alpha_{\rider} \geq 
    \displaystyle\sum_{\driver\in \Driver}
    \alloci \cdot \gamma_{\threeindex}$
    for all $\rider \in \Rider$.
    This inequality is satisfied 
    by the construction of $\alphabf$, $\gammabf$
    and \ref{eq:stationary z} in \Cref{lem:KKT restate}.
    
Finally, to finish the proof, 
it is sufficient for us to show
that for any stage $k\in [K]$
and $(\threeindex)\in E_k$,
\begin{align}
\label{eq:dual feasibility critical}
    \gamma_{\threeindex}
    +
    \sum_{s=1}^k
    \Delta\beta_\driver\sed \geq \cratio{K}\cdot \weight_{\threeindex}
\end{align}
We show inequality \eqref{eq:dual feasibility critical}
holds for any stage $k\in [K - 1]$ and
the last stage $K$ separately.

\vspace{2mm}
Fix an arbitrary stage $k\in[K - 1]$
and $(\threeindex)\in E_k$.
Let $\taup\in [\totalweight]$
be the index that $\weightHat_{\taup}=\weight_{\threeindex}$.
Note that
\begin{align*}
    \gamma_{\threeindex} &= \mui\ked
    \\
    &=
    %  \left(
    \weight_{\threeindex}
    -
    \sum_{\wi\in[\totalweight]:
    \weightHat(\wi) \leq \weight_{\threeindex}}
    \left(\weightHat_\wi - \weightHat_{\wi-1}\right)
    \cdot 
    f_k\left(\cumcomsumptioni\ked\right)
    % -
    % \thetai\ked  
    + \psii\ked  
    \\
    &\geq 
    \weight_{\threeindex}
    -
    \sum_{\wi\in[\totalweight]:
    \weightHat(\wi) \leq \weight_{\threeindex}}
    \left(\weightHat_\wi - \weightHat_{\wi-1}\right)
    \cdot 
    f_k\left(\cumcomsumptioni\ked\right)
    % -
    % \thetai\ked 
    \\
    &=
    \sum_{\wi\leq\taup}
    % \left(
    (\weightHat_{\wi} - \weightHat_{\wi-1})
    \cdot \left(1 - 
    f_k\left(\cumcomsumptioni\ked\right)
    \right)
    % \right)
    % -\thetai\ked
    % \right) 
\end{align*}
Combining with equality \eqref{eq:beta reformulate},
we have 
\begin{align*}
    \gamma_{\threeindex}
    +
    \sum_{s=1}^k
    \Delta\beta_\driver\sed
    \geq 
    \sum_{\wi\leq \taup} 
    (\weightHat_{\wi} - \weightHat_{\wi-1})
    \cdot 
    \left(
    1 - 
    f_k\left(\cumcomsumptioni\ked\right)
    +
    \sum_{s=1}^k
    \left(
    \cumcomsumptioni\ked - \cumcomsumptioni\kminused
    \right)
    \cdot 
    % \left(
   f_k\left(\cumcomsumptioni\ked\right)
    \right)
\end{align*}
Thus, to show
the approximated dual feasibility in
\eqref{eq:dual feasibility critical},
it is sufficient to show
\begin{align*}
    1 - 
    f_k\left(\cumcomsumptioni\ked\right)
    +
    \sum_{s = 1}^k
    \left(
    \cumcomsumptioni\sed - \cumcomsumptioni\sminused
    \right)
    \cdot 
   f_s\left(\cumcomsumptioni\sed\right)
   \geq  \cratio{K}~
   \qquad
   \forall \wi\leq \taup
\end{align*}
which is satisfied as the property of 
$\{f_k\}_{k\in [K]}$ 
by \Cref{lem:polynomial inequality 2} as follows.
% \begin{lemma}[\citealp{FN-21}]
% \label{lem:polynomial inequality}
% Fix any $K\in\N$ and $k\in [K - 1]$.
% For any $0 = a^{(0)} \leq a^{(1)} \leq \dots \leq a^{(k)} \leq 1$,
% % \begin{align*}
% $    1 - f_k\left(a^{(k)}\right) + \sum_{s = 1}^k \left(
%     a\sed - a\sminused
%     \right)
%     \cdot 
%     f_s\left(a\sed\right) 
%     \geq 
%     \cratio{K}.$
% % \end{align*}
% \end{lemma}

\vspace{2mm}
For the last stage $K$, 
fix any $(\threeindex)\in E_K$.
Let $\taup\in [T]$ be the index  
such that $\weightHat_{\taup} = \thetai\KED$.\footnote{Note that 
the optimal solution of the convex program \ref{eq:convex primal} 
as well as the outcome of 
\Cref{alg:opt} 
does not change if we enlarge the revenue level set $\{\weightHat_\wi\}$. 
Thus, it is without loss of generality to assume there exists index
$\taup$ such that $\weightHat_{\taup} = \thetai\KED$.}
% Note that for any $\wi < \taup$,
% we have $y_{\twoindex}\KED = 0$ 
% and thus
% $\pi_{\twoindex}\KED = 0$ 
% if $\comsumption_{\twoindex}\KMINUSED > 0 = y_{\twoindex}\KED$
% because of
% the complementary slackness in 
% program \ref{eq:convex dual K}
% and the definition of index $\taup$.
% Similarly, for any $\wi > \taup$, we have 
% $y_{\twoindex}\KED = \comsumption_{\twoindex}\KMINUSED$
% and thus 
% $\thetai\KED + \pi_{\twoindex}\KED = \weightHat_{\wi}$
% if $y_{\twoindex}\KED = \comsumption_{\twoindex}\KMINUSED > 0$;
% and for $\wi = \taup$, we have $\pi_{\twoindex}\KED = 0$ regardless.
Applying \Cref{lem:dual restate theta K}, we have
\begin{align}
\label{eq:beta K reformulate}
\begin{split}
    \Delta\beta_{\driver}\KED &=
    \thetai\KED 
    - 
    \displaystyle\sum_{\driver\in\Driver,\wi\in[\totalweight]}
    \comsumption_{\twoindex}\KMINUSED
    \cdot (\weightHat_{\wi} - \pi_{\twoindex}\KED)
    \\
    &= 
    \thetai\KED 
    -
    \sum_{\wi \geq \taup}
    \comsumption_{\twoindex}\KMINUSED \cdot \thetai\KED
    - 
    \sum_{\wi< \taup} 
    \comsumption_{\twoindex}\KMINUSED \cdot \weightHat_{\wi}
    \end{split}
\end{align}

We consider two subcases: 
$\weight_{\threeindex} < \thetai\KED$
and
$\weight_{\threeindex} \geq \thetai\KED$.

Suppose $\weight_{\threeindex} < \thetai\KED$. 
Let $\taupp$ be the index such that $\weightHat_\taupp = \weight_{\threeindex}$.
By definition, $\taupp < \taup$.
% Note that in the case, $x_{\threeindex}\KED = 0$
% due to the complementary slackness.
% and for any index $\wi$ such that $\comsumption_{\twoindex}\KMINUSED > 0$
% we have $\wi \geq \taupp$.
\begin{align*}
    &\gamma_{\threeindex} + \sum_{s = 1}^K \Delta\beta_{\driver}\sed
    % - 
    % \cratio{K}\cdot \weight_{\threeindex}  
    \\
    \overset{(a)}{=}~&
    \mu_{\threeindex}\KED + 
    \left(
   \thetai\KED 
    -
    \sum_{\wi \geq \taup}
    \comsumption_{\twoindex}\KMINUSED \cdot \thetai\KED
    - 
    \sum_{\wi < \taup} 
    \comsumption_{\twoindex}\KMINUSED \cdot \weightHat_{\wi}
    \right)
    +
    \sum_{s = 1}^{K-1} \Delta\beta_{\driver}\sed
    % - 
    % \cratio{K}\cdot \weight_{\threeindex}  
    \\
    =~&
    \mu_{\threeindex}\KED + 
    \left(
   \thetai\KED 
    -
    \sum_{\wi \geq \taup}
    \comsumption_{\twoindex}\KMINUSED \cdot \thetai\KED
    - 
    \sum_{\wi =\taupp}^{\taup - 1} 
    \comsumption_{\twoindex}\KMINUSED \cdot \weightHat_{\wi}
    - 
    \sum_{\wi < \taupp} 
    \comsumption_{\twoindex}\KMINUSED \cdot \weightHat_{\wi}
    \right)
    +
    \sum_{s = 1}^{K-1} \Delta\beta_{\driver}\sed
    % - 
    % \cratio{K}\cdot \weight_{\threeindex} 
    \\
    \overset{(b)}{\geq}~&
    \mu_{\threeindex}\KED + 
    \left(
   \thetai\KED 
    -
    \sum_{\wi \geq \taup}
    \comsumption_{\twoindex}\KMINUSED \cdot \thetai\KED
    - 
    \sum_{\wi =\taupp}^{\taup - 1} 
    \comsumption_{\twoindex}\KMINUSED \cdot \thetai\KED
    - 
    \sum_{\wi < \taupp} 
    \comsumption_{\twoindex}\KMINUSED \cdot \weightHat_{\wi}
    \right)
    +
    \sum_{s = 1}^{K-1} \Delta\beta_{\driver}\sed
    % - 
    % \cratio{K}\cdot \weight_{\threeindex} 
    \\
    \overset{}{\geq}~&
    \left(
    1 - \sum_{\wi\geq\taupp}\comsumption_{\twoindex}\KMINUSED
    \right)
    \cdot \left(
    \mu_{\threeindex}\KED + 
   \thetai\KED 
   \right)
    - 
    \sum_{\wi < \taupp} 
    \comsumption_{\twoindex}\KMINUSED \cdot \weightHat_{\wi}
    +
    \sum_{s = 1}^{K-1} \Delta\beta_{\driver}\sed
    % - 
    % \cratio{K}\cdot \weight_{\threeindex} 
    \\
    \overset{(c)}{\geq}~&
    \left(
    1 - \sum_{\wi\geq\taupp}\comsumption_{\twoindex}\KMINUSED
    \right)
    \cdot \weight_{\threeindex}
    - 
    \sum_{\wi < \taupp} 
    \comsumption_{\twoindex}\KMINUSED \cdot \weightHat_{\wi}
    +
    \sum_{s = 1}^{K-1} \Delta\beta_{\driver}\sed
    % - 
    % \cratio{K}\cdot \weight_{\threeindex} 
    \\
    =~&
    \left(
    1 - \cumcomsumption_{\driver,\taupp}\KMINUSED
    \right)
    \cdot \weightHat_{\taupp}
    - 
    \sum_{\wi < \taupp} 
    \comsumption_{\twoindex}\KMINUSED \cdot \weightHat_{\wi}
    +
    \sum_{s = 1}^{K-1} \Delta\beta_{\driver}\sed
    % - 
    % \cratio{K}\cdot \weight_{\threeindex} 
    \\
    =~&
    \sum_{\wi\leq \taupp} 
    \left(\weightHat_{\wi} - \weightHat_{\wi-1}\right)
    \cdot 
    \left(
    1 - \cumcomsumption_{\twoindex}\KMINUSED
    \right)
    +
    \sum_{s = 1}^{K-1} \Delta\beta_{\driver}\sed
    \\
    \overset{(d)}{=}~&
    \sum_{\wi\leq \taupp} 
    \left(\weightHat_{\wi} - \weightHat_{\wi-1}\right)
    \cdot 
    \left(
    1 - f_{K-1}\left(\cumcomsumption_{\twoindex}\KMINUSED\right)
    \right)
    +
    \sum_{s = 1}^{K-1} \Delta\beta_{\driver}\sed
    \\
    \overset{(e)}{\geq}~&
    \cratio{K}\cdot \weightHat_{\taupp}
    =
    \cratio{K}\cdot \weight_{\threeindex}
\end{align*}
where (a) uses the construction of $\gamma_{\threeindex}$
and equality~\eqref{eq:beta K reformulate};
(b) uses the fact that $\weightHat_{\wi} \leq \thetai\KED$
for every $\wi\in[\taupp:\taup - 1]$;
(c) uses the fact that 
$\mu_{\threeindex}\KED + \thetai\KED \geq \weight_{\threeindex}$;
(d) uses the fact that $f_{K-1}(a) = a$ for all $a \in[0,1]$;
and 
(e) uses the similar argument as the one for stage $k\in [K-1]$.

Suppose $\weight_{\threeindex} \geq \thetai\KED$. 
Let $\taupp$ be the index such that $\weightHat_\taupp = \weight_{\threeindex}$.
By definition, $\taupp \geq \taup$.
\begin{align*}
    &\gamma_{\threeindex} + \sum_{s = 1}^K \Delta\beta_{\driver}\sed
    % - 
    % \cratio{K}\cdot \weight_{\threeindex}  
    \\
    \overset{}{=}~&
    \mu_{\threeindex}\KED + 
    \left(
   \thetai\KED 
    -
    \sum_{\wi \geq \taup}
    \comsumption_{\twoindex}\KMINUSED \cdot \thetai\KED
    - 
    \sum_{\wi < \taup} 
    \comsumption_{\twoindex}\KMINUSED \cdot \weightHat_{\wi}
    \right)
    +
    \sum_{s = 1}^{K-1} \Delta\beta_{\driver}\sed
    % - 
    % \cratio{K}\cdot \weight_{\threeindex}  
    \\
    =~&
    \mu_{\threeindex}\KED + 
    \left(
   \thetai\KED 
    -
    \sum_{\wi \geq \taupp}
    \comsumption_{\twoindex}\KMINUSED \cdot \thetai\KED
    - 
    \sum_{\wi =\taup}^{\taupp - 1} 
    \comsumption_{\twoindex}\KMINUSED \cdot \thetai\KED
    - 
    \sum_{\wi < \taup} 
    \comsumption_{\twoindex}\KMINUSED \cdot \weightHat_{\wi}
    \right)
    +
    \sum_{s = 1}^{K-1} \Delta\beta_{\driver}\sed
    % - 
    % \cratio{K}\cdot \weight_{\threeindex} 
    \\
    \overset{(a)}{\geq}~&
    \mu_{\threeindex}\KED + 
    \left(
   \thetai\KED 
    -
    \sum_{\wi \geq \taupp}
    \comsumption_{\twoindex}\KMINUSED \cdot \thetai\KED
    - 
    \sum_{\wi =\taup}^{\taupp - 1} 
    \comsumption_{\twoindex}\KMINUSED \cdot \weightHat_{\wi}
    - 
    \sum_{\wi < \taup} 
    \comsumption_{\twoindex}\KMINUSED \cdot \weightHat_{\wi}
    \right)
    +
    \sum_{s = 1}^{K-1} \Delta\beta_{\driver}\sed
    % - 
    % \cratio{K}\cdot \weight_{\threeindex} 
    \\
    \overset{}{\geq}~&
    \left(
    1 - \sum_{\wi\geq\taupp}\comsumption_{\twoindex}\KMINUSED
    \right)
    \cdot \left(
    \mu_{\threeindex}\KED + 
   \thetai\KED 
   \right)
    - 
    \sum_{\wi < \taupp} 
    \comsumption_{\twoindex}\KMINUSED \cdot \weightHat_{\wi}
    +
    \sum_{s = 1}^{K-1} \Delta\beta_{\driver}\sed
    % - 
    % \cratio{K}\cdot \weight_{\threeindex} 
    % \\
    % \overset{(c)}{\geq}~&
    % \left(
    % 1 - \sum_{\wi\geq\taupp}\comsumption_{\twoindex}\KMINUSED
    % \right)
    % \cdot \weight_{\threeindex}
    % - 
    % \sum_{\wi < \taupp} 
    % \comsumption_{\twoindex}\KMINUSED \cdot \weightHat_{\wi}
    % +
    % \sum_{s = 1}^{K-1} \Delta\beta_{\driver}\sed
    % % - 
    % % \cratio{K}\cdot \weight_{\threeindex} 
    % \\
    % =~&
    % \left(
    % 1 - \cumcomsumption_{\driver,\taupp}\KMINUSED
    % \right)
    % \cdot \weightHat_{\taupp}
    % - 
    % \sum_{\wi < \taupp} 
    % \comsumption_{\twoindex}\KMINUSED \cdot \weightHat_{\wi}
    % +
    % \sum_{s = 1}^{K-1} \Delta\beta_{\driver}\sed
    % % - 
    % % \cratio{K}\cdot \weight_{\threeindex} 
    % \\
    % =~&
    % \sum_{\wi\leq \taupp} 
    % \left(\weightHat_{\wi} - \weightHat_{\wi-1}\right)
    % \cdot 
    % \left(
    % 1 - \cumcomsumption_{\twoindex}\KMINUSED
    % \right)
    % +
    % \sum_{s = 1}^{K-1} \Delta\beta_{\driver}\sed
    % \\
    % \overset{(d)}{\geq}~&
    % \sum_{\wi\leq \taupp} 
    % \left(\weightHat_{\wi} - \weightHat_{\wi-1}\right)
    % \cdot 
    % \left(
    % 1 - f_{K-1}\left(\cumcomsumption_{\twoindex}\KMINUSED\right)
    % \right)
    % +
    % \sum_{s = 1}^{K-1} \Delta\beta_{\driver}\sed
    % \\
    % \overset{(e)}{\geq}~&
    % \cratio{K}\cdot \weightHat_{\taupp}
    % =
    % \cratio{K}\cdot \weight_{\threeindex}
\end{align*}
where (a) uses the fact that $\weightHat_{\wi}\geq \thetai\KED$
for every $\wi\in[\taup:\taupp-1]$,
and the remaining argument is identical to the subcase for $\weight_{\threeindex} < \thetai\KED$ which we omit here.
\hfill\halmos}

\ydrevcolor{
\section{Numerical Experiments}
\label{apx:numerical}
In this section we provide 
numerical justifications for the performance of 
our multi-stage configuration allocation algorithm 
on synthetic instances. 
Our instances are motivated by the display-ad
application.
We compare the performance of our algorithm with both the batched-greedy
algorithm (that at each stage 
greedily picks the best local allocation) 
and the online-greedy
algorithm (that picks an ordering over vertices 
in each batch and greedily allocates), 
as well the optimum offline. 
We observe that the competitive ratio of 
our algorithm beats the
other algorithms in our simulations.

\newcommand{\wtp}{v}
\newcommand{\wtpij}{\wtp_{ij}}
\newcommand{\OnlineGR}{\texttt{Online-GR}}
\newcommand{\BatchedGR}{\texttt{Batched-GR}}

\paragraph{Experimental setup.}
We generate the following configuration allocation
problem instances.
There are 20 offline advertisers (i.e., $\Driver \triangleq [20]$)
and 100 arriving users 
(i.e., $\Rider \triangleq [100]$).\footnote{We also
ran our experiments by varying all parameters in our 
synthetic instances. We obtained similar results and verified
the robustness of our numerical findings.}
Each advertiser $j$ demands $n_j \triangleq 5$ 
number of impressions in total.
For each user $i\in\Rider$ and advertiser $j\in\Driver$,
we denote $\wtpij$ by the willingness-to-pay of advertiser $j$
for an impression from user $i$ (see the paragraph below about how 
we generate $\{\wtpij\}$).
For each user $i\in\Rider$, the platform needs to choose 
an advertising configuration $\config$
which specifies a subset of 
advertisers $\Driver_\config\subseteq\Driver$ 
to participate 
in a second-price auction for the arriving impression from user $i$:
the winning advertiser $j^* = \argmax_{j\in\Driver_\config}\wtpij$
with the highest willingness-to-pay 
receives the impression from user $i$,
and pays the per-impression price
equal to the second highest willingness-to-pay among participating advertisers,
i.e., $\alloci \triangleq \indicator{j = j^*}$ and 
$\weight_{\threeindex} \triangleq \wtp_{ij\primed}$
where $j\primed = \argmax_{j\in\Driver_\config\backslash \{j^*\}} 
\wtpij$.

We generate the willingness-to-pay quantities $\{\wtpij\}$ as follows.
First, we draw $20$ $(\triangleq T)$ i.i.d.\ samples from 
the exponential distribution,
and denote them as $\weightHat_1 \leq \dots \leq \weightHat_{T}$.
Next, we draw $20$ i.i.d.\ samples between $[1, 2]$
uniformly at random, and denote them as $\mu_1 \leq \dots \leq \mu_{20}$. 
For each user $i\in\Rider$ and advertiser $j\in\Driver$,
we set the willingness-to-pay $\wtpij \triangleq \weightHat_{\tau}$
where $\tau = \lceil\tau_{i} + \tau_{ij}\cdot \mu_j\rceil$ 
and 
both
$\tau_i$, $\tau_{ij}$ are drawn i.i.d.\ from $[0,\frac{T}{3}]$
uniformly at random.
Note that from the ex ante perspective, 
the willingness-to-pay $\{\wtpij\}$'s are 
symmetric over all users for each fixed advertiser $j$,
and increasing (i.e., stochastic dominant) with respect to advertiser $j$ (since $\mu_j$ is increasing in $j$) for each fixed user $i$.

We consider both scenarios where arriving users are ex ante symmetric 
and ex ante asymmetric, respectively.
In the symmetric-user (resp.\ asymmetric-user) scenario, 
for each user $i\in\Rider$,
we randomly generate $5$ feasible configurations, 
each of which corresponds to a second-price auction 
with 3 advertisers drawn uniformly at random from $\Driver$
(resp.\ $\Driver \backslash [\lfloor \frac{17\cdot i}{100} \rfloor]$).
We assume that there are $K$ stages, where
each stage $k\in[K]$
contains a batch of users $\Rider_k \triangleq [(k - 1)\cdot \lceil\frac{100}{K}\rceil + 1: \min\{100, k\cdot \lceil\frac{100}{K}\rceil\}]$.
Note that in the asymmetric-user scenario, 
the advertisers who can participate in the auction for user $i$
become more restrictive as $i$ increases.
In this way, policies have incentives to carefully 
preserve enough demands for later users.

\paragraph{Policies.} In our numerical experiments, we compare the expected revenue of four different policies/benchmark:
\begin{enumerate}
    \item \emph{Online greedy policy} ($\OnlineGR$):
    greedily decide the fractional allocation for each user $i = 1, 
    2, \dots, 100$.
    for each user $i = 1, 2, \dots, 100$,
    given the current allocation, $\OnlineGR$
    greedily finds the fractional allocation which maximizes the immediate revenue improvement for user $i$.
    
    \item \emph{Batched greedy policy} ($\BatchedGR$):
    for each stage $k\in[K]$,
    given the current allocation, $\BatchedGR$
    greedily finds the fractional allocation which maximizes the 
    immediate revenue improvement for users in batch~$\Rider_k$.
    
    \item \emph{Polynomial regularized max configuration allocation}
    ($\PRMCA$): this policy is \Cref{alg:opt} designed in
    \Cref{sec:extensions}.
    It attains the optimal competitive ratio $\Gamma(K)$.
    We numerically solve convex program \ref{eq:convex primal}
    in $\PRMCA$
    with the Frank–Wolfe algorithm (100 iterations per convex program,
    \citealp{FW-56}).
    
    \item \emph{Optimum offline solution}: this benchmark is
    the solution of linear program~\ref{eq:LP primal}
    in \Cref{sec:extension-model}. It upperbounds the expected revenue of
    any multi-stage allocation algorithms.
\end{enumerate}

\paragraph{Results.}
First, we randomly generate
100 configuration allocation problem instances 
for both user-symmetric and user-asymmetric scenarios 
with $K = 2$ and $K=5$ stages. 
For each instance, we simulate all three policies defined above,
and compute the ratio between the revenue of each policy
against the revenue of the optimum offline solution.
We summarize 
% the (unweighted) average of 
the competitive ratios for each policy 
in \Cref{table:numerical ratio different policies} and \Cref{fig:numerical ratio different policies}.
It can be observed that both $\BatchedGR$
and $\PRMCA$ which utilize the batch arrival 
outperform $\OnlineGR$, and the ratio between 
better from $K = 5$ batches to $K = 2$ batches.
Furthermore, $\PRMCA$ which hedges the batch-wise allocation
outperforms $\BatchedGR$ which greedily decides the
batch-wise allocation.
Specifically, by switching from $\OnlineGR$ (resp.\ $\BatchedGR$) to $\PRMCA$,
there are strictly more than 5\% revenue improvement 
in all scenarios.
\begin{table}[t]
    \centering
    \caption{The average competitive ratios for 
    different policies.}
    \label{table:numerical ratio different policies}
    \begin{tabular}[t]{lcccc}
    \toprule
    & \multicolumn{2}{c}{user-symmetric} & \multicolumn{2}{c}{user-asymmetric} \\
    \cmidrule{2-5}
     & $K = 2$ & $K = 5$ 
     & $K = 2$ & $K = 5$ 
     \\
     \midrule
     \OnlineGR & 0.913 & 0.913 & 0.849 & 0.849 \\
     \BatchedGR & 0.947 & 0.919 & 0.877 & 0.860 \\
     \PRMCA & 0.982 & 0.961 & 0.937 & 0.903 \\
    \bottomrule
    \end{tabular}
\end{table}

\begin{figure}[ht]
  \centering
       \subfloat[user-symmetric, $K = 2$]
      {\includegraphics[width=0.4\textwidth]{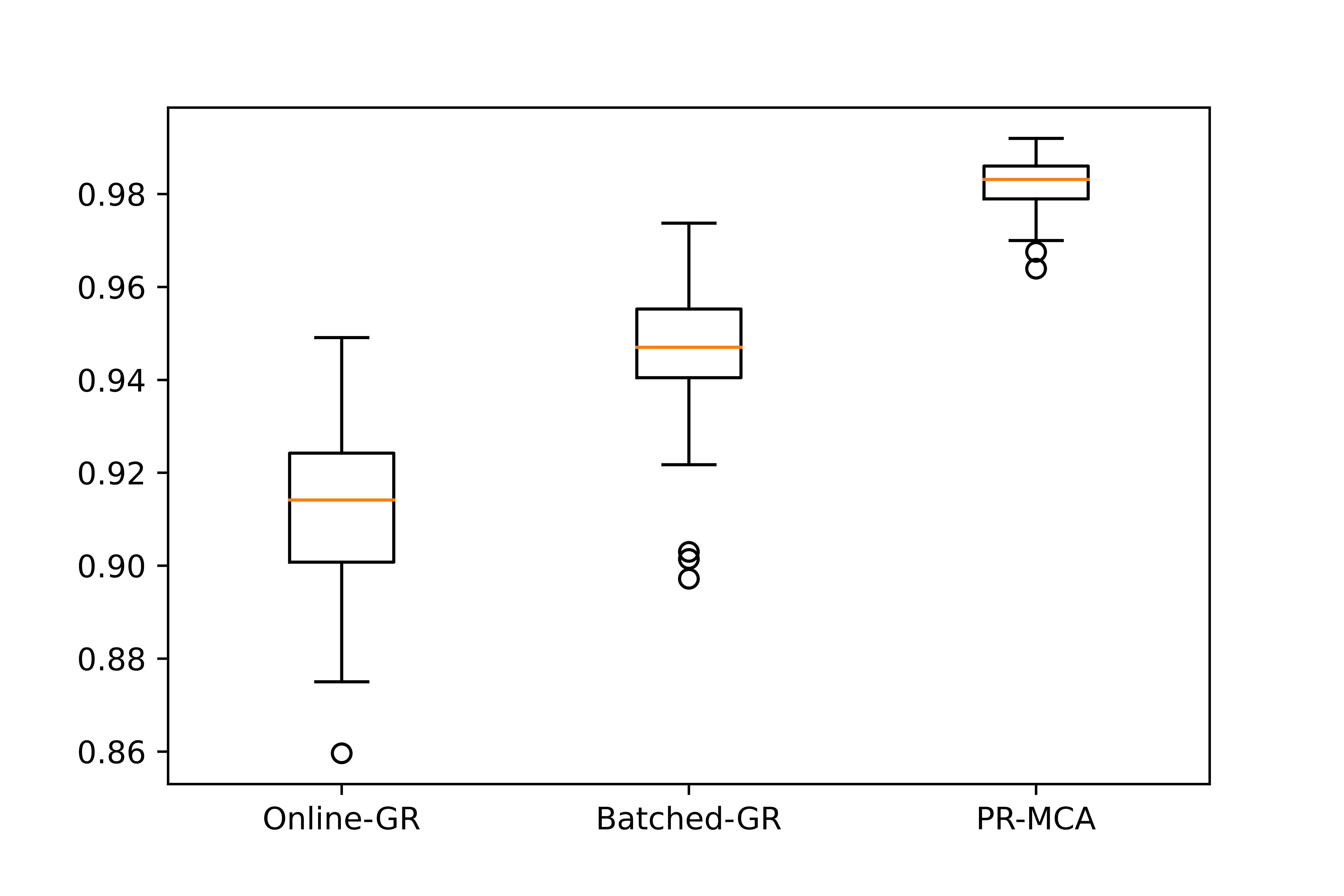}}
      \subfloat[user-symmetric, $K = 5$]
      {\includegraphics[width=0.4\textwidth]{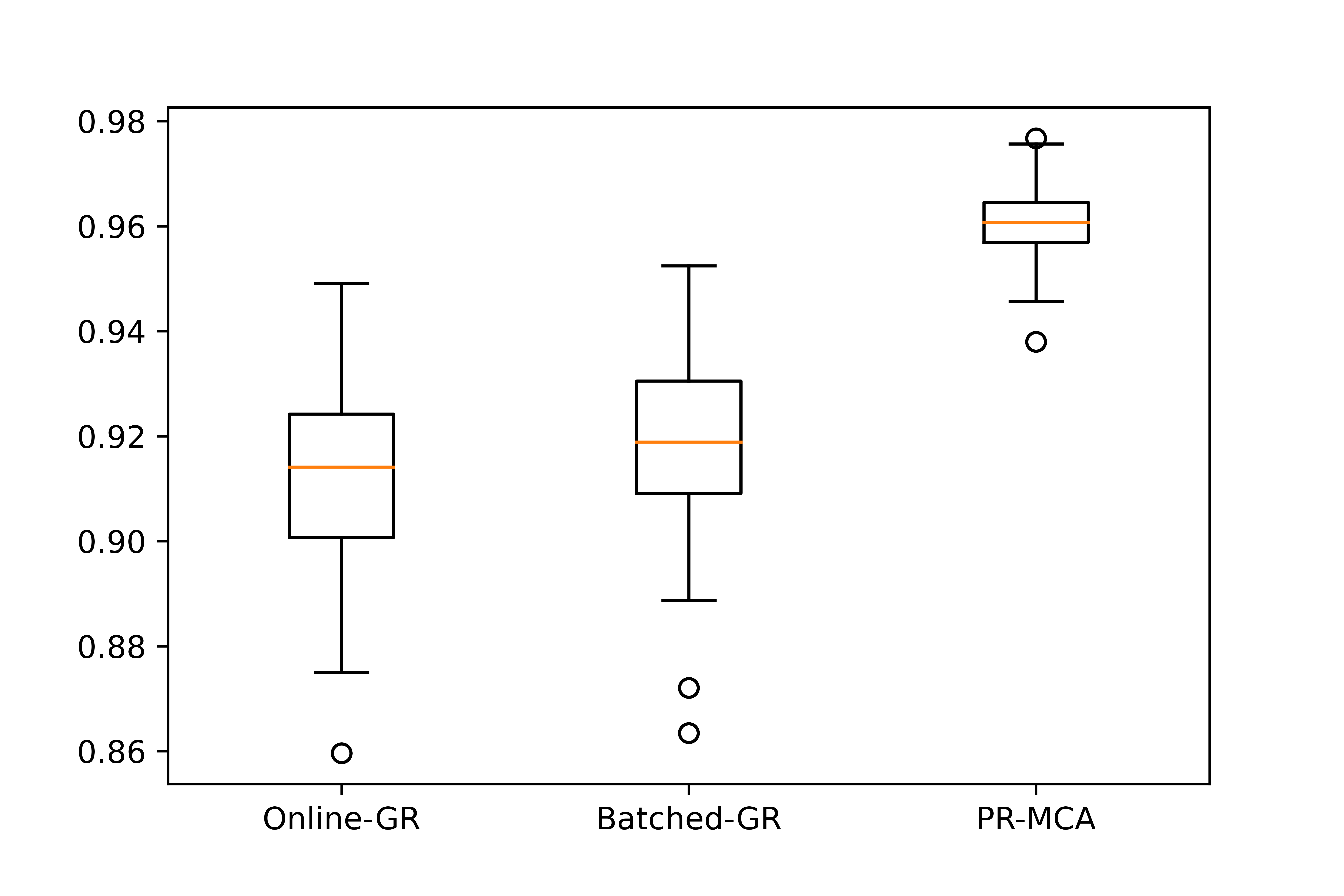}}
      \\
       \subfloat[user-asymmetric, $K = 2$]
      {\includegraphics[width=0.4\textwidth]{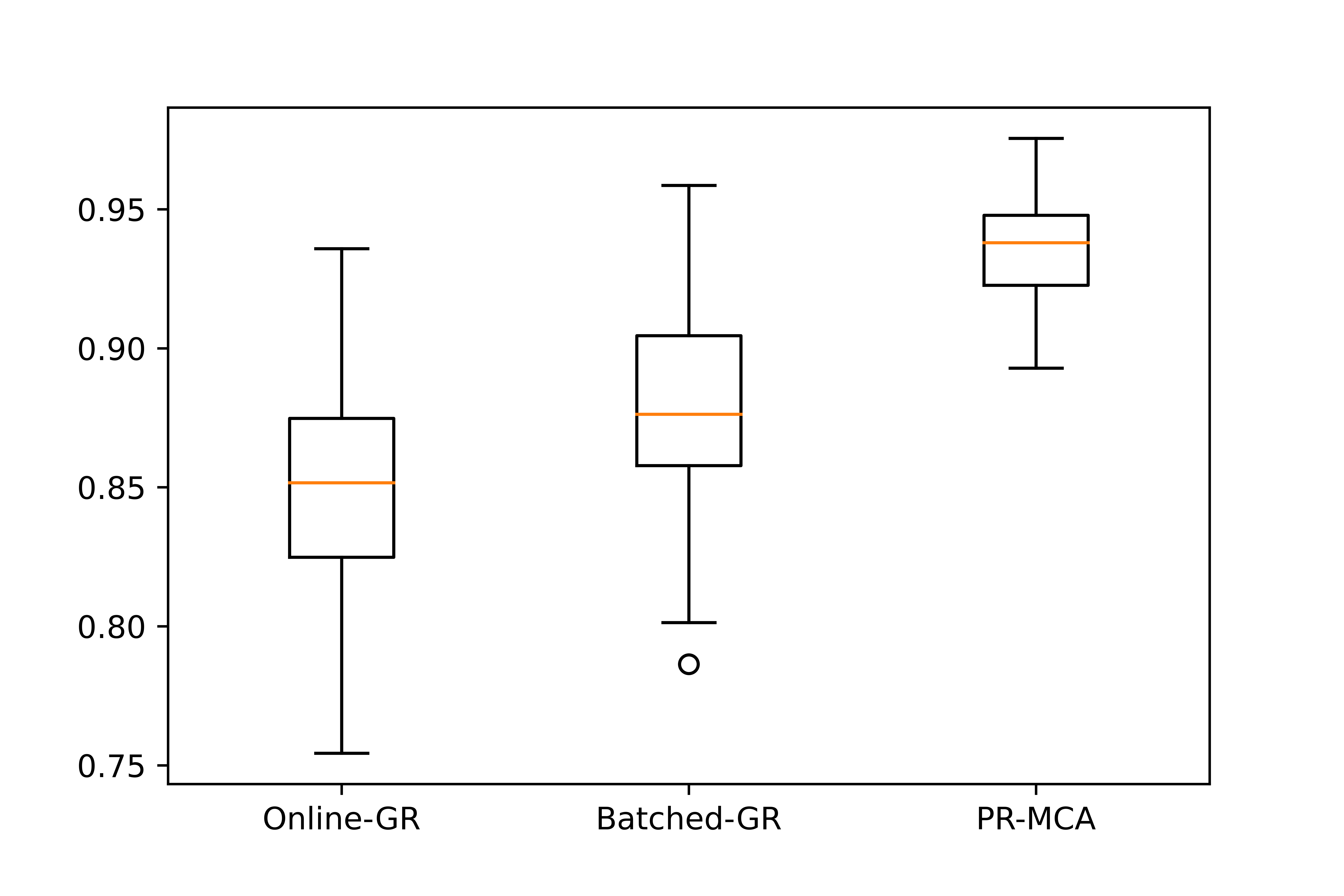}}
      \subfloat[user-asymmetric, $K = 5$]
      {\includegraphics[width=0.4\textwidth]{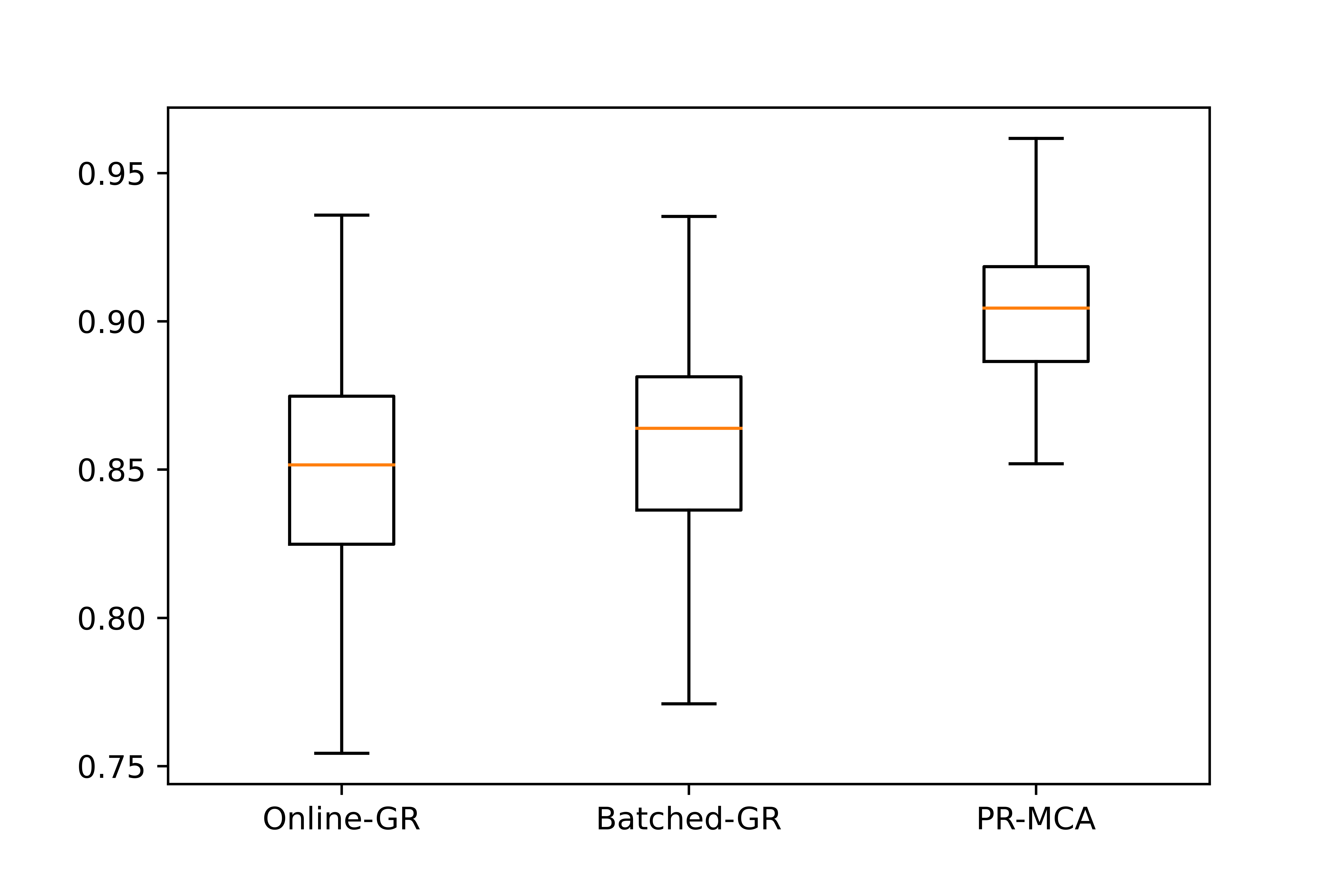}}
      \\
  \caption{Box and whisker comparison of different policies in terms of the
  competitive ratios. Results are based on 100 iterations of Monte-Carlo simulation.
  }
   \label{fig:numerical ratio different policies}
\end{figure}

To illustrate the the benefit of batching, 
we also plots the average 
(over 100 configuration allocation problem instances) 
of the ratios for $\PRMCA$
and $\BatchedGR$ for $K = 1, \dots, 10$ for both 
user-symmetric and user-asymmetric scenario
in \Cref{fig:numerical vary K}. 
It can be observed that the average ratio is decreasing as 
the total number of stages $K$ increases, and 
$\PRMCA$ outperforms $\BatchedGR$ in both scenarios over all $K$. 

\begin{figure}[ht]
  \centering
       \subfloat[user-symmetric]
      {\includegraphics[width=0.4\textwidth]{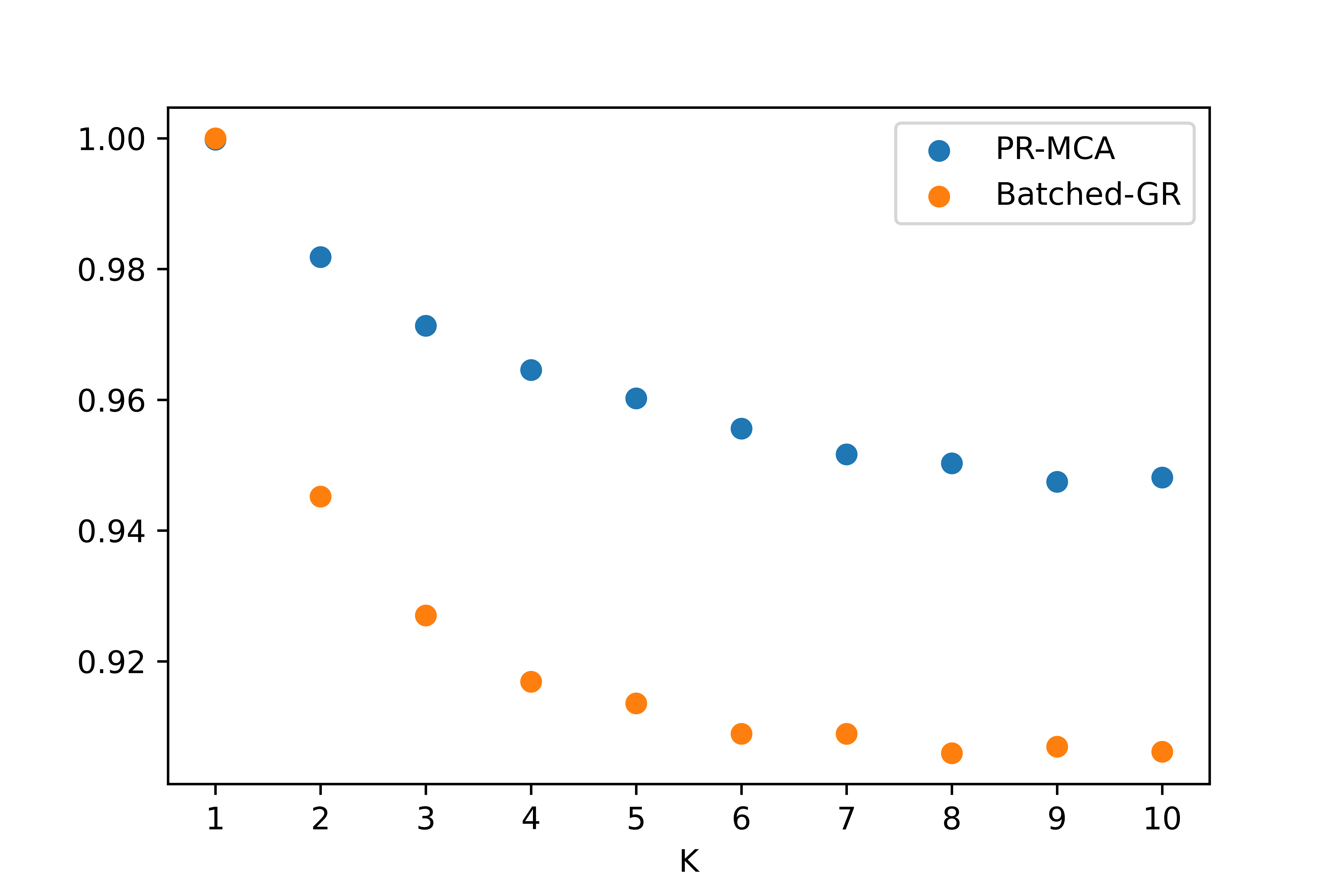}}
      \subfloat[user-asymmetric]
      {\includegraphics[width=0.4\textwidth]{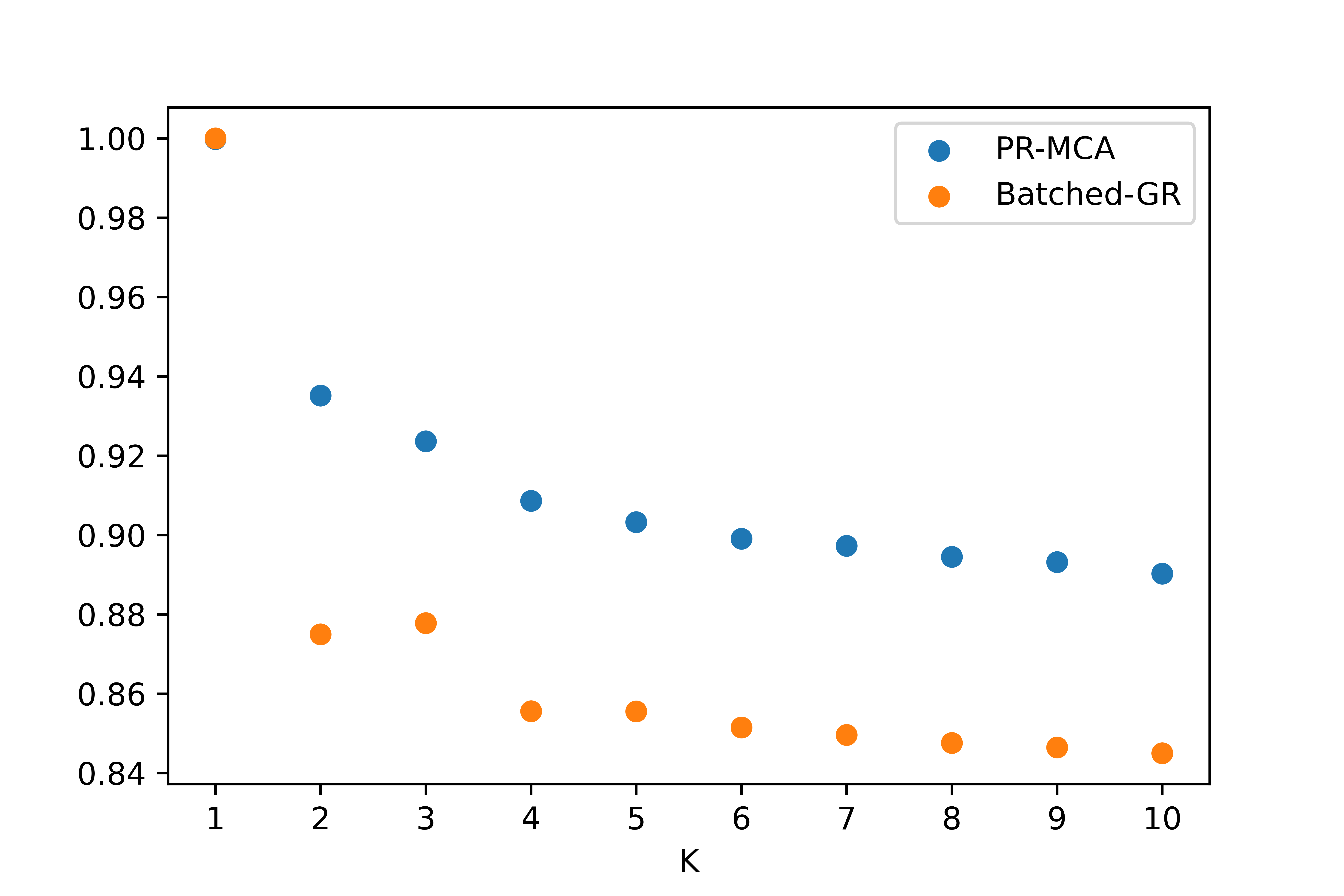}}
  \caption{The average competitive ratios for 
  $\PRMCA$ and
  $\BatchedGR$ as functions of the total number of stages $K$.
  }
   \label{fig:numerical vary K}
\end{figure}

}

\section{Future Directions and Open Problems}
\label{apx:open}
Several open questions arise from this work. 
The first question is to identify the optimal multi-stage algorithm and its competitive ratio
for the multi-stage \emph{integral} bipartite matching problem (i.e., the multi-stage version of the classic problem studied in \citealp{KVV-90}).  Showing whether our competitive ratio can be obtained in the integral matching problem for $K>2$ is still an open problem. 
\revcolor{Second, inspired by the importance of batched allocations in applications such as ride-sharing and cloud computing, one might want to study the multi-stage resource allocation where the resources are \emph{reusable}, similar to some recent work on online allocation of reusable resources \citep{FNS-19, feng2021online, goyal2020onlineb,gong2019online}. In such a setting, a resource after being assigned to a demand request will be available again for reassignment after a certain time durations, which can be deterministic (and fixed over time) or i.i.d. over time. Note that in such a problem formulation, resource usage durations are described in the number of stages it takes for a resource to be available again. Third, our multi-stage matching algorithms assume knowing the number of stages $K$ in advance. One can now investigate the question of having access to a predictor for $K$ and asks the question of obtaining robustness and consistency bounds (in the same style as in the learning-augmented online algorithm literature).}
Fourth, one can consider applying our developed machinery and framework in this paper to \revcolor{multi-stage assortment optimization~\citep{GNR-14},} multi-stage stochastic matching~\citep{MGS-12,JL-14}, or  multi-stage matching with stochastic rewards~\citep{mehta2012online}. \revcolor{We conjecture that our approach can be extended to all of these settings and help with improved competitive ratios. We leave further investigation around these directions and the possibility of extending our approach as open problems.}

\revcolor{Finally, 
it is also worthwhile to study
the benefit of batching under 
environments with additional structure.
Note that the optimal competitive ratio $\Gamma(K)$
converges to $1 - \frac{1}{e}$ at a linear rate as $K$ converges to infinity, i.e., $\Gamma(K)=1-\frac{1}{e}+O(\frac{1}{K})$. In other words, the benefit of batching 
decays quite fast in the worst-case analysis. However, this fast decay might be an artifact of the model and the fact that designer aims to have a robust multi-stage allocation algorithms that performs well under any problem instance.
To resolve this mystery, one can consider more refined models with additional (non-)stochastic structures,
and identify algorithms that benefit from batching, but this time with the goal of highlighting the benefit of the batching in a regime where the number of batches is super constant, but not very large. In such a model, another possible variable that can play a role is the average size of a batch. We leave all these investigations to future work.}

\end{APPENDIX}
\end{document}